\documentclass[12pt,preprint]{aastex}

\shorttitle{Compact Radio Sources in NGC 4945}
\shortauthors{Lenc Tingay}

\newcommand\rah{\mbox{$^{\mathrm h}$}}%
\newcommand\ram{\mbox{$^{\mathrm m}$}}%

\begin{document}
%
%
\title{The Sub-parsec Scale Radio Properties of Southern Starburst Galaxies.
II. Supernova Remnants, the Supernova Rate, and the Ionised Medium in the NGC 4945 Starburst}
%


\author{E. Lenc\altaffilmark{1}}

\affil{Australia Telescope National Facility, CSIRO, PO Box 76, Epping, NSW 1710, Australia}
\email{Emil.Lenc@csiro.au}

\author{S.J. Tingay}
\affil{Department of Imaging and Applied Physics, Curtin University of Technology, Bentley 6845, Western Australia, Australia}
\email{s.tingay@curtin.edu.au}

\altaffiltext{1}{Centre for Astrophysics and Supercomputing, Swinburne University of Technology,
Mail number H39, P.O. Box 218, Hawthorn, Victoria 3122}
%
\begin{abstract}
Wide-field, very long baseline interferometry (VLBI) observations of the nearby starburst galaxy NGC 4945, obtained with the Australian Long Baseline Array (LBA), have produced 2.3 GHz images over two epochs with a maximum angular resolution of 15 mas (0.3 pc). 15 sources were detected, 13 of which correspond to sources identified in higher frequency (3 cm and 12 mm) ATCA images. Four of the sources are resolved into shell-like structures ranging between 60 and 110 mas (1.1 to 2.1 pc) in diameter. From these data the spectra of 13 compact radio sources in NGC 4945 were modelled; nine were found to be consistent with free-free absorbed power laws and four with a simple power law spectrum. The free-free opacity is highest toward the nucleus but varies significantly throughout the nuclear region ($\tau_0\sim 6-23$), implying that the overall structure of the ionised medium is clumpy. Of the 13 sources, 10 have steep intrinsic spectra associated with synchrotron emission from supernova remnants, the remaining sources have flat intrinsic spectra which may be associated with thermal radio emission. A non-thermal source with a jet-like morphology is detected $\sim1\arcsec$ from the assumed location of the AGN. A type II supernova rate upper limit of 15.3 yr$^{-1}$ is determined for the inner 250 pc region of the galaxy at the 95\% confidence level, based on the lack of detection of new sources in observations spanning 1.9 years and a simple model for the evolution of supernova remnants. A type II supernova rate of $>0.1 (v/ 10^{4})$ yr$^{-1}$ is implied from estimates of supernova remnant source counts, sizes and expansion rates, where $v$ is the radial expansion velocity of the supernova remnant in km s$^{-1}$. A star formation rate of $2.4 (v/10^{4}) < SFR(M\geq5M_{\Sun})<370$ M$_{\Sun}$ yr$^{-1}$ has been estimated directly from the supernova rate limits and is of the same order of magnitude as rates determined from integrated FIR (1.5 M$_{\Sun}$ yr$^{-1}$) and radio luminosities ($14.4\pm1.4$ $(Q/8.8)$ M$_{\Sun}$ yr$^{-1}$). The supernova rates and star formation rates determined for NGC 4945 are comparable to those of NGC 253 and M82.
\end{abstract}

\keywords{galaxies: active -- galaxies: individual (NGC 4945) -- galaxies: starburst -- radiation mechanisms:general -- supernovae: general -- techniques: interferometric}

\section{Introduction}
\label{sec:introduction}

This paper is the second of a series of papers on the sub-parsec scale properties of local ($D<10$ Mpc), bright ($S_{1.4}>10$ mJy) southern ($\delta<-20\arcdeg$) starburst galaxies. Starburst galaxies are the result of processes acting on a wide range of scales and themselves drive energetic activity on a wide range of scales. In Paper I \citep{Lenc:2006p6695} of this series we discussed the properties of the nuclear starburst region of NGC 253. Direct observations of supernova remnants enabled an investigation of the star-formation and supernova history. Furthermore, high-resolution radio observations were used to map the ionized gaseous environment of the starburst region. These observations provide a link between the large-scale dynamical effects that feed the starburst region, the activity in the starburst region, and the energetic phenomenon that in turn is driven by the starburst. In the present work (Paper II of the series) we discuss the sub-parsec scale properties of NGC 4945 and implications on the star-formation and supernova rate of this galaxy.

NGC 4945 is a nearly edge-on spiral galaxy associated with the Centaurus group of galaxies at a declination of $\sim-40\arcdeg$ \citep{Webster:1979p9704}. The galaxy has been classified as SB(s)cd or SAB(s)cd and has an optical extent of 17 arcmin \citep{deVaucouleurs:1964p10569, Braatz:1997p10531}. Distance estimates for NGC 4945 have varied between 3.7 \citep{Mauersberger:1996p7843} and 8.1 Mpc \citep{Baan:1985p10435}. More recently, \citet{Karachentsev:2007p7477} have determined the magnitude of stars at the tip of the red giant branch using \emph{Hubble Space Telescope} (\emph{HST}) Advanced Camera for Surveys (ACS) images and arrived at a more reliable distance estimate of $3.82\pm0.31$ Mpc. This new distance estimate will be adopted throughout this paper and implies a spatial scale of $\sim19$ pc arcsec$^{-1}$.

While it is classified as a Seyfert 2 galaxy \citep{Braatz:1997p10531, Madejski:2000p9093}, it exhibits both starburst and Seyfert characteristics \citep[e.g.][]{Whiteoak:1986p9113}. The galaxy is one of the strongest and richest sources of extragalatic molecular lines with $\sim50$ line features detected to date \citep[e.g.][]{Henkel:1994p9082, Curran:2001p10543, Wang:2004p6109}. Evidence of an active galactic nucleus (AGN) comes from strong and variable hard x-ray emission \citep{Iwasawa:1993p6812}. H$_{2}$O megamaser emission has been observed from the nucleus \citep{Greenhill:1997p7509} and is assumed to be distributed in a $\sim3$ pc disk around the AGN.

Although the central region of NGC 4945 is highly obscured by gas and dust at optical wavelengths \citep{Madejski:2000p9093}, it is one of the three brightest IRAS point sources beyond the Magellanic Clouds \citep{Sanders:2003p7880}. Neutral hydrogen observations of NGC 4945 reveal the signature of a bar associated with the disc and can be traced out to a radius of $\sim7$ kpc \citep{Ott:2001p3998}. CO(2$-$1) observations suggest that the bar extends inwards to within $\sim100$ pc of the galaxy centre \citep{Chou:2007p10537}, where it may act to feed gas and dust into fast-rotating disk of cold gas \citep{Dahlem:1993p10544, Mauersberger:1996p7843}. HNC(1-0) maps of the disk reveal a dense inner region \citep{HuntCunningham:2005p10542} that may fuel a circumnuclear starburst ring of similar extent observed by \citet{Marconi:2000p7671} with \emph{HST} PA$\alpha$ images. \citet{Chou:2007p10537} also detect a smaller-scale ($r<27$ pc), high density, kinetically de-coupled component in the nuclear region and suggest that it may be evidence of a circumnuclear torus around the AGN \citep{Antonucci:1993p10522}. 

Associated with the nuclear region, and having its origin in the inner starburst region, is a powerful wind that forms a cone-shaped plume perpendicular to the disk of the galaxy. The plume, extending $\sim500$ pc north-west, is clearly observed in H$\alpha$ spectral line observations \citep{Rossa:2003p10550} and \emph{Chandra} soft x-ray images \citep{Schurch:2002p7434}. Such plumes have been found in other nearby starburst galaxies \citep[e.g.][]{Strickland:2004p9721} and are believed to be driven by supernovae in the starburst region \citep{Doane:1993p10547,Strickland:2004p1157}.

The bar associated with the disk of the galaxy, a fast-rotating disk of cold gas, a starburst ring and the conical nuclear outflow are all characteristics that are observed in other nearby starburst galaxies such as NGC 253 \citep{Lenc:2006p6695} and M82 \citep{ForsterSchreiber:2000p8653}. These features are interpreted in terms of bar-induced gas dynamics and star formation. Even though NGC 4945, NGC 253 and M82 are associated with different galaxy groups and are situated in different locations in the celestial sphere, they share several further similarities. All three galaxies are nearly edge-on, all have distances close to 4 Mpc and all have similar infrared luminosities \citep{Rice:1988p7793}. It is these similarities that make them particularly interesting targets for study and comparison in other respects.

\begin{figure}
\plotone{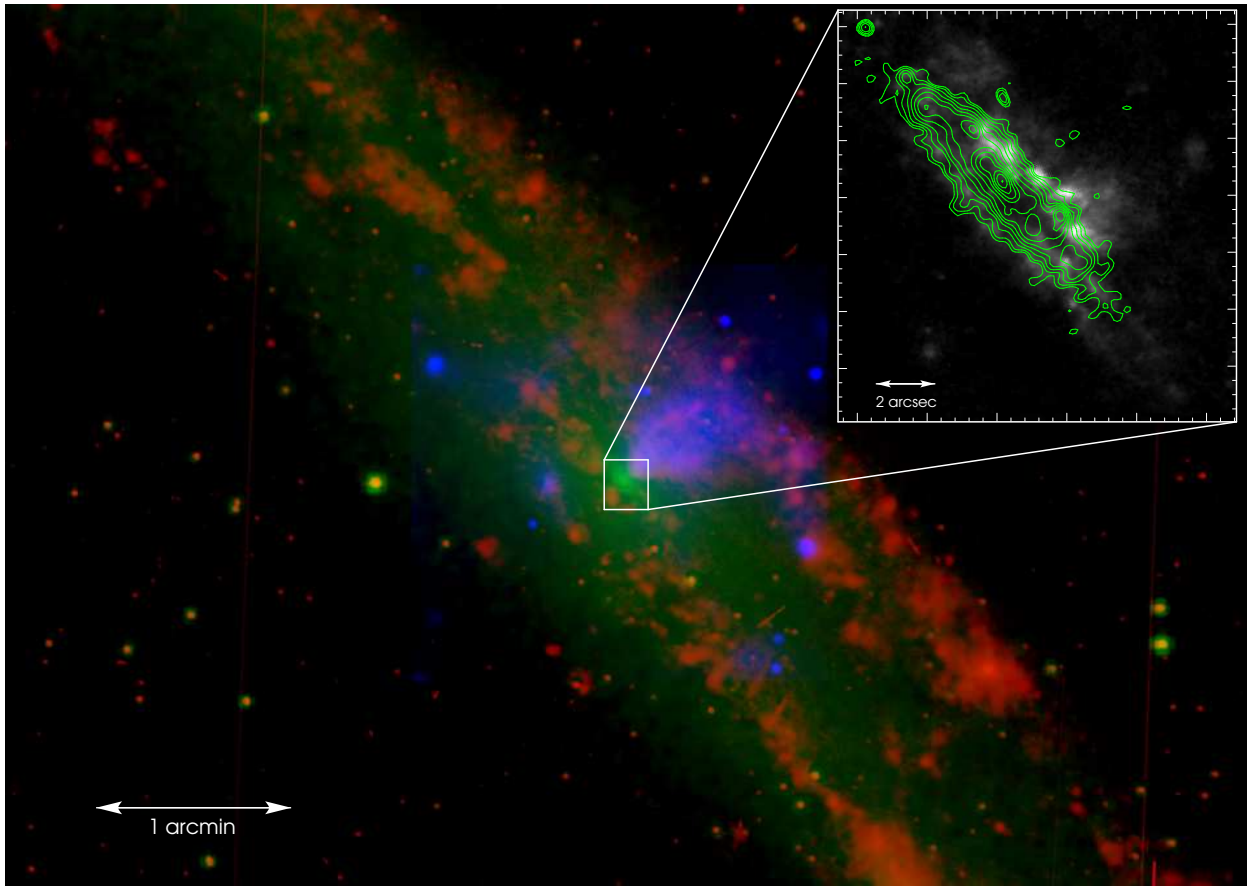}
\caption{A three-colour composite, multi-wavelength view of NGC 4945. Red, green and blue indicate H$\alpha$ \citep{Rossa:2003p10550}, K$-$band Two Micron All Sky Survey IR \citep{Jarrett:2003p8453}, and \emph{Chandra} soft x-ray \citep{Schurch:2002p7434} respectively. Inset: HST Pa$\alpha$ image \citep{Marconi:2000p7671} with ATCA 23 GHz contours overlaid, contours are logarithmic intervals of $2^{1/2}$, beginning at 1.02 mJy beam$^{-1}$.}
\label{fig:figmultiwav}            
\end{figure}

Figure \ref{fig:figmultiwav} presents images of NGC 4945 at various wavelengths, on different spatial scales, to illustrate the relationships between the distribution of cool stars (K-band), ionised hydrogen gas (H$\alpha$), the inner starburst region (radio and Pa$\alpha$), and the wind eminating from that region (H$\alpha$ and x-ray). The supernova remnants that are a primary subject of this paper are concentrated within the inner 250 pc of the galaxy. 

The observation of new supernovae and their associated remnants in the starburst regions of nearby galaxies is rare as these regions are highly obscured by gas and dust. Only a few have been observed in local starbursts and are generally revealed in x-rays where such events are more conspicuous and then followed up with multi-wavelength observations. In a few cases these can be traced back to an earlier optical or infrared outburst, the presumed supernova event, as with SN 1996cr in Circinus \citep{Bauer:2007p10525} and SN 1978K in NGC 1313 \citep{Smith:2007p7796, Ryder:1993p7804}. However, in most cases, where the remnants are embedded deep in the starburst region and only x-ray and radio observations can reveal their presence, they are discovered long after the event and their ages remain uncertain. Such a population of supernova remnants has been discovered in M82 \citep{Pedlar:1999p3534, McDonald:2002p3330} and NGC 253 \citep{Tingay:2004p778, Lenc:2006p6695} using VLBI observations.

Motivated by the existence of compact radio sources and strong indicators of free-free absorption in similar starburst galaxies, such as NGC 253 and M82, we have embarked on a program of VLBI observations of NGC 4945 and other prominant Southern Hemisphere starburst galaxies. This paper reports results from new observations of the nuclear region of NGC 4945 at a range of frequencies between 2.3 GHz and 23 GHz. The resulting data have better resolution and sensitivity than previous observations and we have used them to investigate in detail the free-free absorbed spectra of supernova remnants and \ion{H}{2} regions in the NGC 4945 starburst (\S~\ref{sec:p3ffmodel}). In particular we are interested in probing the structure of the ionised environment of the supernova remnants, constraining the supernova rate in NGC 4945 using our new ATCA and VLBI images (\S~\ref{sec:snrate}) and re-evaluating estimates of the star formation rate based on our results (\S~\ref{sec:sfrate}).

A subsidiary focus of this project is to achieve high sensitivity, high fidelity, high resolution, and computationally efficient imaging over wide fields of view (at least compared to traditional VLBI imaging techniques).  As such, the associated exploration of the relevant techniques is a small first step toward the much larger task of imaging with the next generation of large radio telescopes, for example the Square Kilometre Array (SKA: \citet{Hall:2005p10570, Carilli:2004p10534}).  The SKA will use baselines ranging up to approximately 3000 km and cover fields of view greater than one square degree instantaneously at frequencies of around 1 GHz.  The computational load required for this type of imaging task is vast and will drive the development of novel imaging and calibration algorithms and the use of super-computing facilities.  Investigating the performance of these techniques now, under the most challenging current observing conditions, is therefore a useful activity.

%
%
%
\section{Observations, correlation and data reduction}
\subsection{LBA Observations}
A VLBI observation of NGC 4945 was made on 13/14 May, 2005 using a number of the Long Baseline Array (LBA) telescopes: the 70 m NASA Deep Space Network (DSN) antenna at Tidbinbilla; the 64 m antenna of the Australia Telescope National Facility (ATNF) near Parkes; 5 $\times$ 22 m antennas of the ATNF Australia Telescope Compact Array (ATCA) near Narrabri used as a phased array; the ATNF Mopra 22 m antenna near Coonabarabran; the University of Tasmania's 26 m antenna near Hobart; and the University of Tasmania's 30 m antenna near Ceduna.  The observation utilised the S2 recording system \citep{Cannon:1997p10533} to record 2 $\times$ 16 MHz bands (digitally filtered 2-bit samples) in the frequency ranges: 2252 - 2268 MHz and 2268 - 2284 MHz.  Both bands were upper side band and right circular polarisation.

A second epoch LBA observation was made on 21 March, 2007 using the same set of radio telescopes except that only 3 $\times$ 22 m antennas of the ATCA were used as a phased array to maintain a wide field of view given the longer baseline configuration of the ATCA at that time. The second epoch observation utilised hard disk data recorders (Phillips et al. 2008, in preparation); allowing dual circular polarisation data across $4\times16$ MHz bands to be recorded at Parkes, ATCA, and Mopra, and $2\times16$ MHz bands at the remaining antennas. Observing parameters associated with each of the LBA observations are shown in Table \ref{tab:tabobs}. 


During each of the VLBI observations, three minute scans of NGC 4945 ($\alpha = 13\rah5\ram27\fs50$; $\delta = -49\arcdeg28\arcmin6\farcs00$ [J2000]) were scheduled, alternating with three minute scans of a nearby phase reference calibration source, J1237$-$5046 ($\alpha = 12\rah37\ram15\fs239286$; $\delta = -50\arcdeg46\arcmin23\farcs17260$ [J2000]).

\subsection{LBA correlation}
The first epoch data were correlated using the ATNF Long Baseline Array (LBA) processor at ATNF headquarters in Sydney \citep{Wilson:1992p9290}. The data were correlated using an integration time of 2 seconds and with 32 frequency channels across each 16 MHz band (channel widths of 0.5 MHz) to limit the effects of bandwidth smearing (a form of chromatic aberration) and time-averaging smearing \citep{Cotton:1999p19452,Bridle:1999p10564}. The ATCA primary beam limits the field of view of this observation to a half-width half-maximum (HWHM) of $\sim45\arcsec$. At the HWHM point, bandwidth smearing and time-averaging smearing losses are estimated to be approximately 15\% and 4\%, respectively.

The second epoch data were correlated using the DiFX Software Correlator \citep{Deller:2007p10545}. The data were correlated using an integration time of 2 seconds and with 64 frequency channels across each 16 MHz band (channel widths of 0.25 MHz). The ATCA primary beam limits the field of view of this observation to a HWHM of $\sim1\arcmin$. At the HWHM point, bandwidth smearing and time-averaging smearing losses are estimated to be approximately 8\% and 7\%, respectively.

\subsection{LBA Data Reduction}

The correlated data were imported into the AIPS\footnote{The Astronomical Image Processing System (AIPS) was developed and is maintained by the National Radio Astronomy Observatory, which is operated by Associated Universities, Inc., under co-operative agreement with the National Science Foundation} package for initial processing. The data for the phase reference source were fringe-fit (AIPS task FRING) using a one minute solution interval, finding independent solutions for each of the 16 MHz bands.  The delay and phase solutions for the phase reference source were examined and averaged over each three minute calibrator scan, following editing of bad solutions, before being applied to both the phase reference source and NGC 4945.  Further flagging of the data was undertaken via application of a flag file that reflected the times during which each of the antennas were expected to be slewing, or time ranges that contained known bad data.  Finally, data from the first 30 seconds of each scan from baselines involving the ATCA or Parkes were flagged, to eliminate known corruption of the data at the start of each scan at these two telescopes.

\begin{figure}
\epsscale{0.4}
\begin{center}
\mbox{

\plotone{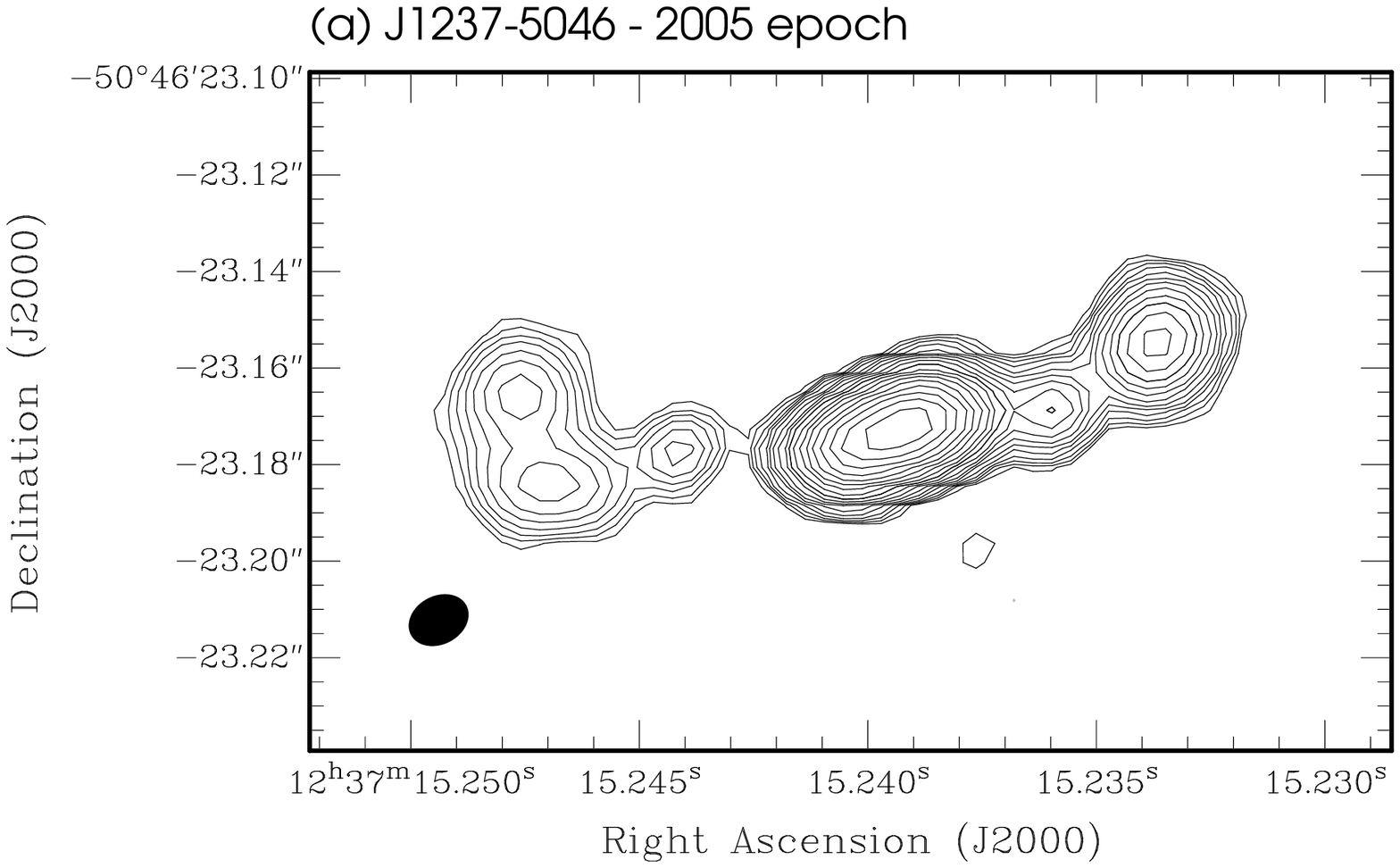} \quad
\plotone{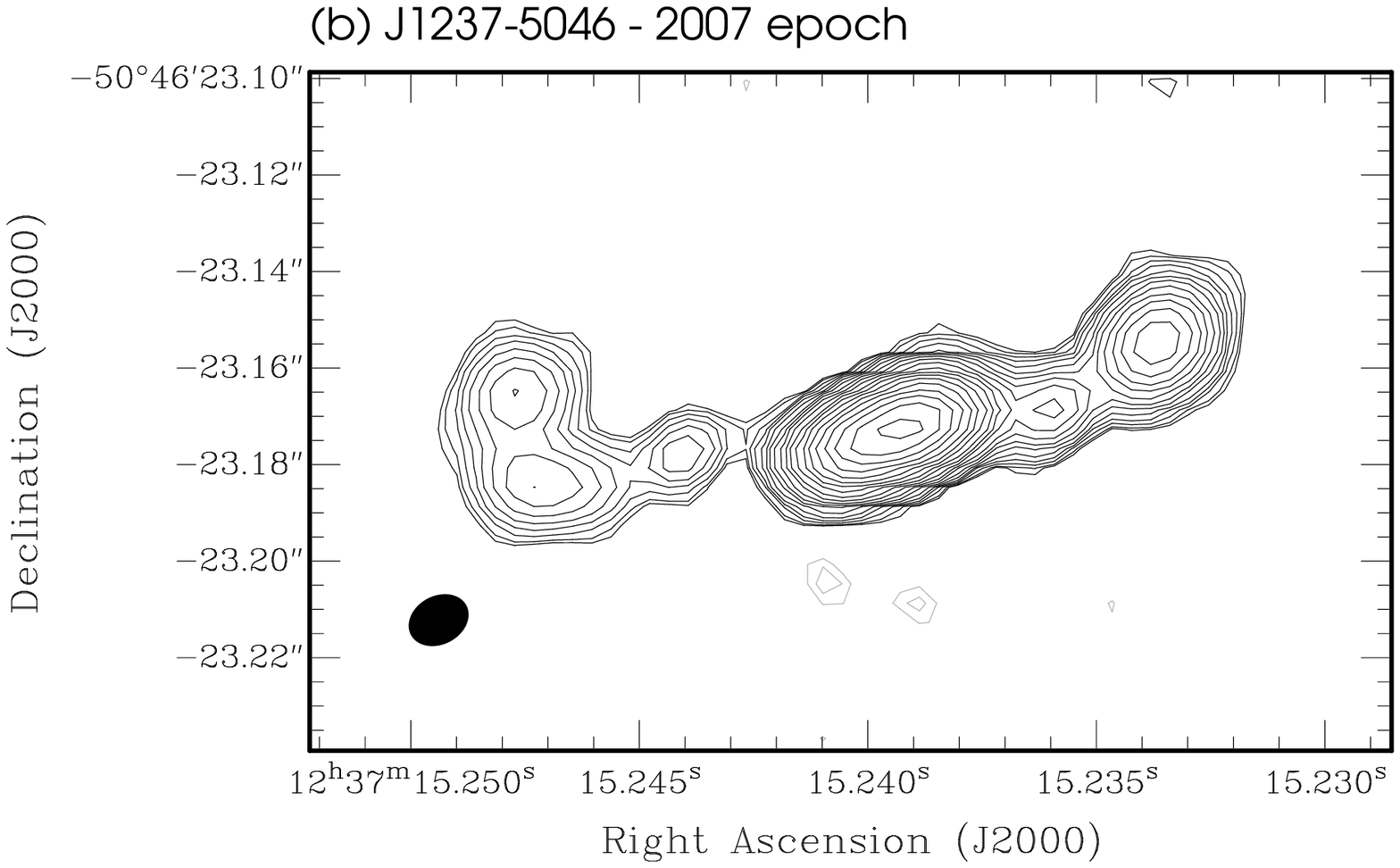}
}
\caption{Uniformly-weighted total-power map of the source used for phase calibration, J1237-5046, as observed with the Australian LBA at 2.3 GHz over two epochs. Map statistics for the individual maps are shown in Table \ref{tab:tabimage}. Contours are drawn at $\pm2^{0}, \pm2^{\frac{1}{2}}, \pm2^{1}, \pm2^{\frac{3}{2}}, \cdots$ times the $3\sigma$ rms noise.}
\label{fig:figJ1237-5046}            
\end{center}
\end{figure}

During correlation, nominal (constant) system temperatures (in Jy) for each antenna were applied to the correlation coefficients in order to roughly calibrate the visibility amplitudes (mainly to ensure roughly correct weights during fringe-fitting).  Following fringe-fitting, the nominal calibration was refined by collecting and applying the antenna system temperatures (in K) measured during the observation, along with the most recently measured gain (in Jy/K) for each antenna. Further refinement in the calibration was complicated by the complex structure of the phase reference source J1237$-$5046. To account for this structure a new DIFMAP \citep{Shepherd:1994p10583} task, \emph{cordump}\footnote{The \emph{cordump} patch is available for DIFMAP at \url{http://astronomy.swin.edu.au/$\sim$elenc/DifmapPatches/}} \citep{Lenc:2006p32}, was developed to enable the transfer of all phase and amplitude corrections made in DIFMAP during the imaging process to an AIPS compatible solution table. The phase reference data were averaged in frequency and exported to DIFMAP where several iterations of modelling and self-calibration of both phases and amplitudes were performed. The resulting contour maps of the phase reference source are shown in Figure \ref{fig:figJ1237-5046} and have a measured one sigma RMS noise of 0.342 and 0.292 mJy beam$^{-1}$ in epochs 1 and 2, respectively. \emph{cordump} was then used to transfer the phase and amplitude corrections back to the unaveraged AIPS data set.

Averaging in frequency was performed in AIPS to increase the channel widths to 2 MHz; this significantly reduced the data set to a manageable size and improved imaging performance. While this also reduced the overall field of view, the starburst region of interest was unaffected by the averaging process. To facilitate imaging in DIFMAP, a ParselTongue\footnote{A Python scripting tool for AIPS. ParselTongue was developed in the context of the ALBUS project, which has benefited from research funding from the European Community's sixth Framework Programme under RadioNet R113CT 2003 5058187. ParselTongue is available for download at \url{http://www.radionet-eu.org/rnwiki/ParselTongue}} script was written to convert the averaged frequency channels into intermediate frequencies (IFs). This conversion allowed DIFMAP to treat the frequency channels independently in the (u,v) plane rather than averaging them together, thus avoiding any further bandwidth smearing effects during the imaging process.

Initial imaging of NGC 4945 was performed at reduced resolution ($80\times32$ mas) by excluding data from the Hobart and Ceduna antennas (the longest baselines); uniform weighting was applied to maximise resolution at the expense of image noise. Several iterations of CLEAN and phase self-calibration were performed in DIFMAP. The resulting images are shown in Figure \ref{fig:figLBANGC4945} and achieve a one sigma RMS noise of 0.152 and 0.082 mJy beam$^{-1}$ for epoch 1 and 2, respectively. The image map parameters for both figures are shown in Table \ref{tab:tabimage}.

A second phase of imaging was repeated at full resolution ($16\times15$ mas) with data from all antennas. The imaging process followed the same procedure to that used in the initial imaging phase except that natural weighting was applied to provide a compromise between the increased resolution available with the full array and the available image noise. Image maps for both epochs were created for each of the sources that fell above the six sigma detection limit. The images from both epochs were then combined to produce a high dynamic range image map for each of the detected sources. The epoch 1, epoch 2 and combined image maps for each of the detected sources are shown in Figure \ref{fig:fighra} and the associated map statistics tabulated in Table \ref{tab:tabimage}. The epoch 1, epoch 2 and combined maps, achieve a one sigma RMS noise of 0.149, 0.110 and 0.075 mJy beam$^{-1}$, respectively. These figures compare favourably with the expected theoretically predicted thermal noise of $\sim0.120$, $\sim0.080$ and $\sim0.070$ mJy beam$^{-1}$, respectively\footnote{Estimated with the ATNF VLBI sensitivity calculator: \url{http://www.atnf.csiro.au/vlbi/calculator/}}.

\subsection{ATCA Observations and Data Reduction}

Observations of NGC 4945 were carried out using all six antennas in the EW367 and 6C configurations of the Australia Telescope Compact Array (ATCA) on the 2006 March 18, 24 and 25. The EW367 configuration is compact with most baselines less than 367 m in length, the 6C configuration is for high resolution observations and has a spread of baselines up to 6 km in length. The flux scale was set using PKS B1934$-$638 and J1326$-$5256 was used as a phase calibrator. During each observing run, we alternated between two setups every 6.5 minutes. The first setup observed at 17 GHz and 19 GHz and the second setup at 21 GHz and 23 GHz. For intermediate frequencies of 4.8 GHz and 8.64 GHz, archive data was obtained from the Australia Telescope Online Archive. The data were observed with the 6A configuration of the ATCA, which has a spread of baselines up to 6 km in length, on 1993 November 19 and 20. The flux scale was set using PKS B1934$-$638 and PKS 1320$-$446 was used as a phase calibrator. A list of all ATCA observations reported in this paper and their associated observing parameters are tabulated in Table \ref{tab:tabobs}.

\begin{figure}
\epsscale{0.4}
\begin{center}
\mbox{
\plotone{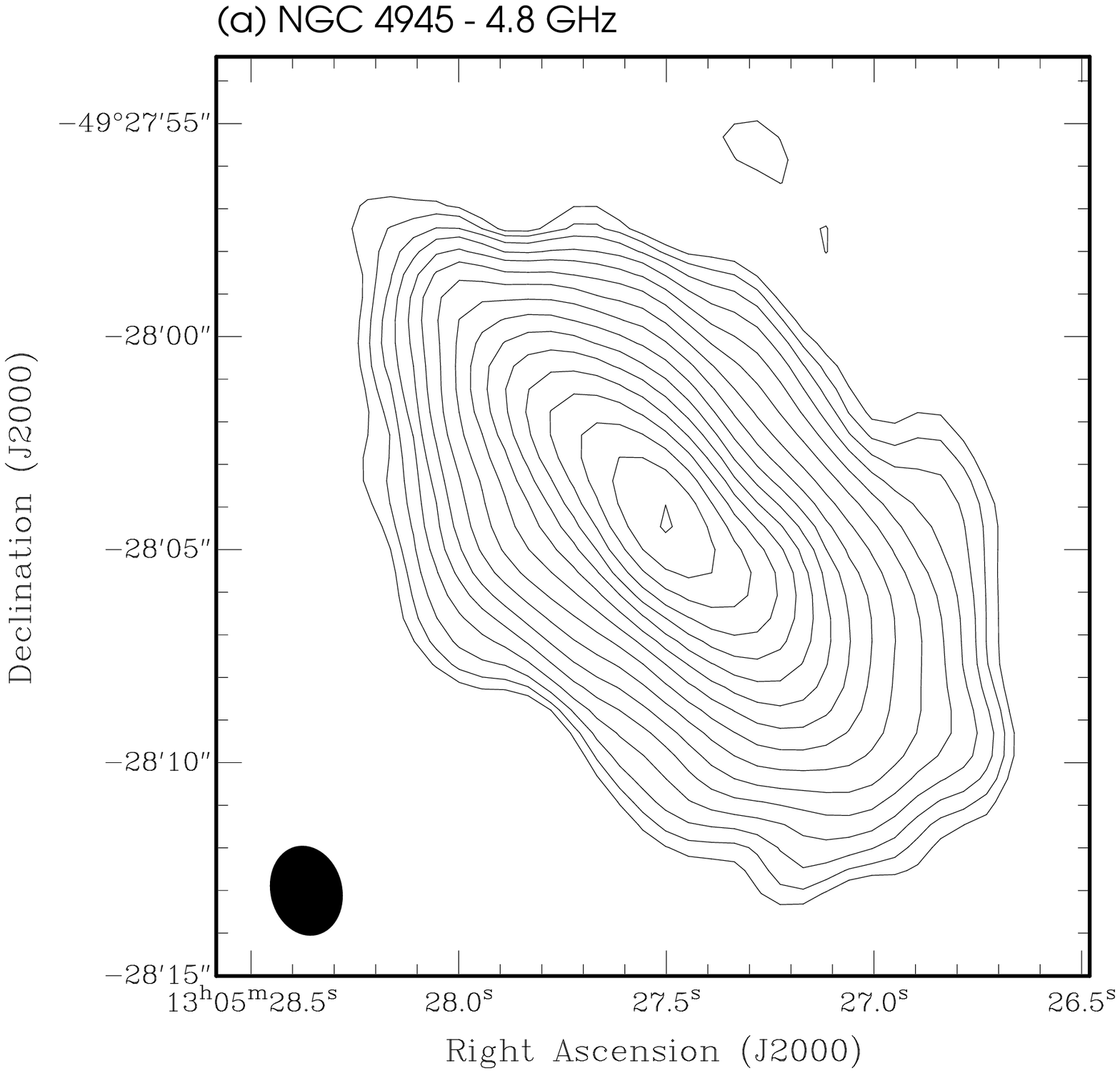} \quad
\plotone{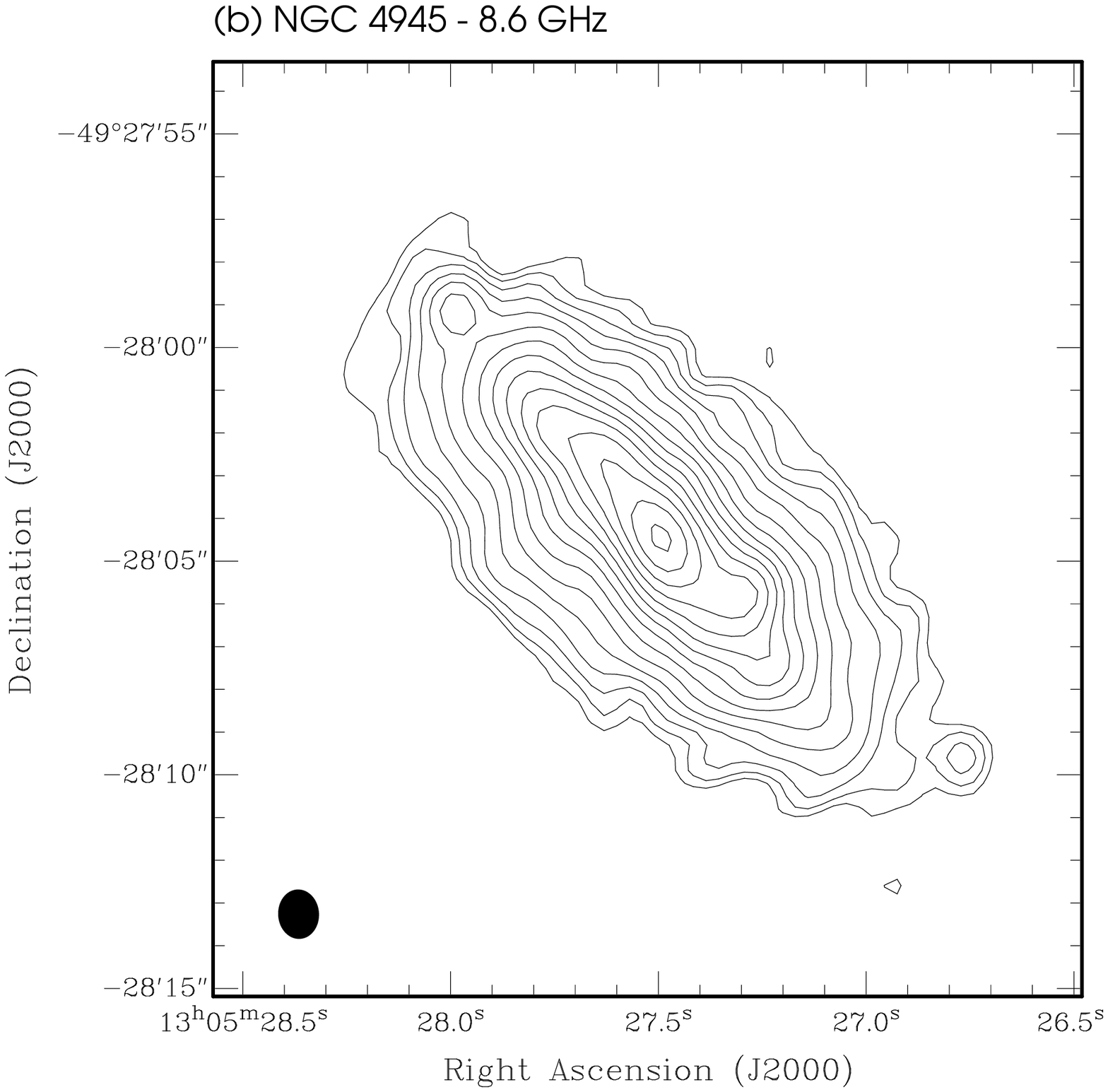}
}
\mbox{
\plotone{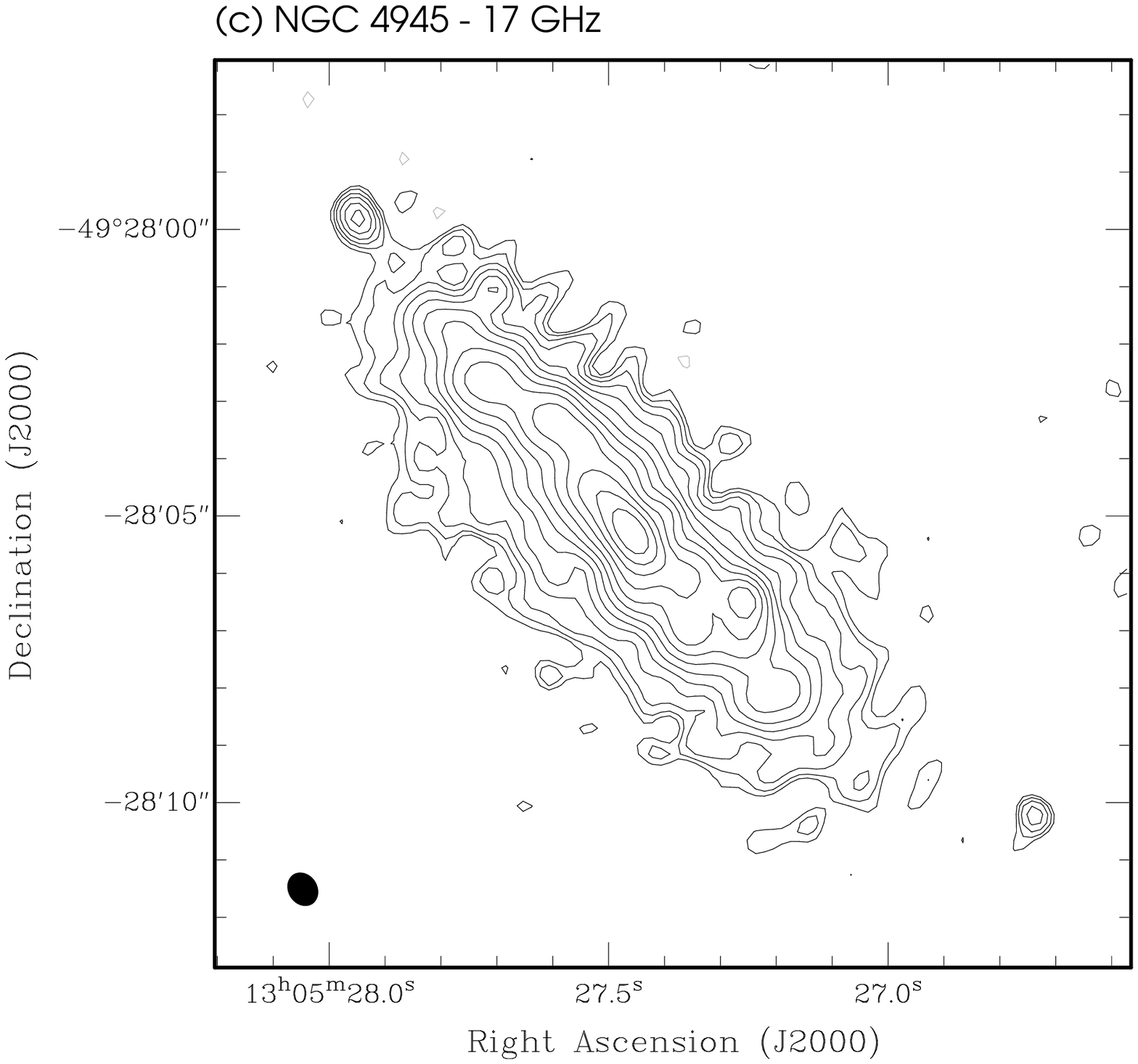} \quad
\plotone{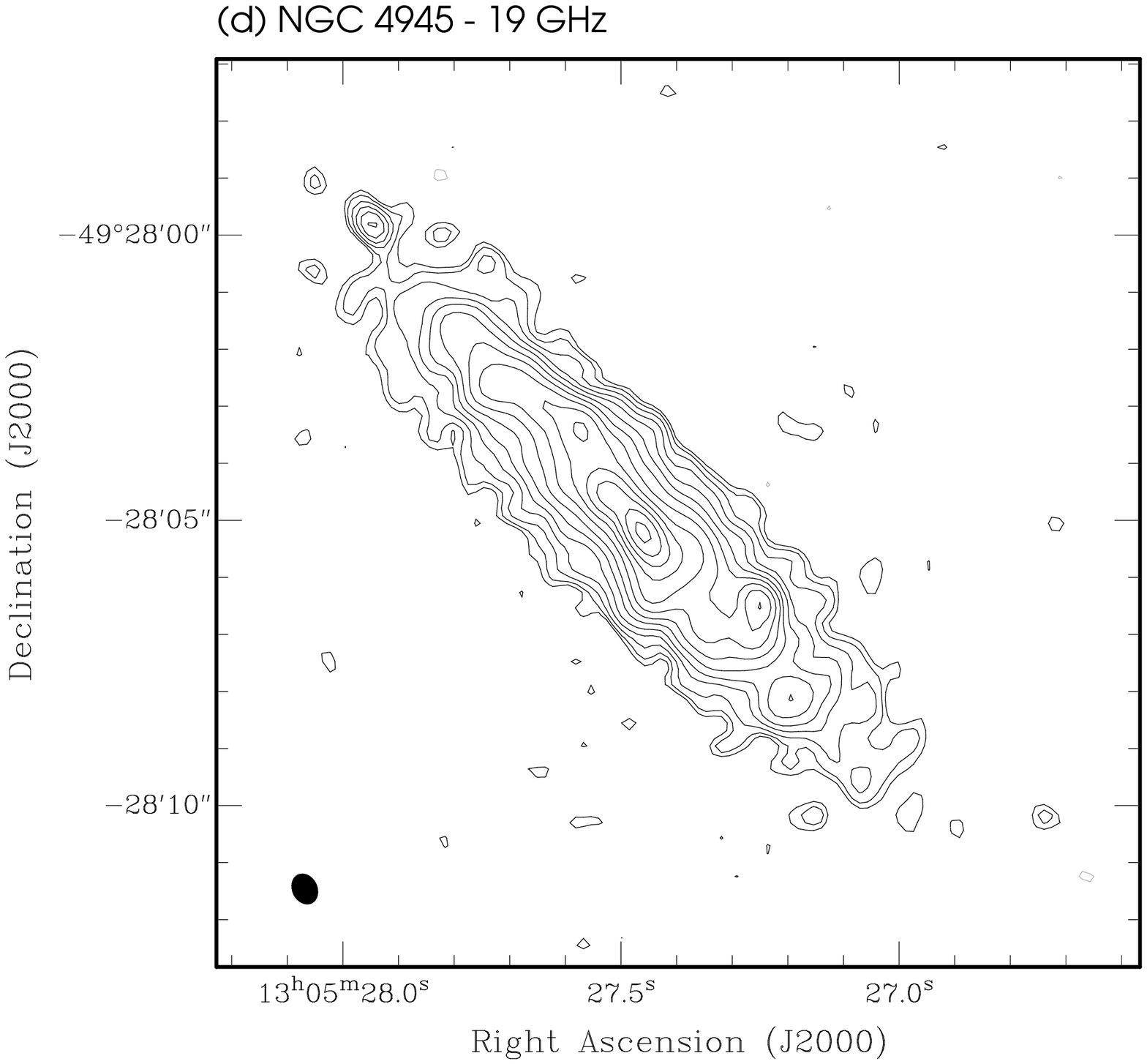}
}
\mbox{
\plotone{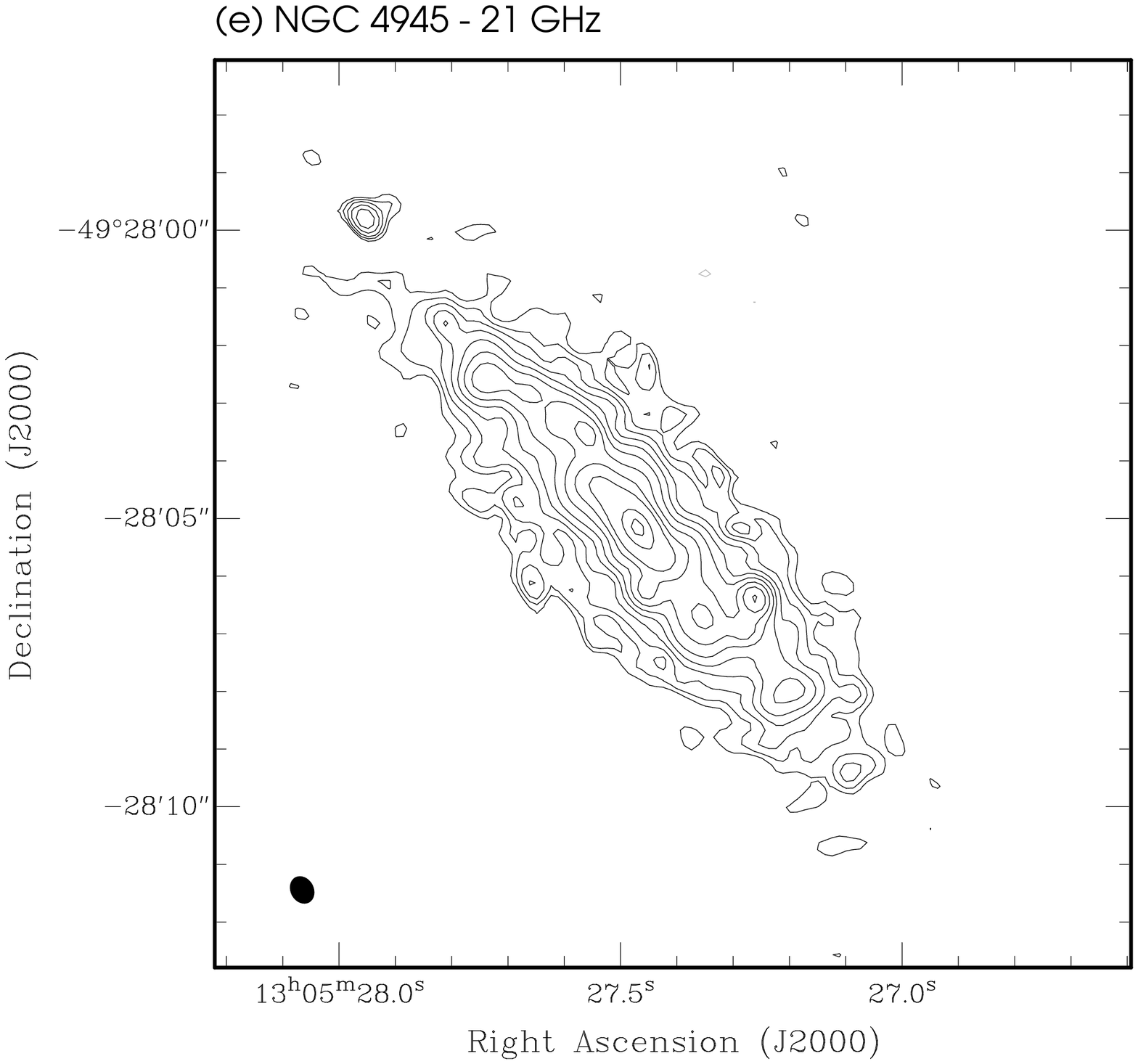} \quad
\plotone{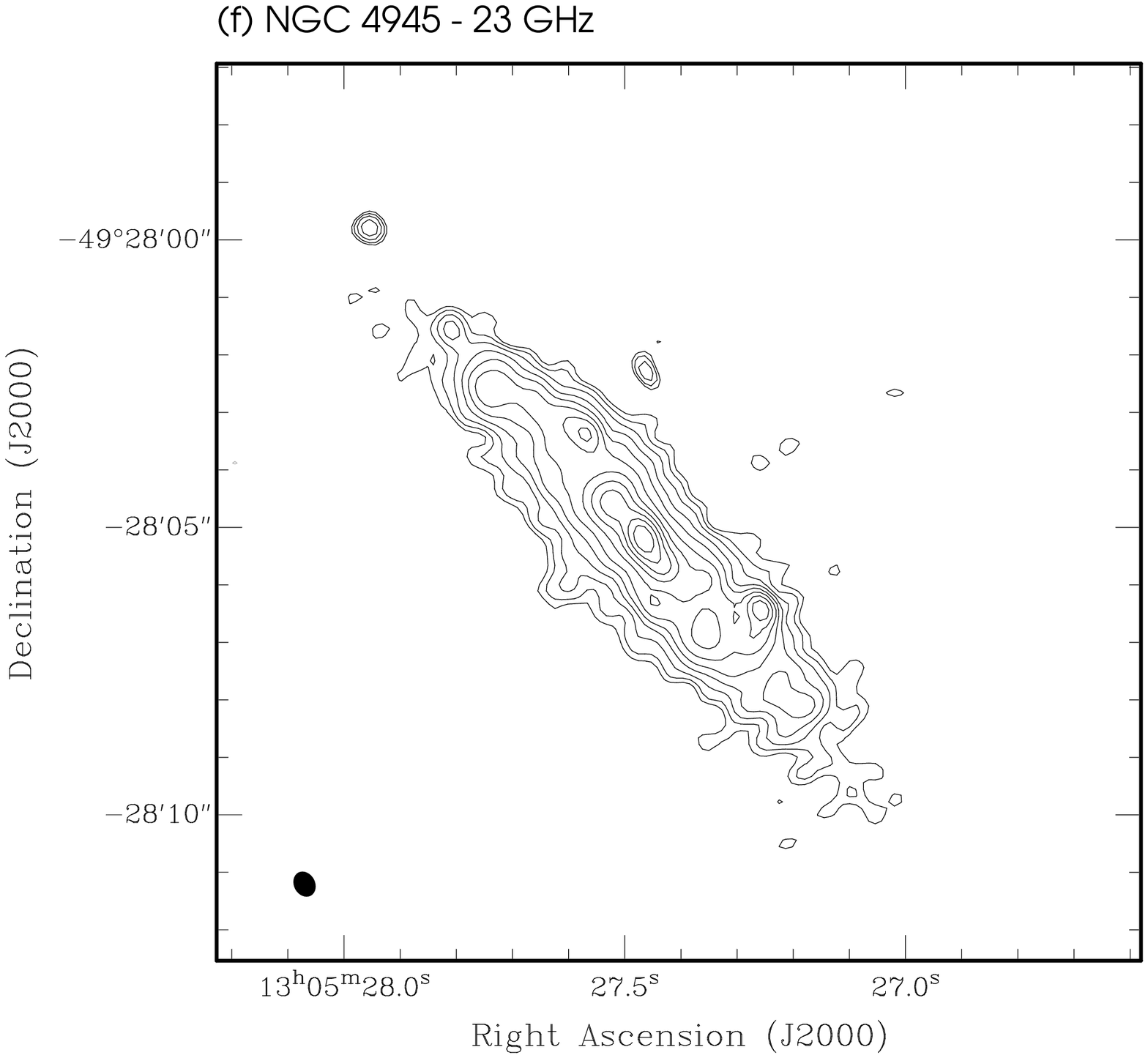}
}
\caption{Uniformly-weighted total-power maps of NGC 4945 as observed with the ATCA at 4.8 GHz, 8.6 GHz, 17 GHz, 19 GHz, 21 GHz and 23 GHz. Map statistics for the individual maps are shown in Table \ref{tab:tabimage}. Contours are drawn at $\pm2^{0}, \pm2^{\frac{1}{2}}, \pm2^{1}, \pm2^{\frac{3}{2}}, \cdots$ times the $3\sigma$ rms noise for all maps except for (a) and (b) where the lowest contour is at $6\sigma$ rms noise.}
\label{fig:figATCANGC4945}            
\end{center}
\end{figure}

All ATCA data were initially calibrated using the MIRIAD software package \citep{Sault:1995p10582}. Subsequent calibration, deconvolution and imaging was performed with the DIFMAP \citep{Shepherd:1994p10583} software package and final images were created using the KARMA software package \citep{Gooch:1996p7263}. The resulting contour maps are shown in Figure \ref{fig:figATCANGC4945} with all associated beam parameters, image rms noise, peak and integrated flux densities tabulated in Table \ref{tab:tabimage}.

\section{Results}
\subsection{Identification of Sources}
\label{sec:sourceid}
A list of sources detected above the $6\sigma$ detection threshold in both epochs of the short-baseline VLBI data is given in Table \ref{tab:tabfluxes}. The $6\sigma$ threshold provides a conservative false detection rate of less than one in 3000. Flux density errors of $\pm10\%$ are listed due to uncertainties in the absolute flux density scale for Southern Hemisphere VLBI \citep{rey94}. The total flux density of each source was determined by summing its CLEAN model components in the image plane. The sources were also modelled with elliptical Gaussian components to provide an approximate measure of the full-width half-maximum (FWHM) size, the fitted size and position angle (PA) of each source is also listed in Table \ref{tab:tabfluxes}. Two of the detected sources, $27.5291-04.631$ and $27.6164-03.373$, exhibit structure that is significantly larger than the beam width and are shown in greater detail in Figure \ref{fig:figlr}. Source $27.5291-04.631$, in particular, appears to exhibit a jet-like morphology and has been divided into three components in Table \ref{tab:tabfluxes}: $27.5291-04.631$, $27.5402-04.720$ and $27.5607-04.779$.

Based on the reduced resolution images, a total of 12 discrete sources fell above the $6\sigma$ threshold (0.91 mJy beam$^{-1}$) of the first epoch observation of NGC 4945, whereas 13 fell above the $6\sigma$ threshold (0.51 mJy beam$^{-1}$) of the second epoch observation. When imaged with the longer baselines included (Hobart and Ceduna), eight of these sources were detected in the first epoch data and nine in the second epoch. While source 27.4646$-$05.174 did not achieve the $6\sigma$ detection threshold in the epoch 1 data set, an image of the residuals at the source location is included as part of Figure \ref{fig:fighra} for completeness sake. No observable position offsets were detected between the sources detected in epoch 1 and those detected in epoch 2.

Comparisons between the VLBI images and the higher frequency ATCA images were complicated by a number of factors. Firstly, source identification in the ATCA images was limited by the large-scale diffuse emission and confusion as a result of the significantly lower resolution of these images compared to that of the VLBI images. Secondly, when compared to the VLBI image, a systematic offset of 900 mas, 160 mas and 205 mas was observed in the 8.6 GHz, 17/19 GHz and 21/23 GHz images, respectively. The 4.8 GHz ATCA image was not used in any further comparisons in this paper as a result of its poor resolution. The large offset observed at 8.6 GHz results from a 945 mas error in the phase calibrator position of the original 1993 observation. The $\sim200$ mas offset observed between the high frequency ATCA images and the VLBI images is well below the ATCA beam size and is typical given atmospheric conditions at these frequencies and the $5\arcdeg$ separation between the phase calibrator and target source.

To aid in discrete source identification, four large elliptical Gaussian components were used to model and effectively subtract the large-scale diffuse emission from each of the ATCA images. The full-width half-maximum (FWHM) size, position angle (PA) and flux density of these components, as determined separately for each frequency observed with the ATCA, are listed in \ref{tab:tabfluxes}. To compare the resulting ATCA residual sources against the VLBI sources, the elliptical Gaussian components used in the VLBI model were transferred to each of the ATCA data sets and their positions adjusted accordingly to account for the systematic offsets previously mentioned. Each of the model components were significantly unresolved with respect to even the smallest ATCA beam. With the size, position angle and location fixed, the flux density of each model component was varied to fit the diffuse-emission-subtracted ATCA data. The modelled flux densities for each source and data set are listed in Table \ref{tab:tabfluxes}. No significant residuals were observed in the ATCA image or visibilties after the large-scale structure and source components were subtracted from the image.

\section{Discussion}

\subsection{Radio spectra and free-free absorption modelling of the compact sources}
\label{sec:p3ffmodel}
\citet{Lenc:2006p6695} and \citet{Tingay:2004p778} showed that the spectra of compact radio sources in NGC 253 exhibited a sharp downturn at 1.4 GHz, and to a lesser degree at 2.3 GHz, as a result of free-free absorption. To test a similar scenario quantitatively with the compact radio sources detected in NGC 4945, an analysis was performed using the data tabulated in Table \ref{tab:tabfluxes}. In this analysis, the second epoch VLBI data was used to model the low frequency end of the spectrum and the ATCA data was used to model the high frequency end of the spectrum. Archival 8.64 GHz data were also included to model the mid-range spectrum. Observations of compact radio sources in M82 \citep{Ulvestad:1994p1041} and NGC 253 \citep{Ulvestad:1997p907} have shown that for existing sources, little or no change in flux is observed over periods of order a decade. Here we have assumed that if a source existed in both the archive 8.64 GHz data and our current ATCA or VLBI data then it has not significantly faded or brightened in the intervening years between the observations. Where no 8.64 GHz data exist we do not constrain our models to a lower limit based on the detection threshold at this frequency as we do not know if the source existed at that point in time.

In our analysis we used the same models and techniques described in \citet{Lenc:2006p6695}. However, for sources that were detected in only two observing bands a simple power-law spectrum was used to model the source spectrum, whereas the more highly sampled sources were also modelled with a free-free absorbed-power-law spectrum and a self-absorbed bremsstrahlung (free-free absorbed) spectrum. For the large-scale emission, lower limits were not applied at the low frequency end of the spectrum as this emission is totally resolved by our high-resolution VLBI observations.

The spectrum for each source was tested against each of the models using a reduced $\chi^{2}$ criterion to determine the best fit. $S_{0}$, $\tau_{0}$ and $\alpha$ from the model fits are listed in Table \ref{tab:tabsrcff}. In all cases, the sources were modelled best by either a simple power-law spectrum or a free-free absorbed-power-law spectrum. The model for each source is shown against the corresponding measured spectrum in Figure \ref{fig:figff}. Error bars of 10\% are shown for the flux densities at 23 GHz, 21 GHz and 2.3 GHz, whereas slightly larger errors of 15\%, 15\% and 20\% are estimated for the 19 GHz, 17 GHz and 8.64 GHz measures, respectively. The larger errors associated with the lower frequency ATCA observations take into consideration the effects of confusion and the increased levels of diffuse emission. The data shown in Table \ref{tab:tabsrcff} and Figure \ref{fig:figff} show that at least 9 of the 13 compact radio sources modelled are free-free absorbed.

Using the same criteria as \citet{Lenc:2006p6695}, of the 13 compact sources for which spectral indices could be determined in Table \ref{tab:tabsrcff}, three have flat intrinsic power law spectra ($\alpha > -0.4$) indicative of \ion{H}{2} regions dominated by thermal radio emission (T), the 10 remaining sources have steep intrinsic spectra (S), as normally associated with supernova remnants \citep{McDonald:2002p3330}. The proportion of synchrotron sources (77\%) compared to thermal sources (23\%) in NGC 4945 is similar to that found in other nearby starburst galaxies. \citep{McDonald:2002p3330} found 65\% of the compact radio sources in M82 were synchrotron sources and 35\% thermal, whereas \citep{Lenc:2006p6695} found that 60\% of the detections were synchrotron sources and 40\% thermal (after correcting source 5.805$-$38.92 with a spectral index of -1.2). The comparatively lower proportion of thermal sources detected in NGC 4945 compared to M82 and NGC 253 is believed to be a result of the lower resolution of the ATCA compared to the VLA. Thermal sources are more prominent at the high frequency end of the spectrum, however, even at 23 GHz the ATCA does not have sufficiently long baselines to easily resolve thermal sources from each other and from the large-scale diffuse emission.

\begin{figure}
\epsscale{0.6}
\begin{center}
\plotone{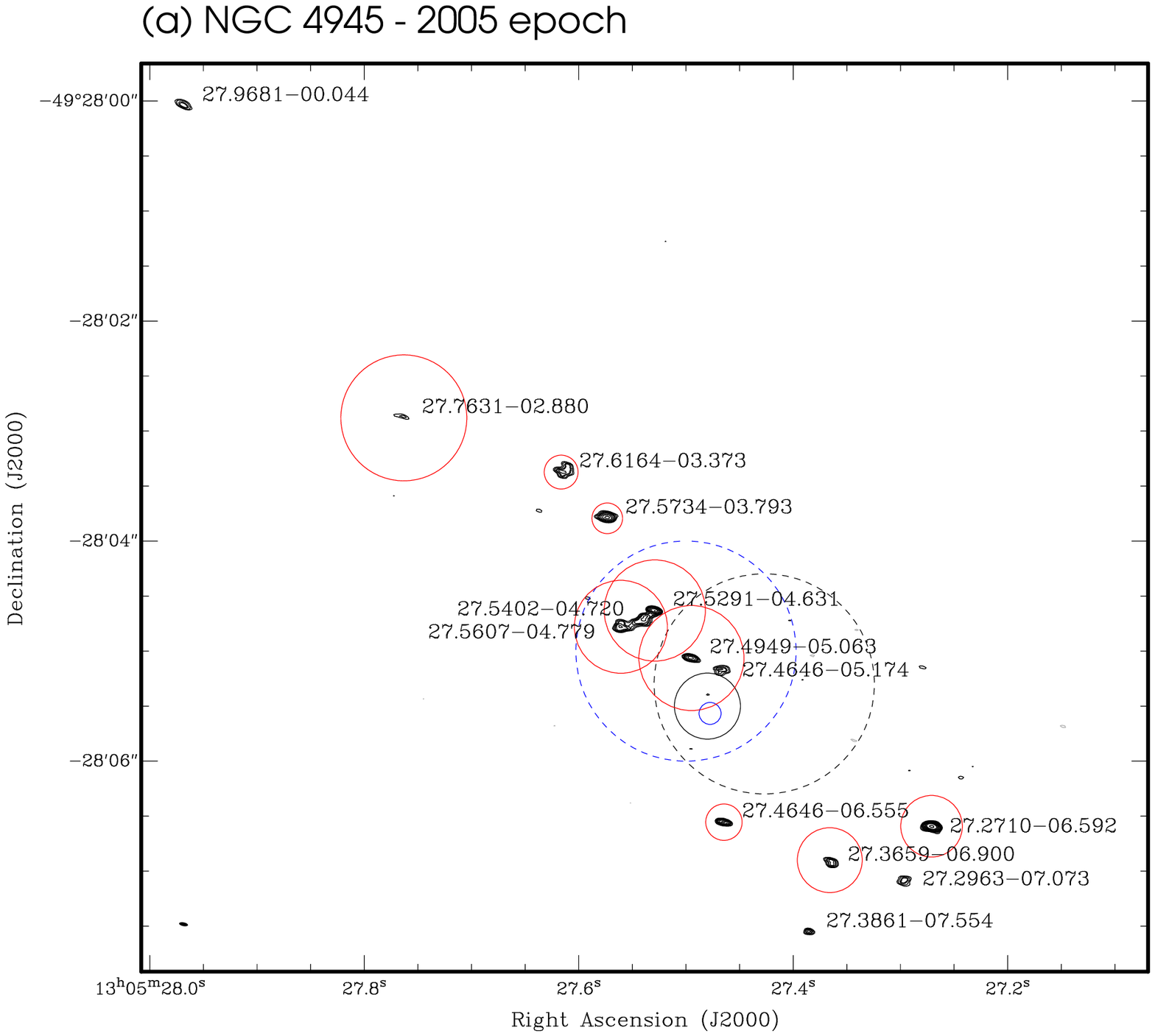}
\plotone{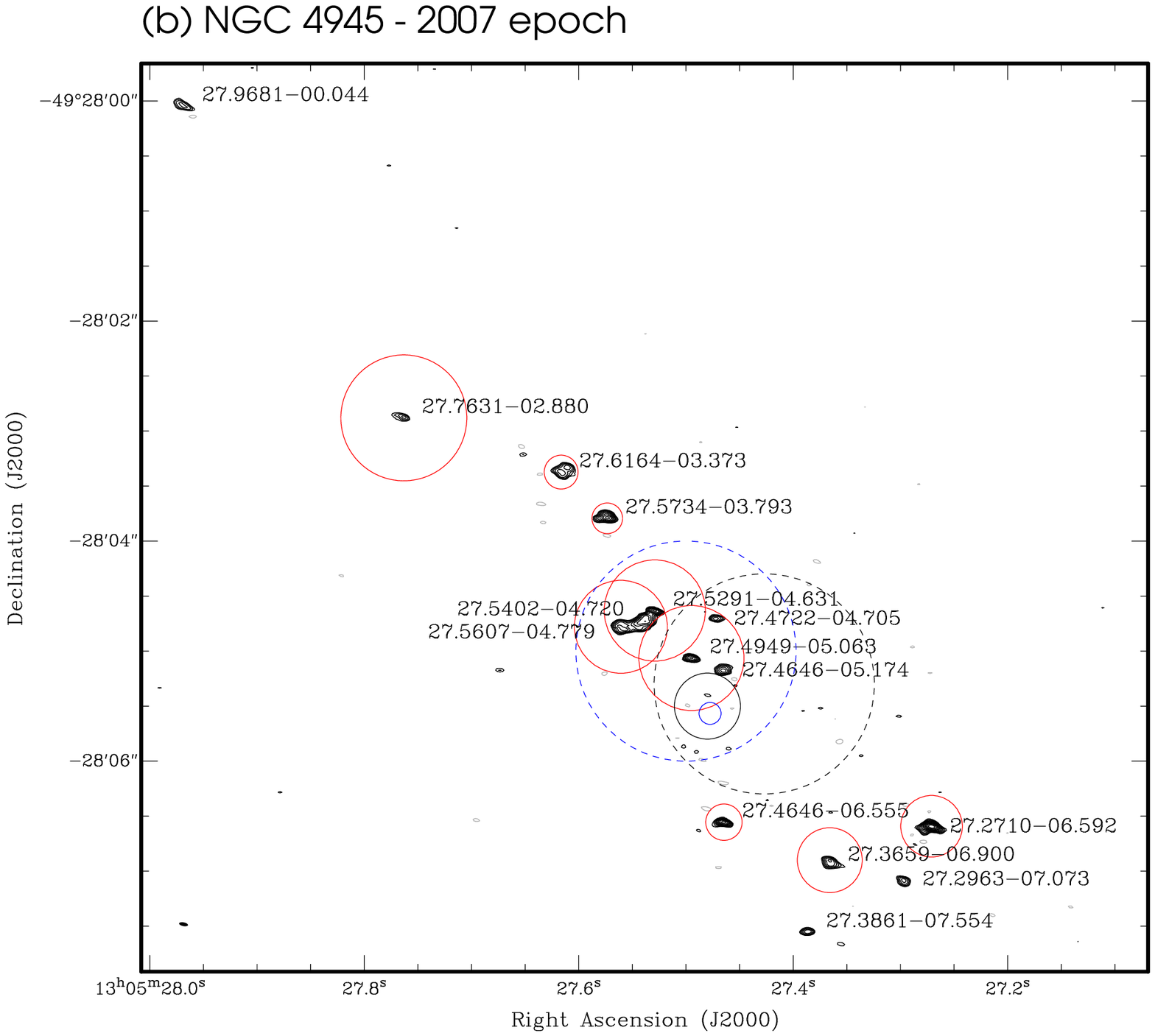}
\caption{Uniformly-weighted total-power map of NGC 4945 as observed with the Australian LBA at 2.3 GHz over two epochs. Map statistics for the individual maps are shown in Table \ref{tab:tabimage}. Contours are drawn at $\pm2^{\frac{1}{2}}, \pm2^{1}, \pm2^{\frac{3}{2}}, \cdots$ times the $3\sigma$ rms noise. The overlaid circles give an indication of the degree of absorption in the vicinity of the source, with the diameter of the circle being directly proportional to $\tau_{0}$. The dashed blue circle and dashed black circle indicate the positional error of the Chandra x-ray \citep{Schurch:2002p7434} and HST k-band \citep{Marconi:2000p7671} peak, respectively. The solid blue circle and solid black circle mark the positional error of the H$_{2}$O megamaser \citep{Greenhill:1997p7509} and the centre of the HNC cloud \citep{HuntCunningham:2005p10542}.}
\label{fig:figLBANGC4945}            
\end{center}
\end{figure}

The red circles in Figures \ref{fig:figLBANGC4945}(a) and \ref{fig:figLBANGC4945}(b) show how the free-free opacity varies across the field of the nuclear region of NGC 4945. Although there appears to be significantly greater free-free absorption toward the nuclear region of NGC 4945, the ionised medium appears to be clumpy. As we are viewing the galaxy almost edge-on this may be the result of seeing sources in front of the disk, embedded within the disk, or behind the disk, rather than being indicative of variation in the ionised medium. However, it is also possible that the absorption may be localised around the source itself, such as has been observed in young, dense, heavily-embedded clusters \citep{Turner:1998p22446, Kobulnicky:1999p28204, Johnson:2003p28201}.

Based on the model parameters, the flux density of each source at 1.4 GHz, $S_{1.4}$, is calculated and listed in Table \ref{tab:tabsrcff}. It is expected that up to eight sources may be detected at 1.4 GHz assuming a detection threshold of approximately 1 mJy beam$^{-1}$. Follow-up observations of NGC 4945 have been proposed with the LBA to further constrain the free-free parameters of the sources at lower frequencies. Such an observation would also help to highlight if any of the source flux density measures have been adversely affected by resolution effects since the LBA beam at 1.4 GHz would be substantially larger than at 2.3 GHz.


\subsection{Source variation}
\label{sec:fluxvar}

The mean of the individual source flux density ratios between the two VLBI epochs is 1.06, while the median is 0.98. If only relatively strong sources are considered ($S_{2.3}>6$ mJy), to exclude those that are significantly affected by deconvolution errors and measurement uncertainties, we observe a mean flux density ratio of $1.00$ and a median of $0.985$ between epochs. This suggests that there are no significant systematic differences between the flux-density scales or general source fading or brightening between the two epochs. None of the eight sources in this partially restricted sample exhibited flux variations above the 10\% errors of our measures. If we take the median ratio as an upper limit on the median fading of supernova remnants in NGC 4945 then this would equate to a 1.5\% reduction in flux density over the 1.9 yr baseline between the two VLBI epochs, or a fading of $<0.8$\% yr$^{-1}$. Further observations, over a longer period of time, would be required to confirm whether this fading is indeed real.

\subsection{Comments on individual compact radio sources}
\subsubsection{\ion{H}{2} regions}
The sources 27.9681$-$00.044, 27.4722$-$04.705 and 27.3861$-$07.554 are flat-spectrum sources that may be associated with \ion{H}{2} regions. All three sources are comparatively weak at 2.3 GHz and are resolved in the low resolution VLBI images. Only one of the sources, 27.3861-07.554, is detected in the high resolution VLBI images. The source 27.4722-04.705 is detected only in the second, more sensitive, epoch image and is not believed to be a new source.

It should be noted that the classification of the three sources as \ion{H}{2} regions is not entirely consistent with their brightness temperatures ($\sim2\times10^{5}$ K) at the low frequency end of the spectrum. Significant errors in the measured spectral index of these particular sources may have resulted in their misclassification. These errors arise from poor sampling in the spectral energy distribution (SED), scatter in the SED as a result source modelling errors and underestimates in the total flux density of extended sources in VLBI images. Another possibility is that the VLBI detections are of one or more supernova remnants embedded within \ion{H}{2} regions. These possibilities could be tested with further observations at intermediate and lower frequencies by improving the modelling of the SED of these sources.

None of the flat-spectrum sources appear to be significantly free-free absorbed, even those that appear to be relatively close to the nuclear centre. Similar characteristics are also observed in NGC 253 where only two flat-spectrum sources are detected at 2.3 GHz, both of which are only weakly free-free absorbed at this frequency \citep{Lenc:2006p6695}. Any curvature due to free-free absorption is difficult to detect in these sources since the low frequency VLBI observations are only sensitive to high brightness temperature sources, furthermore, the absorption acts in the same direction as the intrinsic spectrum.

\subsubsection{Steep-spectrum sources}

A total of ten steep-spectrum sources have been identified in NGC 4945. Two of the synchrotron sources, $27.5291-04.631$ and $27.5607-04.779$, in combination with $27.5402-04.720$, resemble a jet-like source and may be associated with a relativistic jet near the nucleus of NGC 4945 (section \S~\ref{sec:jet}). Both of the synchrotron sources are significantly affected by free-free absorption and to a similar degree ($\tau_{0}=17-18$) suggesting that they are in close proximity of each other and deeply buried behind an ionised screen.

Three of the synchrotron sources $27.6164-03.373$, $27.5734-03.793$ and $27.3659-06.900$, have shell-like structures which may be associated with supernova remnants and have shell diameters of 110 mas, 100 mas and 60 mas, respectively. Assuming an average expansion velocity of $v=$10,000 km s$^{-1}$ relative to the shell centre, similar to that measured for the supernova remnant 43.31+592 in M82 \citep{Pedlar:1999p3534}, this implies ages of $\sim100(10^{4}/v)$ yr, $\sim90(10^{4}/v)$ yr and $\sim50(10^{4}/v)$ yr for these remnants, respectively. The source $27.4646-05.174$ also has a shell-like structure in the VLBI image and is assumed to be steep spectrum source since it is not detected at higher frequecencies. This source has a diameter of 70 mas and, if it is associated with a supernova remnant, has an age of $\sim60(10^{4}/v)$ yr.

Two of the synchrotron sources $27.7631-02.880$ and $27.2963-07.073$ are comparatively weak and are fully resolved in the high resolution VLBI images. The remaining sources, $27.4646-06.555$, $27.4949-05.063$ and the brightest source, 27.2710-06.592, all are partially resolved and of similar size ($30\times25$ mas) but exhibit no significant structure. These may be young remnants with ages of $\sim25(10^{4}/v)$ yr.


\subsection{A relativistic jet at the nucleus of NGC 4945?}
\label{sec:jet}

\begin{figure}
\epsscale{0.5}
\begin{center}
\mbox{
\plotone{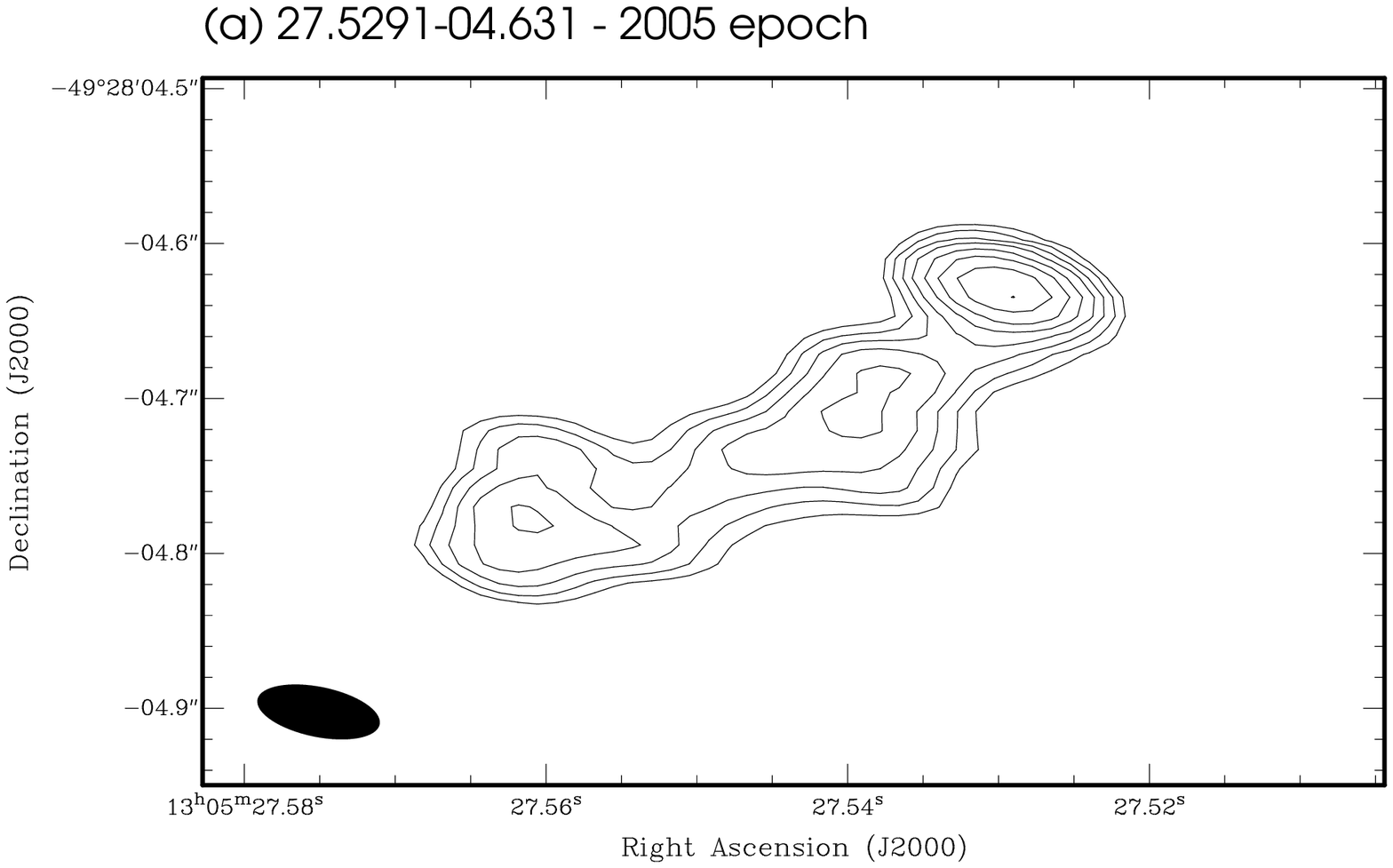} \quad
\plotone{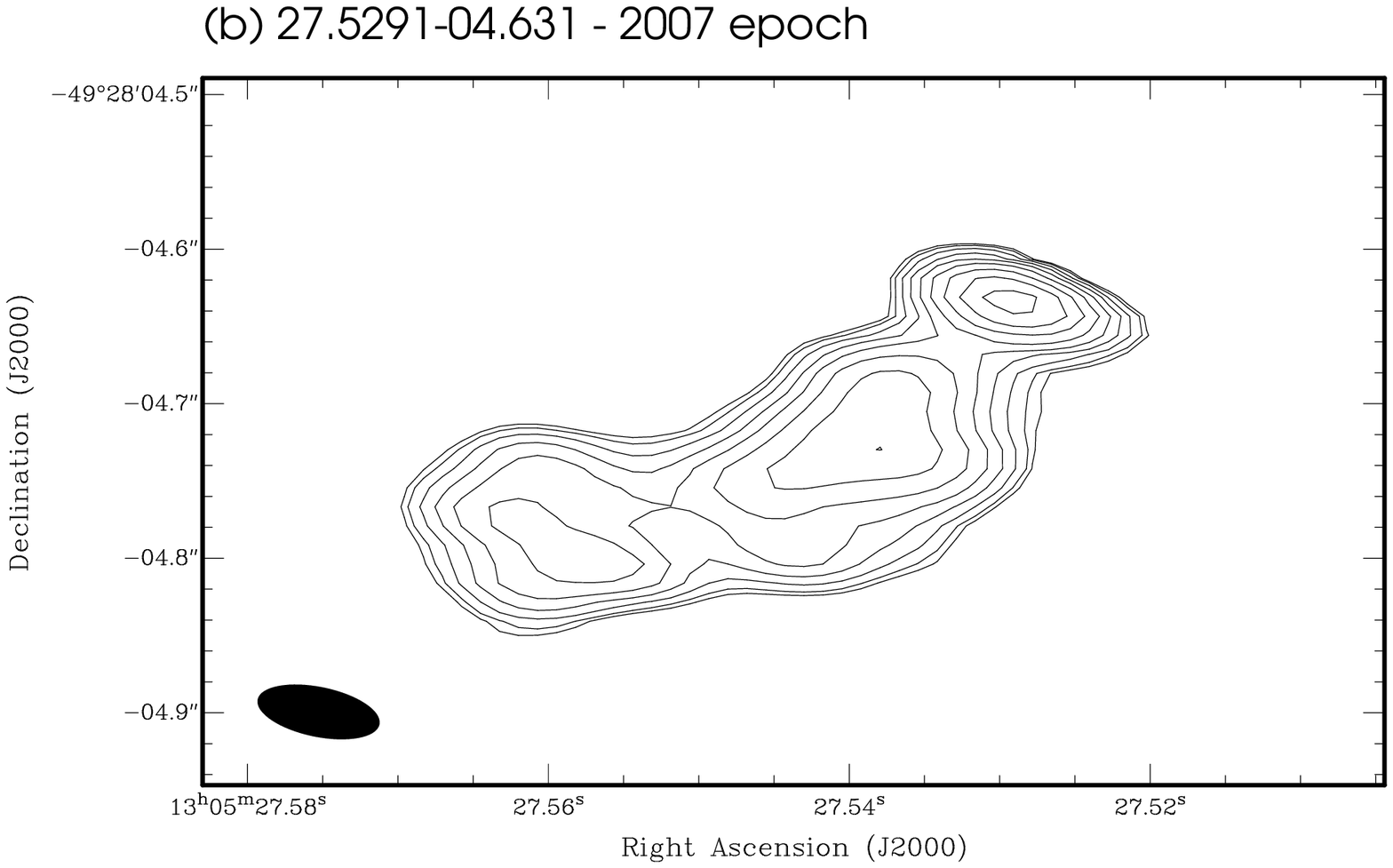}
}
\mbox{
\plotone{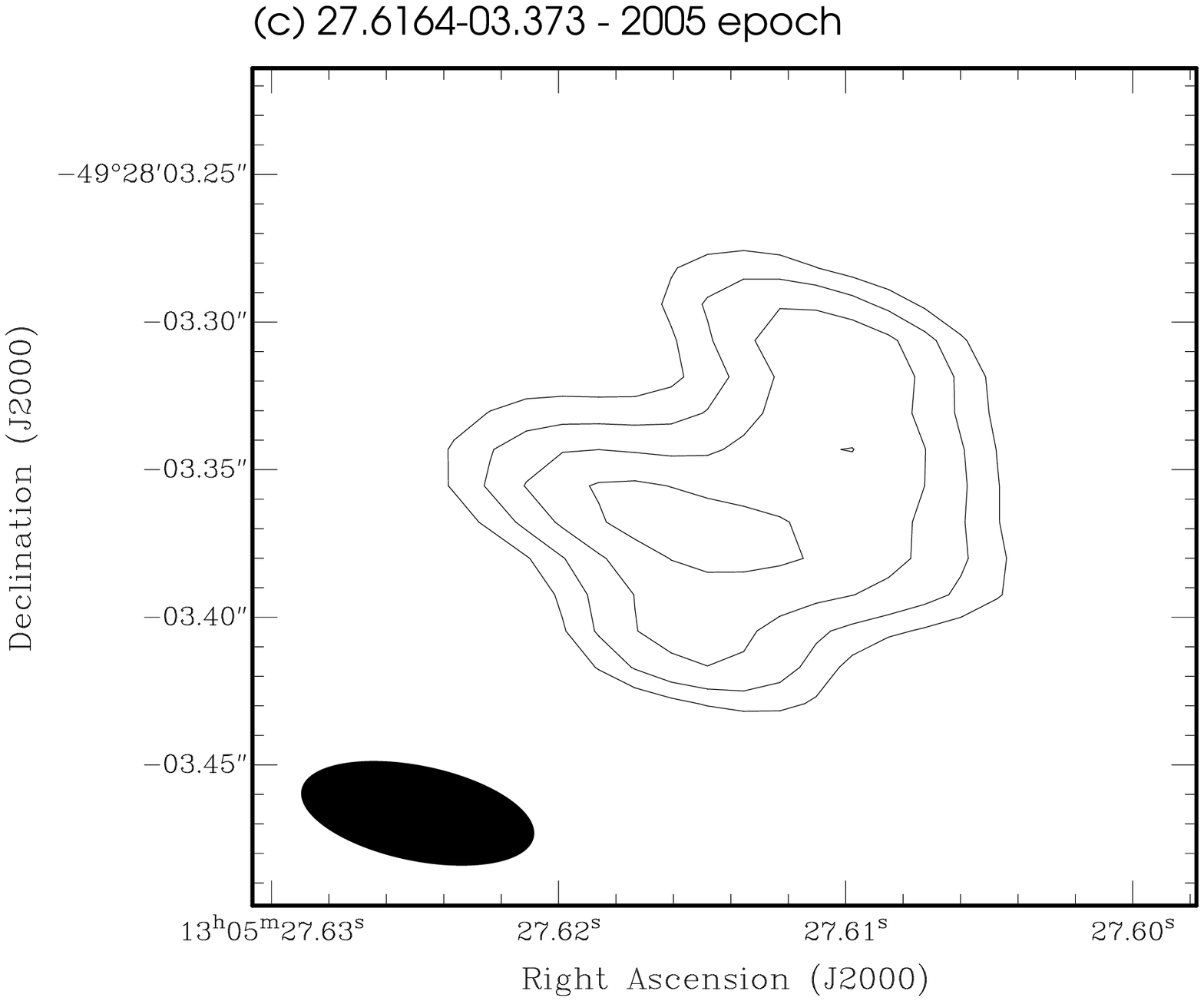} \quad
\plotone{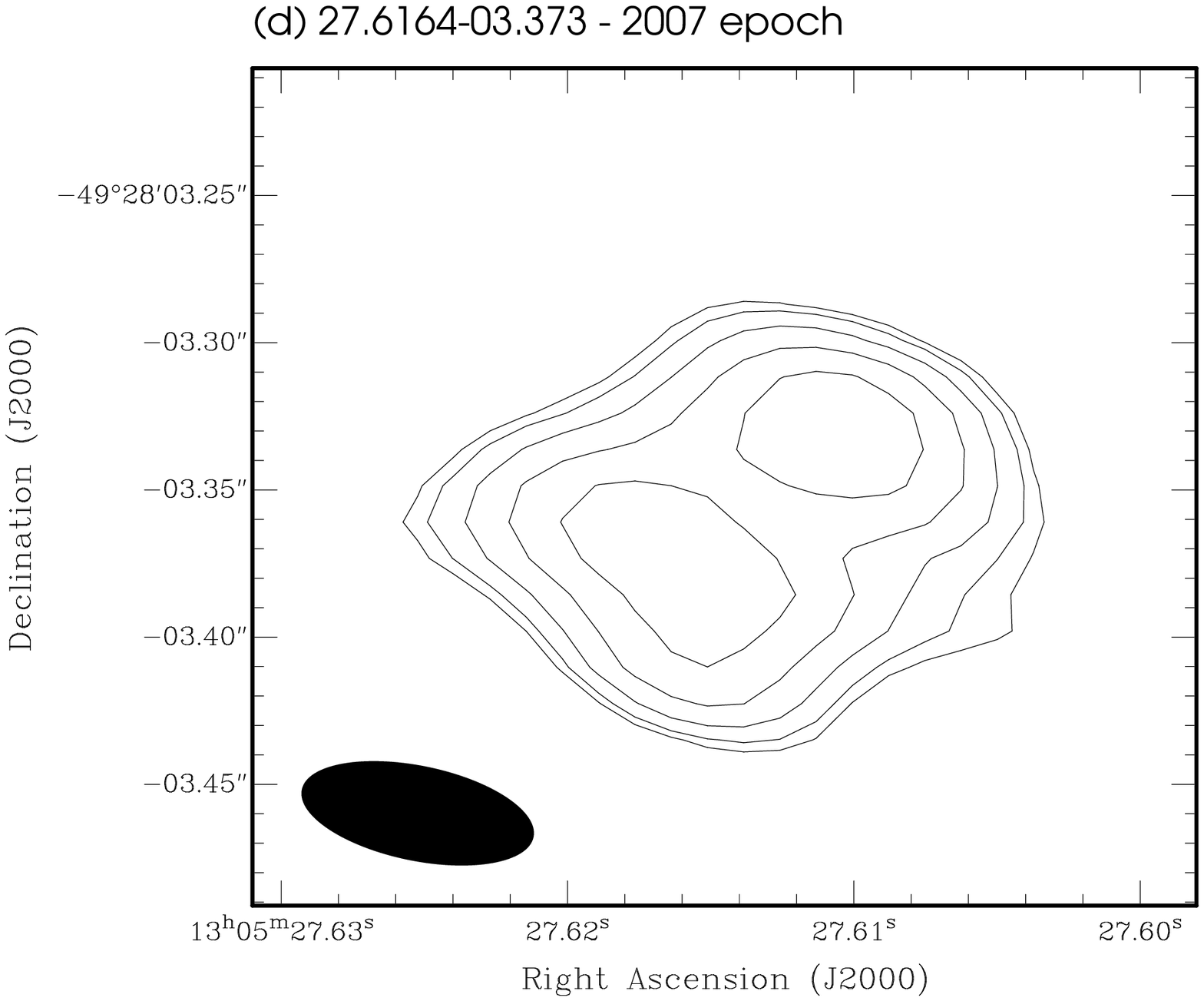}
}
\caption{Uniformly-weighted total-power maps of 27.5291-04.631 and 27.6164-03.373 as observed with the Australian LBA at 2.3 GHz over two epochs. Map statistics for the individual maps are shown in Table \ref{tab:tabimage}. Contours are drawn at $\pm2^{\frac{1}{2}}, \pm2^{1}, \pm2^{\frac{3}{2}}, \cdots$ times the $3\sigma$ rms noise.}
\label{fig:figlr}            
\end{center}
\end{figure}

%
%
An intriguing feature of our VLBI images of NGC 4945 is the jet-like morphology of $27.5291-04.631$, seen in Figures \ref{fig:figlr}(a) and \ref{fig:figlr}(b) and, at higher resolution in Figures \ref{fig:fighra}(g), \ref{fig:fighra}(h) and \ref{fig:fighra}(i).  The morphology of this source is unlike that of any of the other radio sources seen in our images of the NGC 4945 starburst.  Rather than being a compact or somewhat resolved discreet source, $27.5291-04.631$ has an appearance strongly suggestive of the core-jet morphologies seen in arcsecond and milliarscsecond scale radio images of powerful AGN - radio galaxies and quasars. $27.5291-04.631$ consists of a bright and unresolved component, located at the north-east end of an elongated structure of approximately 250 mas angular length and $<80$ mas angular width (5 pc and 1.5 pc, respectively, at the 3.82 Mpc distance of NGC 4945).  The elongated structure consists of at least 4 regions of enhanced brightness when imaged at low resolution (Figures \ref{fig:figlr}(a) and \ref{fig:figlr}(b)).  In the high resolution images in Figures \ref{fig:fighra}(g), \ref{fig:fighra}(h) and \ref{fig:fighra}(i), the elongated structure appears to be almost more resolved, with no particularly bright, compact components standing out, aside from the compact component at the north-eastern end of the structure. This indicates a near-uniform surface brightness distribution along the length of the elongated structure.  The images at the two observing epochs are also highly consistent in the features detected.  In particular, high resolution images from the data at both epochs show a highly linear elongated structure close to the compact north-east component, but a clear apparent bend in the structure at the south-west end of the structure.

What is the nature of $27.5291-04.631$?  As noted above, the morphological similarity of the source to a typical AGN core-jet source is striking.  We explore this possibility below.  Another possibility is that $27.5291-04.631$ is the result of a chance alignment along the line of sight of multiple supernova remnants, similar to the other remnants detected in our VLBI image.  While a possibility, this interpretation does not appear overwhelmingly likely.  For example, having detected $\sim10$ remnants in an angular area of 25 sq. arcseconds (Figure \ref{fig:figLBANGC4945}, and assuming a uniform distribution of remnants in this area, the chance that 3 remnants (probably the minimum required to explain the $27.5291-04.631$ structure) occur in an area of $\sim0.1$ sq. arcseconds is $\sim0.01$.  Further, the high resolution observations in Figure \ref{fig:fighra} provide a further filter on the data, only detecting emission from the highest brightness temperature sources in the field.  Along with $27.6164-03.373$ and $27.5734-03.793$, $27.5291-04.631$ was one of only three high brightness temperature sources detected.  If we revise the above calculation with a total of 5 high brightness temperature sources detected, and again assume a uniform distribution of these sources over the area of the starburst region, the requirement that 3 of these 5 lie together to explain the structure of $27.5291-04.631$ has a probability of only $\sim0.001$.  Even assuming an increase in the density of remnants toward the centre of the starburst does not make the chance alignment argument a plausible one.

The interpretation of $27.5291-04.631$ as a jet implies that the compact component at the north-east end of the elongated structure would coincide with a compact, massive object that provides the engine for the jet, a massive black hole being the likely candidate.  In AGN, the supermassive black holes coincide with the dynamical centroid of the host galaxy.  This does not seem to be the case in NGC 4945, as can be seen from Figure \ref{fig:figLBANGC4945}.

The one-sided appearance of the jet suggests that the jet may be relativistic.  We measure the jet to counterjet surface brightness ratio and find $R > 22$.  Assuming the continuous slab jet model \citep{Blandford:1977p21002}, a constraint can be placed on the product $\beta Cos \theta$, where $\beta$ is the speed of the radio emitting jet material (as a fraction of the speed of light) and $\theta$ is the angle that the motion of the jet material makes to our line-of-sight. In this calculation we assume optically thin emission and a jet spectral index of $\alpha=-0.7$ where $S_{\nu}\propto\nu^{\alpha}$. We estimate $\beta Cos \theta > 0.52$, using a value of $R > 22$. Since we only have a limit on $R$, it is not possible to derive hard constraints on $\beta$, given $\theta$, or vice versa.  A value of $\beta=0.9$ implies $\theta<55^{\circ}$ whereas $\beta=0.52$ implies $\theta=0\arcdeg$ (i.e. perfectly aligned with our line of sight).

An alternative explanation of the one-sided appearance of the assumed jet is that the counterjet is heavily free-free absorbed. Our modelling of the jet SED suggests that it is significantly affected by free-free absorption ($\tau=17-18$), the counterjet would presumably be affected to an even greater degree. In our images, the extent of the jet feature ($\sim5$ pc) is significantly shorter than similar features in other Seyfert galaxies, such as NGC 6240 \citep{Gallimore:2004p24805}, Mrk 348 \citep{Peck:2003p24764}, NGC 4151 \citep{Mundell:1995p24903} and NGC 1068 \citep{Gallimore:1996p24849,Gallimore:1996p24868} which have jet features 9 pc, 50 pc, 260 pc and 1.3 kpc in extent, respectively. It is thus feasible that the counterjet is completely obscured by free-free absorption behind a dense ionised torus. Evidence for free-free absorption in gaseous disks has been found in other Seyferts \citep{Gallimore:1999p24742} and would be consistent with the observations in NGC 4945.

If the jet interpretation is correct then this would be the nearest detection of such a feature to date. No such structures are seen in the two prototypical starburst galaxies, M82 \citep{Pedlar:1999p3534,McDonald:2002p3330} and NGC 253 \citep{Lenc:2006p6695}. The next nearest detection of a Seyfert jet is in NGC 4151 at a distance of 13.3 Mpc \citep{Mundell:1995p24903}. This presents the opportunity to study jet physics at the highest possible spatial resolution.

The jet interpretation for 27.5291-04.631 is easily testable with further observations.  If monitored every 1$-$2 years, the motion of components away from the core may be detectable.  In particular, two bright components appear separated from the core by approximately 30 mas, at both epochs shown in Figure \ref{fig:fighra}.  We have measured the component separations from the core for these components and can put a limit on their motion, relative to the core, of $\sim10$ mas yr$^{-1}$.  For an intrinsic motion of $\beta=0.9$ and an angle to the line of sight of $45^{\circ}$, we calculate an apparent motion of $\sim25$ mas yr$^{-1}$.  Given this ballpark calculation, several years of monitoring would be required to confidently detect any motion of jet components.  We are continuing to monitor NGC 4945 in order to test the jet hypothesis by directly detecting the transport of energy from the core along the jet.

\subsection{The supernova rate in NGC 4945}
\label{sec:snrate}

A lower limit on the supernova rate for type II supernovae associated with massive stars can be estimated, based on the number of detected remnants, their size and an assumed expansion rate. We detect of order 10 compact radio sources with sizes limits of $0.5-2$ pc, based on a distance of 3.82 Mpc to NGC 4945, that exhibit steep intrinsic spectra often associated with synchrotron emission from type II supernova remnants. Assuming an expansion rate of 10,000 km s$^{-1}$ (rate of expansion of FWHM of a Gaussian model), we estimate a supernova rate of $>0.1-0.4$ yr$^{-1}$. This is a lower limit because sources may have been missed as a result of being intrinsically weak, significantly free-free absorbed, or fully resolved. As our VLBI observations of NGC 4945 currently span a period of less than two years, no significant evolution is detected in the supernova remnants and their mean expansion rate remains unclear. It is therefore appropriate to rewrite the supernova rate limit as $\nu_{SN}>0.1 (v/10^{4})$ yr$^{-1}$, where $v$ is the shell radial expansion velocity in km s$^{-1}$.

It is interesting to note that three times as many VLBI sources are detected in NGC 4945 compared to NGC 253. Yet it is assumed, by considering the effects of confusion at 8.3 GHz in VLA observations, that there may be of order 100 compact radio sources in NGC 253 \citep{Ulvestad:1994p1041}, approximately 10 times the number detected in NGC 4945 at a similar frequency with the ATCA. In section \S~\ref{sec:p3ffmodel} we discussed the detrimental effects of confusion and diffuse emission on the thermal source counts in NGC 4945. The same problematic effects limit counts of compact steep spectrum sources with the ATCA but even more so as these sources will be significantly weaker at higher frequencies. So it is likely that there are a great deal more sources in NGC 4945 than our observations lead us to believe. How this will effect the lower limit of the supernova rate is unclear and will depend on both the age and number of sources in this population.

An upper limit on the type II supernova rate may be estimated by statistically modelling the effect of non-detections. \citet{Lenc:2006p6695} developed a Monte Carlo simulation based on a hypothetical population of supernova remnants with a uniformly distributed peak luminosity between 5 and 20 times that of Cas A \citep{Weiler:1989p17306,Ulvestad:1991p1008}. This model has been adapted for NGC 4945 by scaling the flux density of the supernova remnants to the distance of this galaxy, 3.82 Mpc. Similarly, simulation parameters were set accordingly for the two epochs of VLBI observations and are summarised in Table \ref{tab:tabsnrate}. Our simulations suggest that our VLBI observations should have detected all of the hypothetical supernovae that were simulated for between the two epochs assuming there were no significant absorption effects. Given this simple model we determine a supernova rate of $<1.62$ yr$^{-1}$ at a 95\% confidence level. However, it is known that the nuclear region of NGC 4945 is highly obscured and it has been shown, section \S~\ref{sec:p3ffmodel}, that the nuclear region is also situated behind a screen of ionized gas. In a second, more complete simulation, the effects of free-free absorption on the upper limit of the supernova rate are taken into consideration. For this test a median value of $\tau_{0}=11.8$ was taken from the free-free modelling in \S~\ref{sec:p3ffmodel}. With this simulation only 11\% of the hypothetical supernovae were detectable in the second epoch observation and the estimated supernova rate increased to $<15.3$ yr$^{-1}$.

A further method of estimating the supernova rate is to estimate the median age of the supernova remnant population based on the observed flux variation between two epochs. In section \S~\ref{sec:fluxvar} we observed no significant variation in the detected sources above our error limits. However, if we assume that the median variation in the strongest of sources is real then we find an upper limit of $\sim0.8$\% yr$^{-1}$ to the fading rate in NGC 4945. If the supernovae are assumed to fade in time as the $-0.7$ power of time \citep{Weiler:1986p9393} then we estimate that the median age of the supernova remnant population of $\sim85$ years. This would imply a supernova rate of $\sim0.12$ yr$^{-1}$, a value that falls within the limits determined above.

\subsection{The star formation rate in NGC 4945}
\label{sec:sfrate}

The star-formation rate (SFR) for stars with mass $M\geq5M_{\Sun}$ of a star-forming galaxy can be directly related to its radio luminosity $L_{\nu}$ at wavelength $\nu$ \citep{Condon:1992p10540, Haarsma:2000p10556}. This relation is understood purely from radio considerations by calculating the contribution of synchrotron radio emission from supernova remnants and of thermal emission from \ion{H}{2} regions to the observed radio luminosity. For consistency, we have taken the same form of this relation and the same assumptions described in \citet{Lenc:2006p6695}. Based on the total integrated flux densities measured in the ATCA observations between 8.64 GHz and 23 GHz, we derive a SFR of 14.4$\pm$1.4 $(Q/8.8)$ M$_{\Sun}$ yr$^{-1}$ for the inner $\sim$300 pc of the nuclear region of NGC 4945, where the factor $Q$ accounts for the mass of stars in the interval $0.1-100M_{\sun}$ and has a value of 8.8 if a Saltpeter IMF ($\gamma=2.5$) is assumed \citep{Haarsma:2000p10556}. Similar measurements based on the 24 GHz continuum of NGC 253 give a comparable result of $10.4\pm1.0$ M$_{\Sun}$ yr$^{-1}$ \citep{Lenc:2006p6695}.


Estimates of the SFR can be made based on the measured FIR luminosity $L_{FIR}$ since a large proportion of the bolometric luminosity of a galaxy is absorbed by interstellar dust and reemitted in thermal IR. \citet{Condon:1992p10540} determined the relationship between the SFR of a galaxy and the FIR luminosity by modelling the total energy emission of a massive star and assuming that the contribution of dust heating by old stars was negligible. In NGC 4945 this may be complicated by the presence of an AGN which, based on observations of the relatively high HCN/CO ratio, may contribute as much as 50\% of the bolometric luminosity \citep{Spoon:2000p7502, Curran:2001p10543}. However, more recent observations of CO spectral lines in the nuclear region suggest that star formation, rather than AGN activity, is the primary heating agent \citep{Chou:2007p10537} and so the SFR/FIR relationship should still provide a good estimate of the star formation rate.

For NGC 4945, the far-infrared luminosity, $L_{FIR}$, is $1.69\times10^{10}$ $L_{\sun}$ when adjusted for a distance of 3.82 Mpc \citep{Rice:1988p7793}. Based on \citet{Condon:1992p10540}, this leads to a star formation rate of 1.5 $M_{\sun}$ yr$^{-1}$. This is significantly lower than earlier estimates (8.3 $M_{\sun}$ yr$^{-1}$) based on total FIR luminosity \citep{Dahlem:1993p10544} as a result of the revised distance to NGC 4945. However, the new estimate is comparable to the value $2-8$ $M_{\sun}$ yr$^{-1}$ derived from the mass of ionised gas inferred from radio recombination line observations from the galaxy \citep{2005AIPC..783..303R}. This estimate is also compatible with the lower limit on the star formation rate, $>0.19$ $M_{\sun}$ yr$^{-1}$, derived from H$\alpha$ emission \citep{Strickland:2004p9721}.

For most galaxies it is also possible to directly relate the star formation rate with the supernova rate \citep{Condon:1992p10540}. Using the relation derived by \citet{Condon:1992p10540}, the supernova rate limits of $>0.1 (v/10^{4})\nu_{SN}<15.3$ yr$^{-1}$ (\S~\ref{sec:snrate}) give the limits $2.4 (v/10^{4}) < SFR(M\geq5M_{\Sun})<370$ M$_{\Sun}$ yr$^{-1}$ on the star formation rate. While the upper limit is poorly constrained due to the limited number of epochs thus far observed, it is of the same order of magnitude as the nuclear SFR determined from FIR emissions (1.5 M$_{\Sun}$ yr$^{-1}$) and in agreement with that determined from radio luminosities alone ($14.4\pm1.4$ $(Q/8.8)$ M$_{\Sun}$ yr$^{-1}$). The higher radio estimate of the SFR, compared to the FIR-based estimate, may suggest that the $Q$ factor has been over-estimated. A lower $Q$ would suggest an increased proportion of massive star formation and would also be consistent with the high densities associated with the nuclear region. Nonetheless, it is encouraging that the independent estimates of SFR are only a factor of a few different.

\subsection{Comparisons to the nearby starbursts M82 and NGC 253}

With the recent improvements in distance measurements, it appears that the distances to NGC 4945, M82 and NGC 253 are all within a few percent of 3.9 Mpc. This is a dramatic change, by up to a factor of two, from previous estimates. Such a change in distance has an significant effect on luminosity estimates, which are proportional to the square of the distance, and to a lesser degree on spatial measures, which are linearly proportional to the distance. With the three galaxies now converging on a similar distance it is interesting that the star formation rates and supernova rates are also converging. Based on FIR luminosities, we find that the star formation rates of NGC 4945, NGC 253 and M82 are now 1.5 $M_{\sun}$ yr$^{-1}$, 1.8$-$2.8 M$_{\Sun}$ yr$^{-1}$ \citep{Lenc:2006p6695} and $\sim2.0$ M$_{\Sun}$ yr$^{-1}$ \citep{Pedlar:2003p3309}, respectively. Similarly, the supernova rates are $>0.1 (v/ 10^{4})$ yr$^{-1}$, $>0.14 (v/10^{4})$ yr$^{-1}$ \citep{Lenc:2006p6695} and $\sim0.07$ yr$^{-1}$ \citep{Pedlar:2003p3309}, for NGC 4945, NGC 253 and M82, respectively, based on source counts and sizes.

\section{Summary}

We have imaged NGC 4945 at 2.3 GHz over two epochs using the LBA to produce the highest resolution images of the nuclear starburst region of this galaxy to date. The VLBI observations have been complemented with additional data between 8.64 and 23 GHz observed with the ATCA. We find the following results:
\begin{itemize}
\item Fifteen compact radio sources were detected with the LBA and 13 were identified higher frequency ATCA images.
\item In the highest resolution image (15 mas beam), four supernova remnants are resolved into shell-like structures ranging in size between 60 and 110 mas (1.1 to 2.1 pc) in diameter. Assuming an average radial expansion velocity of $v=10,000$ km s$^{-1}$, the remnants are estimated to be approximately $25 (10^{4}/v)$ to $100 (10^{4}/v)$ years of age.
\item By combining flux density measurements from 2.3 GHz LBA and high frequency ATCA observations, the spectra for 13 compact radio sources were determined. The spectra of 9 sources were found to be consistent with a free-free absorbed power law and 4 with a simple power law spectrum.
\item Ten of the 13 sources have steep intrinsic spectra normally associated with supernova remnants. The three remaining sources have flat intrinsic power law spectra ($\alpha>-0.4$) indicative of \ion{H}{2} regions but may have embedded supernova remnants owing to their high brightness temperature ($\sim2\times10^{5}$ K) at 2.3 GHz.
\item Based on the modelled free-free opacities of 9 sources, the morphology of the ionised medium in the central region of NGC 4945 is complex and clumpy in nature.
\item A type II supernova rate upper limit of 15.3 yr$^{-1}$ in the inner 250 pc region of NGC 4945 was derived from the absence of any new sources between epochs, taking into consideration the improved distance estimate for the galaxy, a median free-free opacity, and the sensitivity limits of two observations over a period of $\sim2$ years. A type II supernova rate of $>0.1 (v/ 10^{4})$ yr$^{-1}$ has been estimated based on an estimate of supernova remnant source counts, their sizes and their expansion rates.
\item A star formation rate of $2.4 (v/10^{4}) < SFR(M\geq5M_{\Sun})<370$ M$_{\Sun}$ yr$^{-1}$ has been estimated directly from supernova rate limits for the inner 250 pc region of the galaxy.
\item The supernova rates and star formation rates determined for NGC 4945 are, to within a factor of two, similar to those observed in NGC 253 and M82.
\item A jet-like source is detected that is offset by $\sim$ 1050 mas from the assumed location of the AGN based on H$_{2}$O megamaser emission, the HNC cloud centroid, the K-band peak and the hard x-ray peak.
\end{itemize}

We need to more stringently constrain the free-free parameters by observing a further epoch using VLBI at 1.4 GHz. Frequent observations will further constrain the upper limit on the supernova rate by increasing the proportion of supernova events that may be captured between epochs. Observations over a longer period of time ($\sim10$ years) will also further constrain the lower limit of the supernova rate by enabling the supernova shell expansion velocity to be measured, and allow us to search for motion in the suggested jet, thereby confirming its nature.

%
%





\section*{Acknowledgments}

We thank Tim Roberts and Nick Schurch for \emph{Chandra} X-ray images and Alessandro Marconi for HST P$\alpha$ images. E.L. acknowledges support from a Swinburne University of Technology Chancellor's Research Scholarship, a CSIRO Postgraduate Student Research Scholarship and ATNF co-supervision. The Australia Telescope is funded by the Australian Commonwealth Government for operation as a national facility managed by the CSIRO. This research has made use of the NASA/IPAC Extragalactic Database (NED), which is operated by the Jet Propulsion Laboratory, California Institute of Technology, under contract with the National Aeronautics and Space Administration. We thank an anonymous referee for their constructive comments on the manuscript.

\begin{deluxetable}{lcccccccc}
\tabletypesize{\scriptsize}
\tablecolumns{9}
\tablecaption{Summary of observations.}
\tablehead{
   \colhead{Observatory}                    &
   \colhead{Frequency}                      &
   \colhead{$\alpha$}                       &
   \colhead{$\delta$}                       &
   \colhead{Date}                           &
   \colhead{Config.}                        &
   \colhead{Duration}                       &
   \colhead{Bandwidth}                      &
   \colhead{$\Delta$t}                      \\
   \colhead{}                               &
   \colhead{MHz}                            &
   \colhead{(J2000)}                        &
   \colhead{(J2000)}                        &
   \colhead{}                               &
   \colhead{}                               &
   \colhead{(h)}                            &
   \colhead{(MHz)}                          &
   \colhead{(sec)}                          
}
\startdata
ATCA    & 4800.0  & $13\rah5\ram28\fs00$ & $-49\arcdeg28\arcmin12\farcs00$ & 19/20 NOV 1993 & 6A      & 11  & 128 & 15 \\
\nodata & 8640.0  & \nodata              & \nodata                         & \nodata        & \nodata & 11  & 128 & 15 \\
\hline
LBA\tablenotemark{a} & 2252.0  & $13\rah5\ram27\fs50$ & $-49\arcdeg28\arcmin6\farcs00$  & 13/14 MAY 2005 & \nodata & 9   & 16  & 2  \\
\nodata & 2268.0  & \nodata              & \nodata                         & \nodata        & \nodata & 9   & 16  & 2  \\
\hline
ATCA    & 17000.0 & $13\rah5\ram27\fs50$ & $-49\arcdeg28\arcmin6\farcs00$  & 18 MAR 2006    & EW367   & 3 & 128 & 15 \\
\nodata & 19000.0 & \nodata              & \nodata                         & \nodata        & \nodata & 3 & 128 & 15 \\
\nodata & 21000.0 & \nodata              & \nodata                         & \nodata        & \nodata & 3.5 & 128 & 15 \\
\nodata & 23000.0 & \nodata              & \nodata                         & \nodata        & \nodata & 3.5 & 128 & 15 \\
\hline
ATCA    & 17000.0 & $13\rah5\ram27\fs50$ & $-49\arcdeg28\arcmin6\farcs00$  & 24/25 MAR 2006 & 6C      & 9 & 128 & 15 \\
\nodata & 19000.0 & \nodata              & \nodata                         & \nodata        & \nodata & 9 & 128 & 15 \\
\nodata & 21000.0 & \nodata              & \nodata                         & \nodata        & \nodata & 9 & 128 & 15 \\
\nodata & 23000.0 & \nodata              & \nodata                         & \nodata        & \nodata & 9 & 128 & 15 \\
\hline
LBA\tablenotemark{b} & 2269.0  & $13\rah5\ram27\fs50$ & $-49\arcdeg28\arcmin6\farcs00$  & 21 MAR 2007    & \nodata & 11   & 16  & 2 \\
\nodata & 2285.0  & \nodata              & \nodata                         & \nodata        & \nodata & 11   & 16  & 2 \\
\nodata & 2301.0  & \nodata              & \nodata                         & \nodata        & \nodata & 11   & 16 & 2 \\
\nodata & 2317.0  & \nodata              & \nodata                         & \nodata        & \nodata & 11   & 16 & 2 \\

\enddata
\tablenotetext{a}{Australian LBA observation included 9 hours at ATCA and Mopra, Tidbinbilla, Hobart and Ceduna, and 6 hours at Parkes.}
\tablenotetext{b}{Australian LBA observation included 11 hours at ATCA and Mopra, Hobart and Ceduna, 10 hours at Parkes and 5 hours at Tidbinbilla.}
\label{tab:tabobs}
\end{deluxetable}

\begin{deluxetable}{lcccccccc}
\tabletypesize{\scriptsize}
\tablecolumns{7}
\tablecaption{Map Statistics.}
\tablehead{
   \colhead{Figure}                    &
   \colhead{Source}                    &
   \colhead{Frequency}                 &
   \colhead{Synthesized Beam}          &
   \colhead{$\sigma$}                  &
   \colhead{Peak Flux}                 &
   \colhead{Integrated Flux}           \\
   \colhead{}                          &
   \colhead{}                          &
   \colhead{GHz}                       &
   \colhead{(mas)}                     &
   \colhead{(mJy beam$^{-1}$)}         &
   \colhead{(mJy beam$^{-1}$)}         &
   \colhead{(mJy)}                          
}
\startdata
\ref{fig:figJ1237-5046}(a)  & J1237$-$5046   & 2.3 & $13\times10$     & 0.342 & 520  & 1100 \\
\ref{fig:figJ1237-5046}(b)  & J1237$-$5046   & 2.3 & $13\times10$     & 0.292 & 530  & 1100 \\
\ref{fig:figLBANGC4945}(a)  & NGC 4945       & 2.3 & $80\times32$     & 0.152 & 24  & 150 \\
\ref{fig:figLBANGC4945}(b)  & NGC 4945       & 2.3 & $80\times32$     & 0.085 & 23  & 160 \\
\ref{fig:figlr}(a)           & $27.5291-04.631$ & 2.3 & $80\times32$     & 0.152 & 7.3  & 30 \\
\ref{fig:figlr}(b)           & $27.5291-04.631$ & 2.3 & $80\times32$     & 0.085 & 6.8  & 36 \\
\ref{fig:figlr}(c)           & $27.6164-03.373$ & 2.3 & $80\times32$     & 0.152 & 2.1  & 8.5 \\
\ref{fig:figlr}(d)           & $27.6164-03.373$ & 2.3 & $80\times32$     & 0.085 & 1.8  & 8.2 \\
\ref{fig:figATCANGC4945}(a) & NGC 4945       & 4.8 & $2132\times1680$ & 0.520 & 580 & 2700  \\
\ref{fig:figATCANGC4945}(b) & NGC 4945       & 8.6 & $1155\times955$  & 0.202 & 170 & 1300  \\
\ref{fig:figATCANGC4945}(c) & NGC 4945       & 17  & $611\times510$   & 0.250 & 67 & 940  \\
\ref{fig:figATCANGC4945}(d) & NGC 4945       & 19  & $560\times443$   & 0.270 & 58 & 840  \\
\ref{fig:figATCANGC4945}(e) & NGC 4945       & 21  & $495\times399$   & 0.320 & 48 & 800  \\
\ref{fig:figATCANGC4945}(f) & NGC 4945       & 23  & $453\times363$   & 0.340 & 46 & 780  \\
\ref{fig:fighra}(a)          & $27.6164-03.373$ & 2.3 & $16\times15$     & 0.149 & 1.3 & 10 \\
\ref{fig:fighra}(b)          & $27.6164-03.373$ & 2.3 & $16\times15$     & 0.110 & 0.76 & 9.0 \\
\ref{fig:fighra}(c)          & $27.6164-03.373$ & 2.3 & $16\times15$     & 0.075 & 1.0 & 9.6 \\
\ref{fig:fighra}(d)          & $27.5734-03.793$ & 2.3 & $16\times15$     & 0.149 & 1.8 & 11 \\
\ref{fig:fighra}(e)          & $27.5734-03.793$ & 2.3 & $16\times15$     & 0.110 & 2.2 & 12 \\
\ref{fig:fighra}(f)          & $27.5734-03.793$ & 2.3 & $16\times15$     & 0.075 & 1.9 & 12 \\
\ref{fig:fighra}(g)          & $27.5291-04.631$ & 2.3 & $16\times15$     & 0.149 & 3.5 & 38 \\
\ref{fig:fighra}(h)          & $27.5291-04.631$ & 2.3 & $16\times15$     & 0.110 & 3.0 & 30 \\
\ref{fig:fighra}(i)          & $27.5291-04.631$ & 2.3 & $16\times15$     & 0.075 & 3.3 & 34 \\
\ref{fig:fighra}(j)          & $27.4949-05.063$ & 2.3 & $16\times15$     & 0.149 & 2.2 & 5.7 \\
\ref{fig:fighra}(k)          & $27.4949-05.063$ & 2.3 & $16\times15$     & 0.110 & 2.4 & 5.8 \\
\ref{fig:fighra}(l)          & $27.4949-05.063$ & 2.3 & $16\times15$     & 0.075 & 2.3 & 5.7 \\
\ref{fig:fighra}(m)          & $27.4646-06.555$ & 2.3 & $16\times15$     & 0.149 & 2.5 & 5.9 \\
\ref{fig:fighra}(n)          & $27.4646-06.555$ & 2.3 & $16\times15$     & 0.110 & 3.0 & 7.8 \\
\ref{fig:fighra}(o)          & $27.4646-06.555$ & 2.3 & $16\times15$     & 0.075 & 2.8 & 6.9 \\
\ref{fig:fighra}(p)          & $27.4646-05.174$ & 2.3 & $16\times15$     & 0.149 & 0.66 & 2.3 \\
\ref{fig:fighra}(q)          & $27.4646-05.174$ & 2.3 & $16\times15$     & 0.110 & 0.82 & 4.1 \\
\ref{fig:fighra}(r)          & $27.4646-05.174$ & 2.3 & $16\times15$     & 0.075 & 0.70 & 3.2 \\
\ref{fig:fighra}(s)          & $27.3861-07.554$ & 2.3 & $16\times15$     & 0.149 & 1.1 & 1.0 \\
\ref{fig:fighra}(t)          & $27.3861-07.554$ & 2.3 & $16\times15$     & 0.110 & 0.91 & 2.3 \\
\ref{fig:fighra}(u)          & $27.3861-07.554$ & 2.3 & $16\times15$     & 0.075 & 0.97 & 1.6 \\
\ref{fig:fighra}(v)          & $27.3659-06.900$ & 2.3 & $16\times15$     & 0.149 & 1.1 & 3.1 \\
\ref{fig:fighra}(w)          & $27.3659-06.900$ & 2.3 & $16\times15$     & 0.110 & 0.84 & 5.7 \\
\ref{fig:fighra}(x)          & $27.3659-06.900$ & 2.3 & $16\times15$     & 0.075 & 0.93 & 4.4 \\
\ref{fig:fighra}(y)          & $27.2710-06.592$ & 2.3 & $16\times15$     & 0.149 & 11 & 37 \\
\ref{fig:fighra}(z)          & $27.2710-06.592$ & 2.3 & $16\times15$     & 0.110 & 9.1 & 30 \\
\ref{fig:fighra}(aa)         & $27.2710-06.592$ & 2.3 & $16\times15$     & 0.075 & 9.9 & 33 \\






\enddata
\label{tab:tabimage}
\end{deluxetable}

\begin{deluxetable}{lccccccccc}
\tabletypesize{\scriptsize}
\tablecolumns{10}
\tablecaption{Source flux densities.}
\tablehead{
   \colhead{Source}            &
   \colhead{FWHM Size\tablenotemark{a}} &
   \colhead{P.A.\tablenotemark{b}}      &
   \colhead{S$_{2.3} (2005)$}  &
   \colhead{S$_{2.3} (2007)$}  &
   \colhead{S$_{8.64}$}         &
   \colhead{S$_{17}$}          &
   \colhead{S$_{19}$}          &
   \colhead{S$_{21}$}          &
   \colhead{S$_{23}$}          \\
   \colhead{}                  &
   \colhead{(mas)}                      &
   \colhead{(degrees)}                  &
   \colhead{(mJy)}             &
   \colhead{(mJy)}             &
   \colhead{(mJy)}             &
   \colhead{(mJy)}             &
   \colhead{(mJy)}             &
   \colhead{(mJy)}             &
   \colhead{(mJy)}                          
}
\startdata
$27.9681-00.044$ & $155\times43$                     & 68  & 6.8    & 5.1    & \nodata & 3.4    & 3.6    & \nodata  & \nodata \\
$27.7631-02.880$ & $91\times44$                      & 73  & 1.6    & 2.9    & 17    & 8.2    & 6.8    & 5.1     & 4.0    \\
$27.6164-03.373$ & $105\times54$                     & 44  & 13    & 12    & 8.4    & 6.0    & 3.9    & 3.0     & 2.7    \\
$27.5734-03.793$ & $87\times56$                      & 83  & 18   & 18    & 13    & 6.3    & 7.2    & 6.6     & 4.9    \\
$27.5607-04.779$\tablenotemark{c} & $97\times62$                      & 45  &9.6    & 8.6    & 20    & 8.1    & 6.6    & 5.6     & 6.7    \\
$27.5402-04.720$\tablenotemark{c} & $156\times12$                     & -55 &9.7    & 15    & \nodata & \nodata & \nodata & \nodata & \nodata \\
$27.5291-04.631$\tablenotemark{c} & $72\times37$                      & 73  &15    & 14    & \nodata & 17    & 16    & 13     & 11    \\
$27.4949-05.063$ & $67\times29$                      & 72  & 8.3    & 8.2    & 38    & 20    & 19    & 14     & 16    \\
$27.4722-04.705$ & $85\times37$                      & 90  & \nodata & 2.9    & 16    & \nodata & \nodata & \nodata  & \nodata \\
$27.4646-05.174$ & $77\times62$                      & 59  & 9.1    & 9.0    & \nodata & \nodata & \nodata & \nodata  & \nodata \\
$27.4646-06.555$ & $79\times33$                      & 77  & 11    & 12    & 15    & 8.9    & 9.0    & 9.7     & 8.8    \\
$27.3861-07.554$ & $84\times39$                      & -84 & 3.4    & 4.0    & 2.8    & 2.3    & 1.3    & 4.5     & 6.1    \\
$27.3659-06.900$ & $79\times35$                      & 72  & 6.3    & 6.4    & 14    & 7.5    & 7.9    & 7.6     & 6.4    \\
$27.2963-07.073$ & $78\times39$                      & 79  & 5.2    & 4.1    & \nodata & 1.3    & 1.7    & 2.4     & 1.2    \\
$27.2710-06.592$ & $54\times24$                      & 83  & 36    & 33    & 34    & 13    & 12    & 12     & 10    \\
\hline
$27.4833-04.570$\tablenotemark{d} & $5764\times982$\tablenotemark{e}  & 42  & \nodata & \nodata & 470     & 300     & 280     & 240      & 230     \\
$27.4602-05.845$\tablenotemark{d} & $422\times422$\tablenotemark{e}   & -51 & \nodata & \nodata & 17      & 19      & 16      & 18       & 18      \\
$27.4525-05.245$\tablenotemark{d} & $566\times311$\tablenotemark{e}   & 34  & \nodata & \nodata & 130     & 66      & 59      & 62       & 57      \\
$27.4371-05.695$\tablenotemark{d} & $9485\times3059$\tablenotemark{e} & 41  &\nodata & \nodata & 600     & 430     & 390     & 390      & 370     \\
\enddata
\tablenotetext{a}{Gaussian component sizes measured against short-baseline 2.3 GHz LBA data except where noted otherwise.}
\tablenotetext{b}{Position angle of major axis measured east of north.}
\tablenotetext{c}{$27.5291-04.631$, $27.5402-04.720$ and $27.5607-04.779$ are the apparent source, jet and termination of a jet like source.}
\tablenotetext{d}{Sources used to model diffuse emission observed in ATCA data.}
\tablenotetext{e}{Sizes of large-scale diffuse emission sources measured against 17 GHz ATCA data.}
\label{tab:tabfluxes}
\end{deluxetable}

\begin{deluxetable}{lccccc}
\tablecolumns{6}
\tablecaption{Parameters of the free-free absorption models for all detected sources.}
\tablehead{
   \colhead{Source}                     &
   \colhead{$S_{0}$}                    &
   \colhead{$\alpha$}                   &
   \colhead{$\tau_{0}$}                 &
   \colhead{Type\tablenotemark{a}}      &
   \colhead{$S_{1.4}$\tablenotemark{b}} \\
   \colhead{}                           &
   \colhead{(mJy)}                      &
   \colhead{}                           &
   \colhead{}                           &
   \colhead{}                           &
   \colhead{(mJy)}
}
\startdata
$27.9681-00.044$ &  5.99    & -0.19   & \nodata & T       & 5.62    \\
$27.7631-02.880$ & 534     & -1.49   & 22.9    & S       & 0.004   \\
$27.6164-03.373$ &  84.5    & -1.03   & 6.16    & S       & 2.87    \\
$27.5734-03.793$ &  101     & -0.93   & 5.57    & S       & 4.74    \\
$27.5607-04.779$ &  551     & -1.47   & 16.9    & S       & 0.081   \\
$27.5402-04.720$ &  \nodata & \nodata & \nodata & \nodata & \nodata \\
$27.5291-04.631$ &  1110    & -1.45   & 18.4    & S       & 0.079   \\
$27.4949-05.063$ &  608     & -1.19   & 19.1    & S       & 0.033   \\
$27.4722-04.705$ &  0.99    & 1.29    & \nodata & T       & 1.52    \\
$27.4646-05.174$ &  \nodata & \nodata & \nodata & \nodata & \nodata \\
$27.4646-06.555$ & 67.1    & -0.66   & 6.62    & S       & 2.05    \\
$27.3861-07.554$ &  3.31    & -0.02   & \nodata & T       & 3.33    \\
$27.3659-06.900$ &  103     & -0.88   & 11.8    & S       & 0.23    \\
$27.2963-07.073$ &  5.72    & -0.42   & \nodata & S       & 4.97    \\
$27.2710-06.592$ &  745     & -1.38   & 11.2    & S       & 1.84    \\
\hline
$27.4833-04.570$ &  2150    & -0.70   & \nodata & \nodata & 1690    \\
$27.4602-05.845$ &  16.9    &  0.02   & \nodata & \nodata & 17.0    \\
$27.4525-05.245$ &  875     & -0.89   & \nodata & \nodata & 648     \\
$27.4371-05.695$ &  1710    & -0.49   & \nodata & \nodata & 1450    \\
\enddata
\tablenotetext{a}{As discussed in the text. Thermal sources are shown as type T, and non-thermal synchrotron sources are shown as type S.}
\tablenotetext{b}{Estimated flux density at 1.4 GHz based on model parameters.}
\label{tab:tabsrcff}
\end{deluxetable}

\begin{deluxetable}{llllllll}
\tablecolumns{8}
\tablecaption{The supernova rate upper limit based on Monte Carlo simulations run over two observing epochs.}
\tablehead{
   \colhead{Epoch}                      &
   \colhead{Time}                       &
   \colhead{$\nu_{obs}$}                &
   \colhead{Sensitivity}                &
   \colhead{$\beta_{SN}$\tablenotemark{a}}                           &
   \colhead{$\nu_{SN}$\tablenotemark{a}}                 &
   \colhead{$\beta_{SN}$\tablenotemark{b}}                           &
   \colhead{$\nu_{SN}$\tablenotemark{b}}                 \\
   \colhead{}                           &
   \colhead{(yr)}                       &
   \colhead{(GHz)}                      &
   \colhead{(mJy)}                      &
   \colhead{}               &
   \colhead{(yr$^{-1}$)}                &
   \colhead{}               &
   \colhead{(yr$^{-1}$)}
}
\startdata
1     & $\cdots$ & 2.3         & 0.91         & $\cdots$     & $\cdots$                     & $\cdots$     & $\cdots$     \\  
2     & 1.9      & 2.3         & 0.51         & 1.0          & $<1.62$                      & 0.11         & $<15.3$      \\
\enddata
\tablenotetext{a}{Monte Carlo test run with no modelling of free-free absorption.}
\tablenotetext{b}{Monte Carlo test run with a median free-free opacity of $\tau_{0}=11.8$ applied.}
\tablecomments{The supernova rate upper limit based on Monte Carlo simulations run over two observing epochs. The time between epochs, the observing frequency and sensitivity of the observation are listed. At the end of the second epoch the proportion of supernova remnants detected ($\beta_{SN}$) in that epoch is listed together with an estimate of the supernova rate upper limit based on all observations prior to and including that epoch. }
\label{tab:tabsnrate}
\end{deluxetable}

\clearpage

\begin{figure}
\epsscale{0.28}
\begin{center}
\mbox{
\plotone{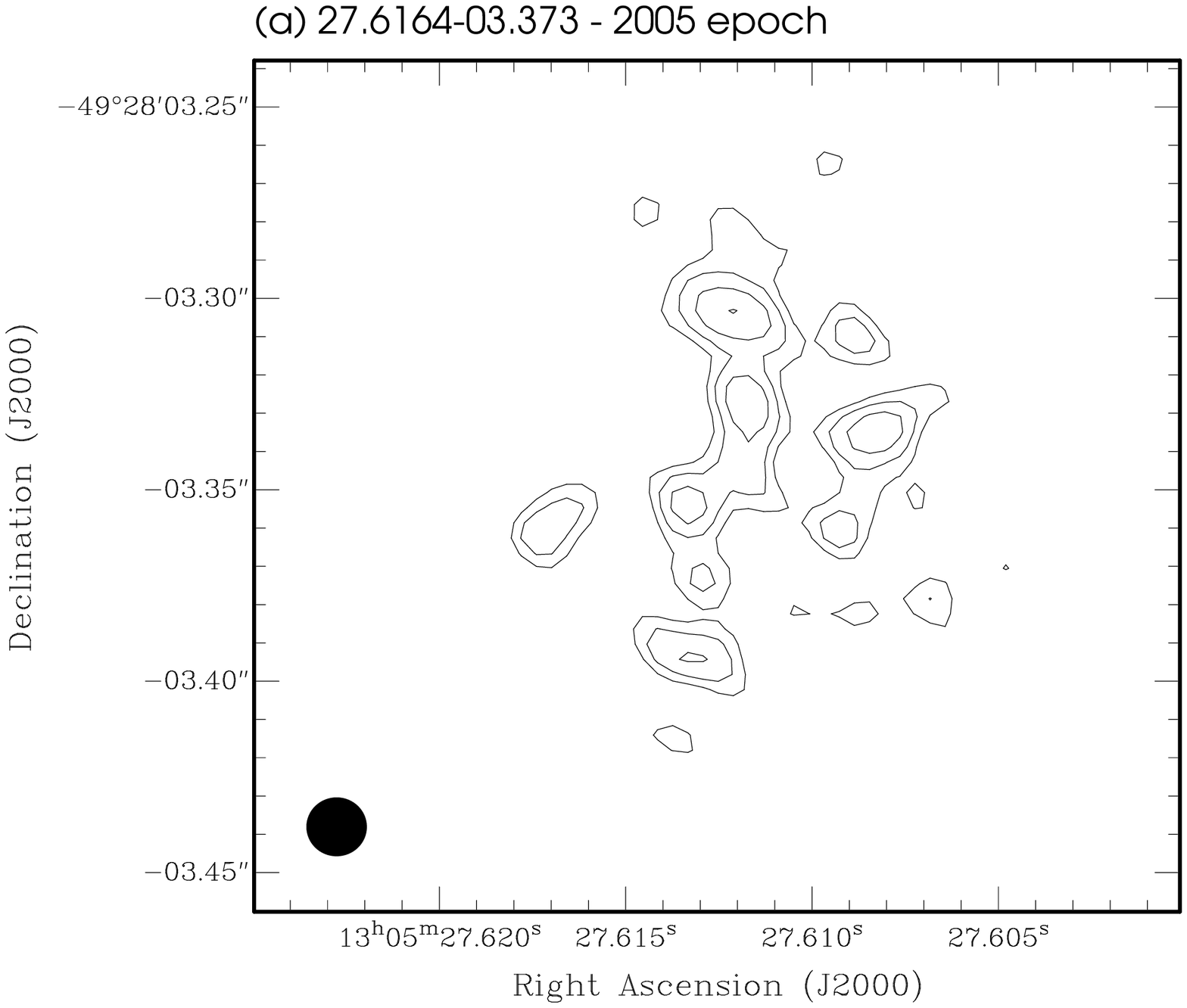} \quad
\plotone{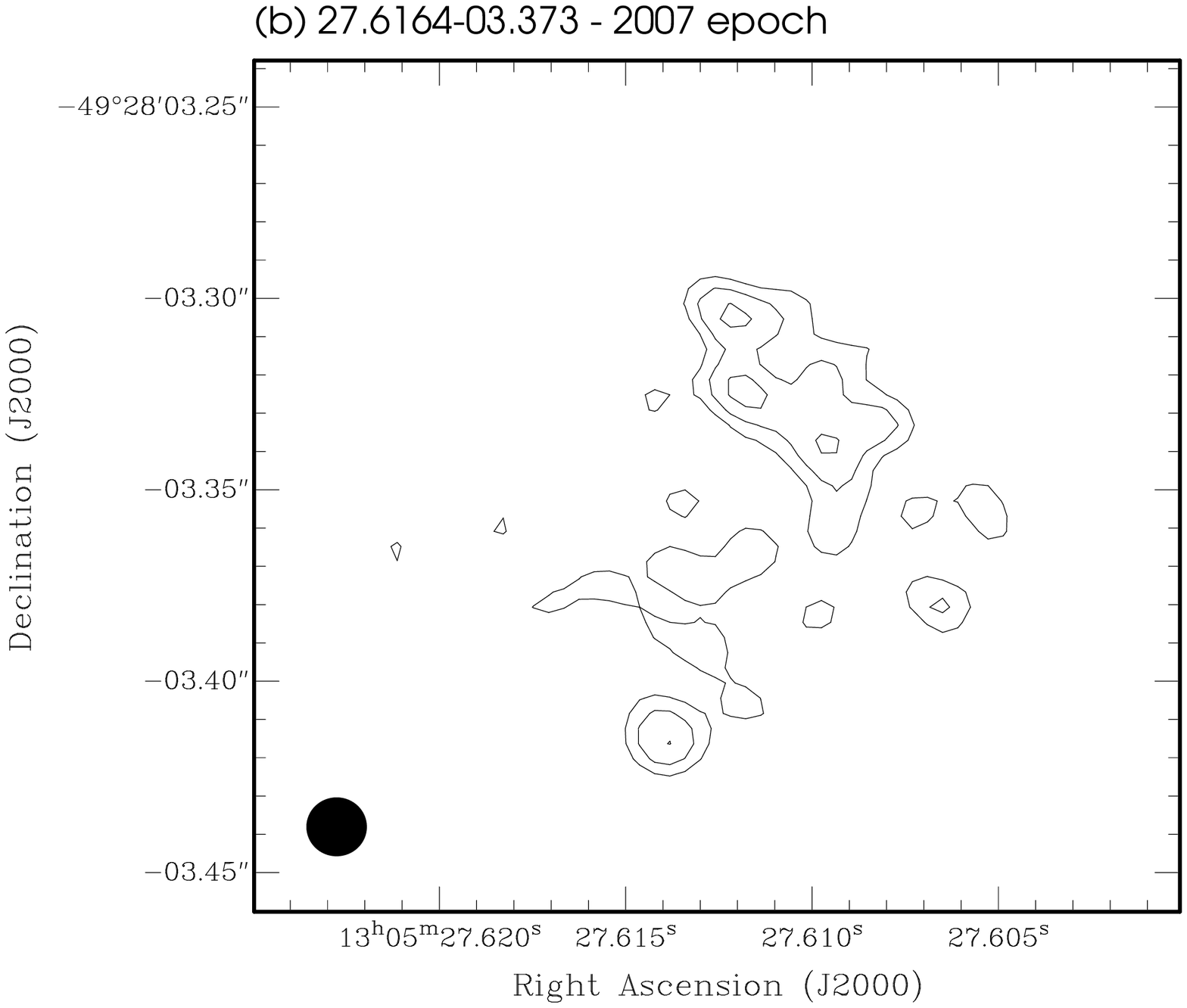} \quad
\plotone{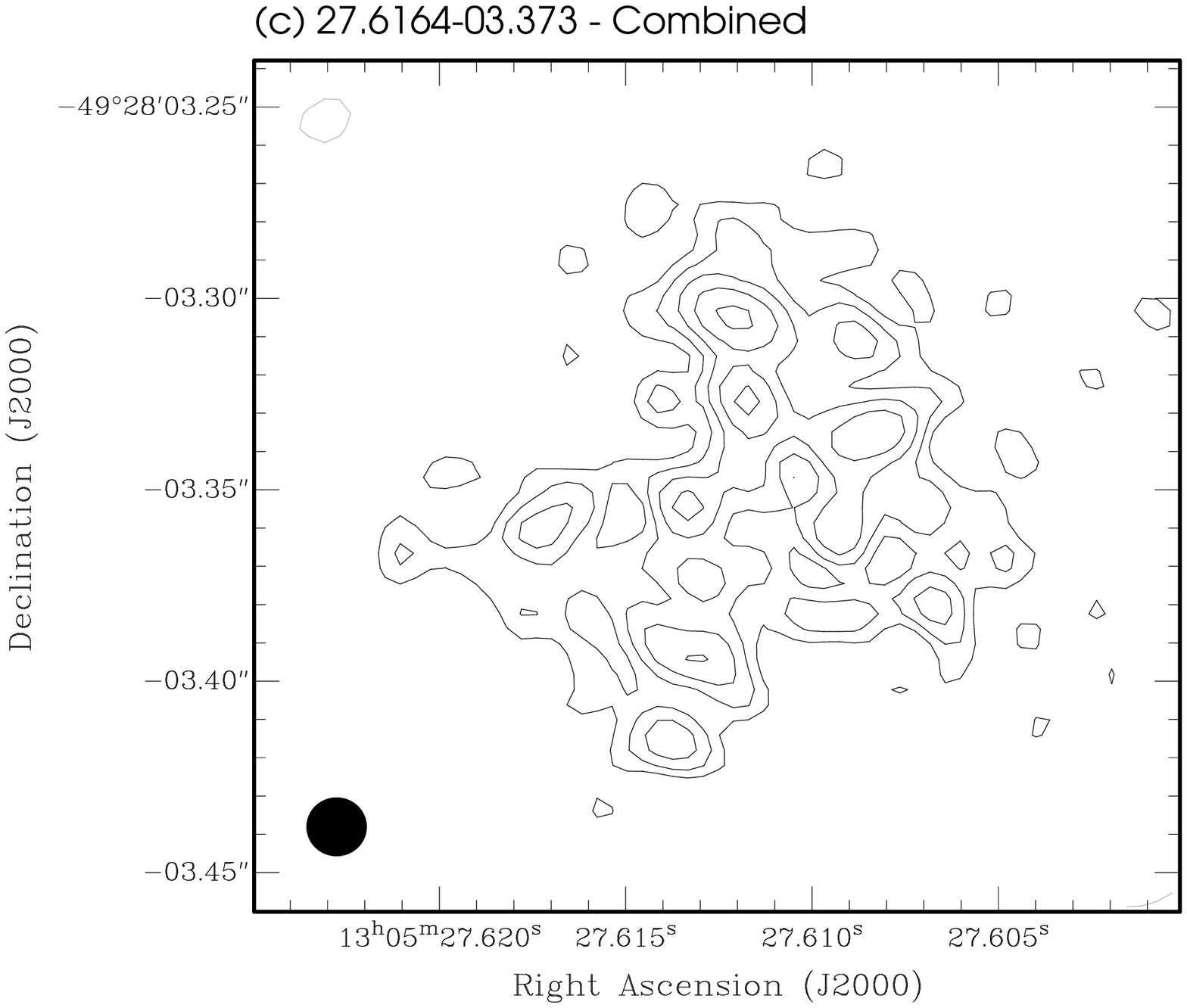}
}
\mbox{
\plotone{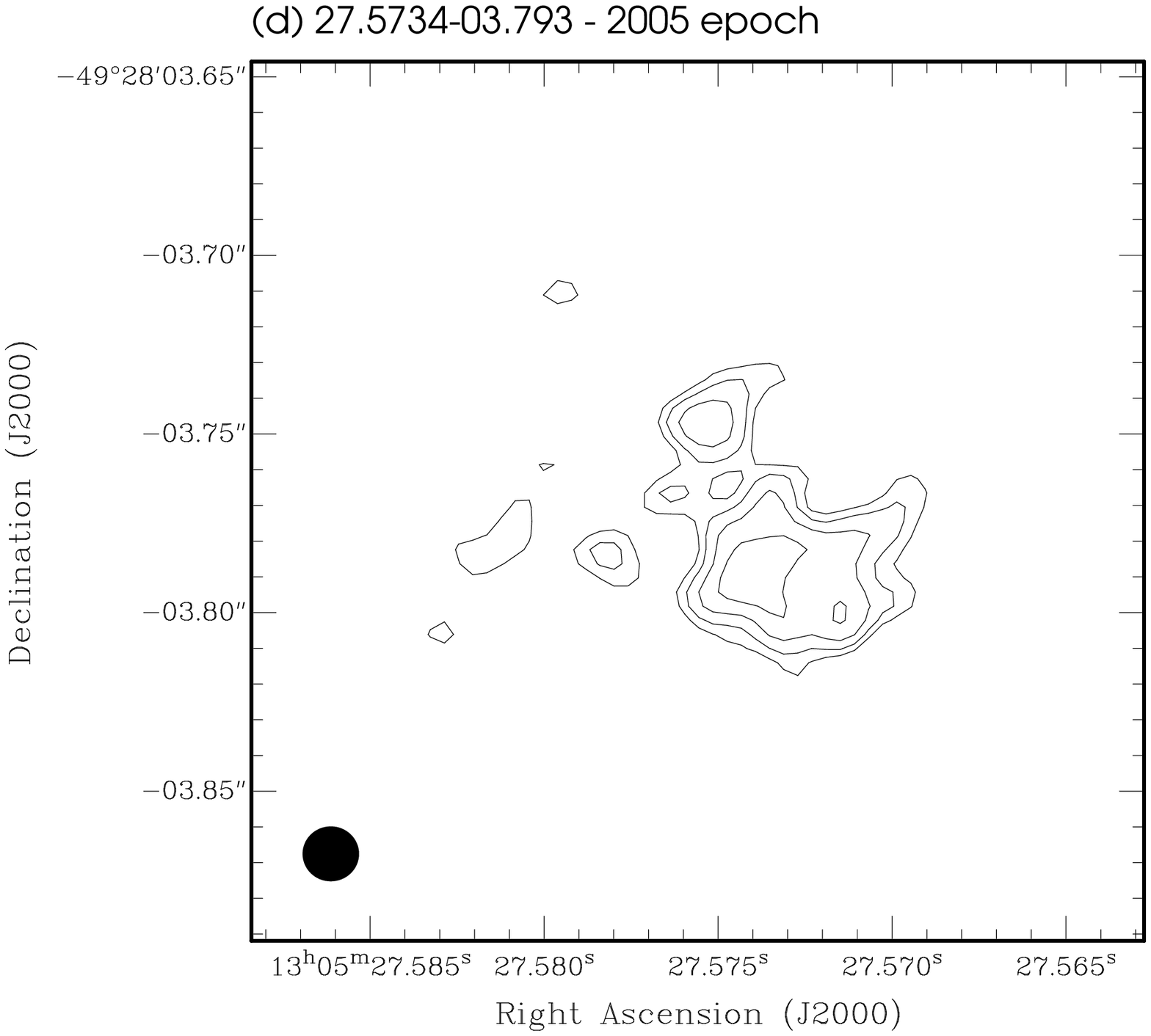} \quad
\plotone{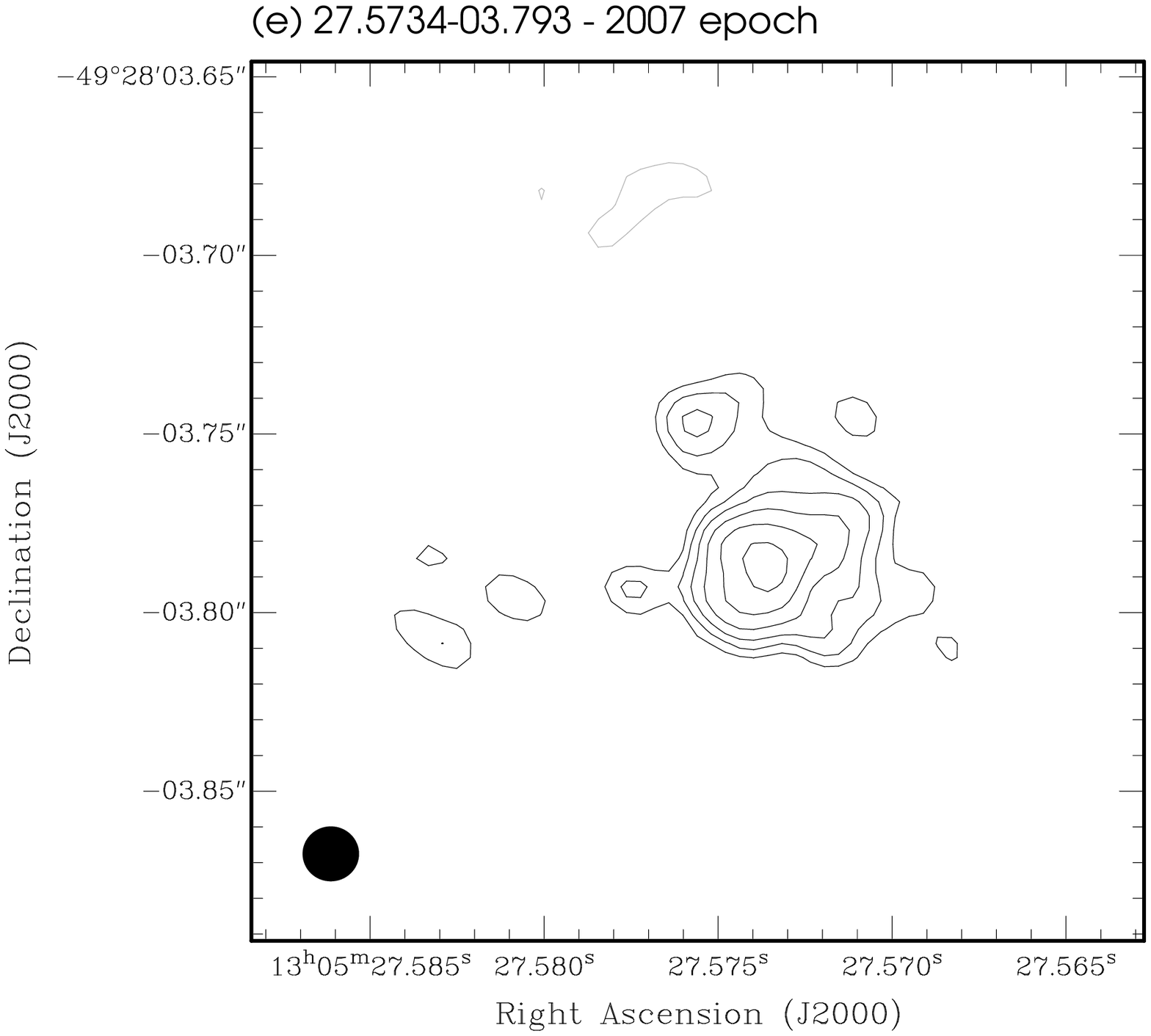} \quad
\plotone{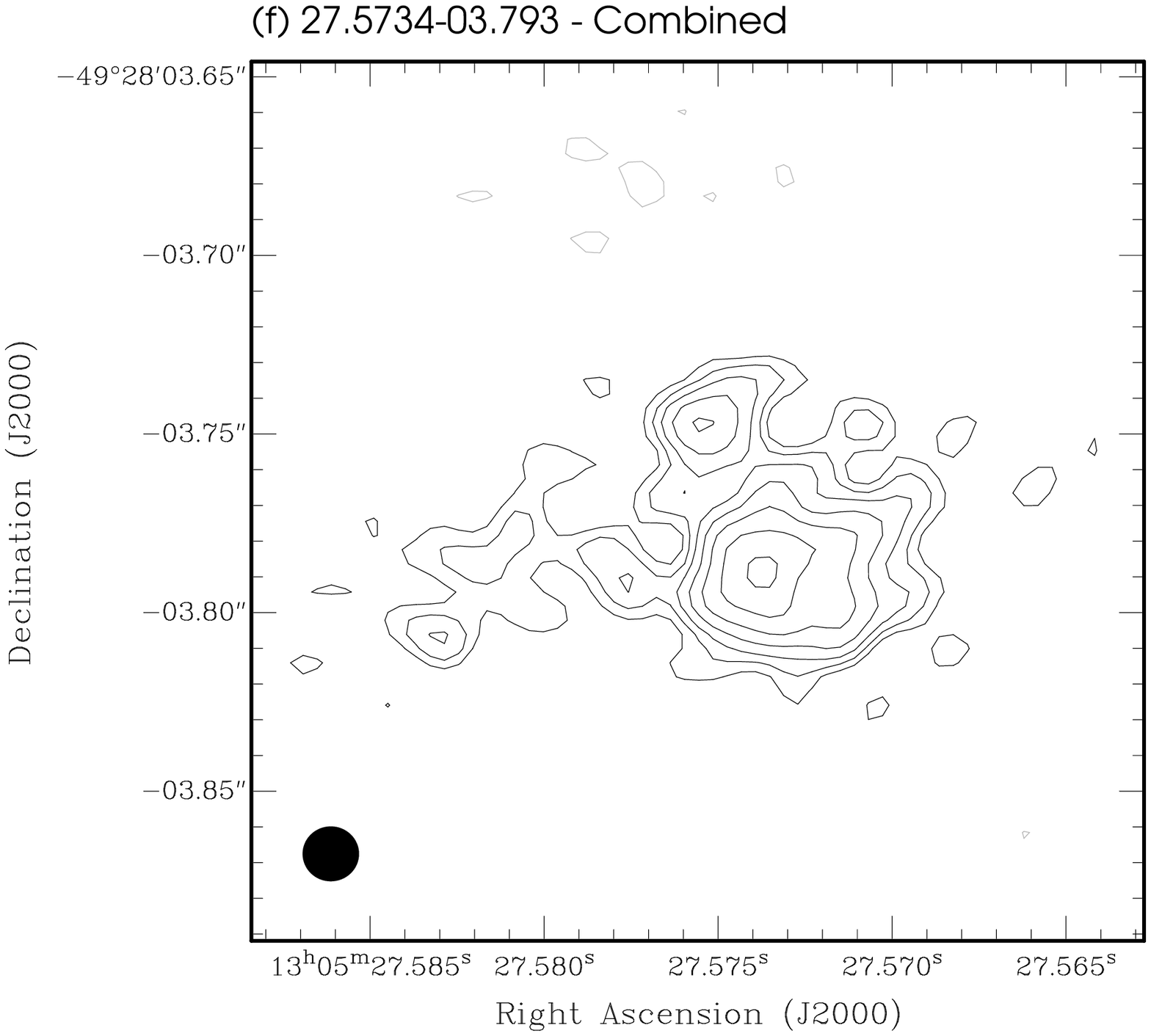}
}
\mbox{
\plotone{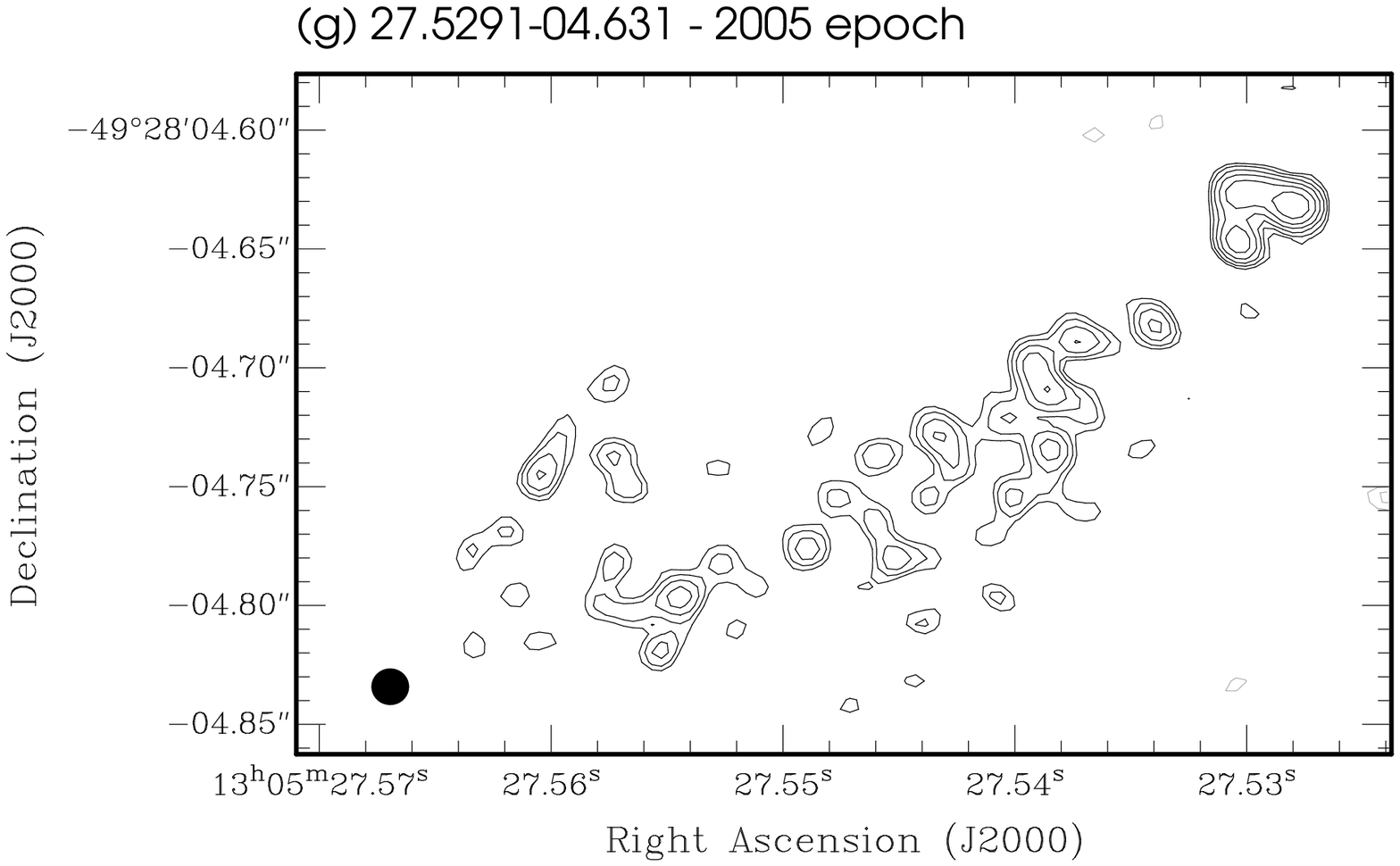} \quad
\plotone{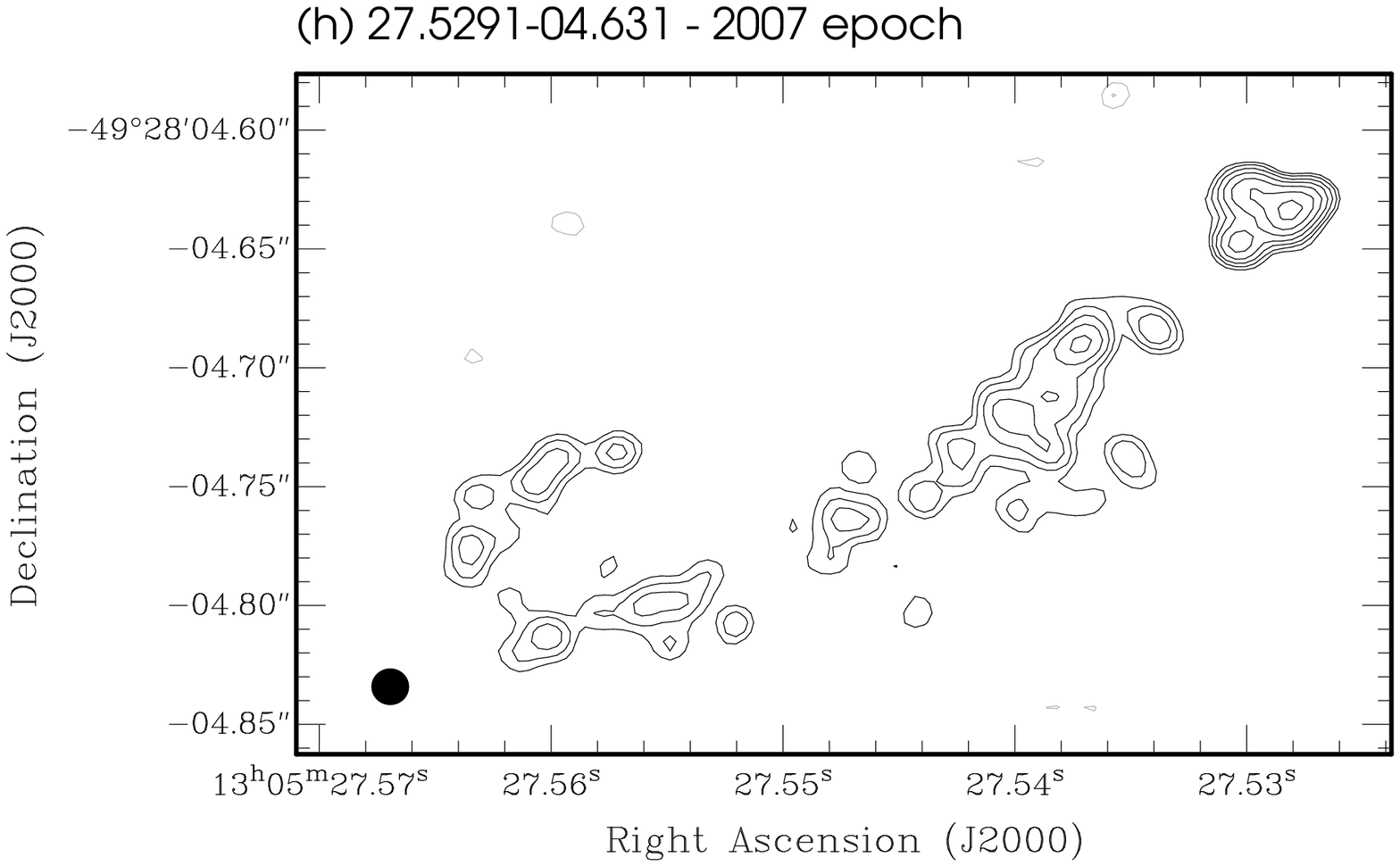} \quad
\plotone{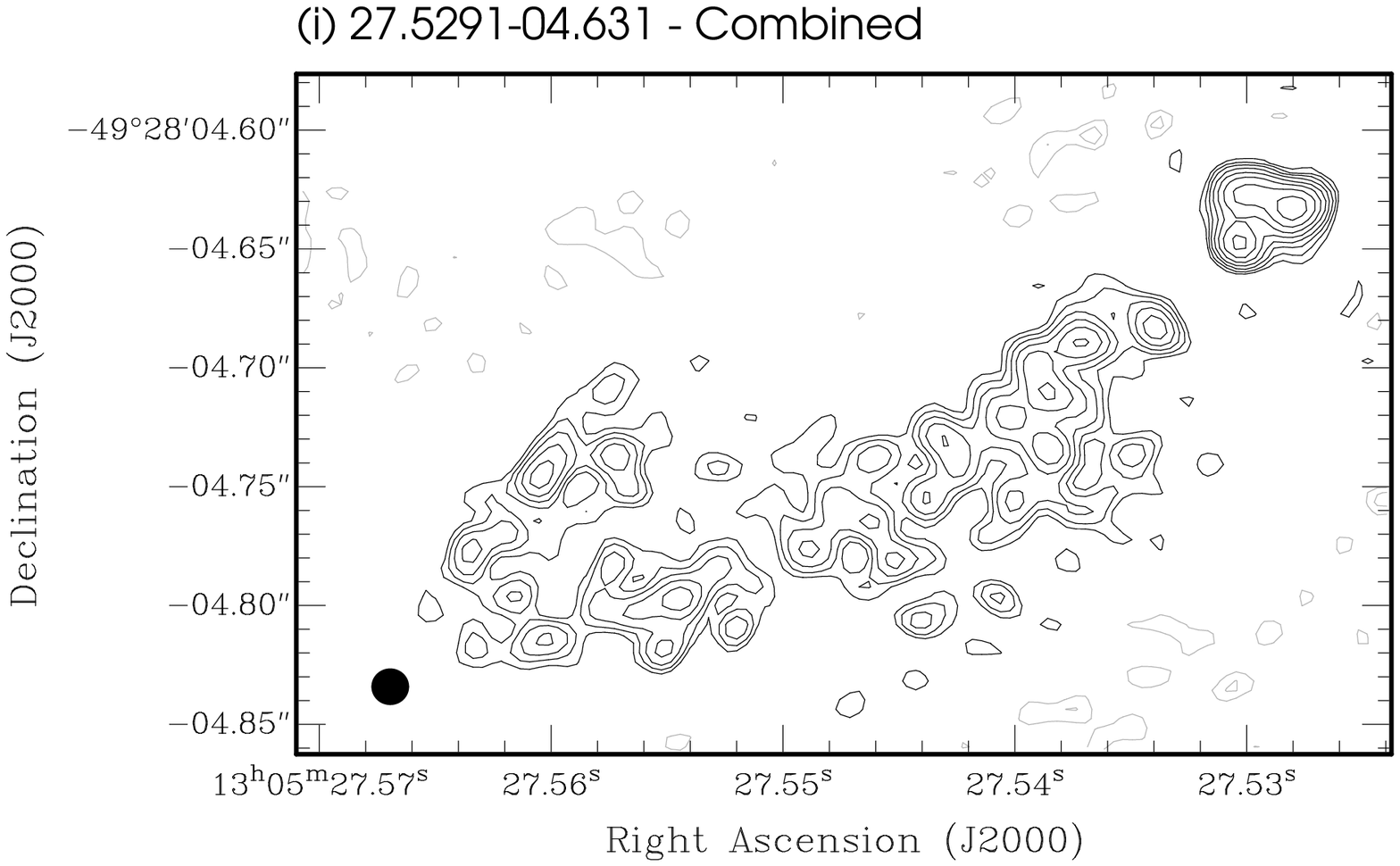}
}
\mbox{
\plotone{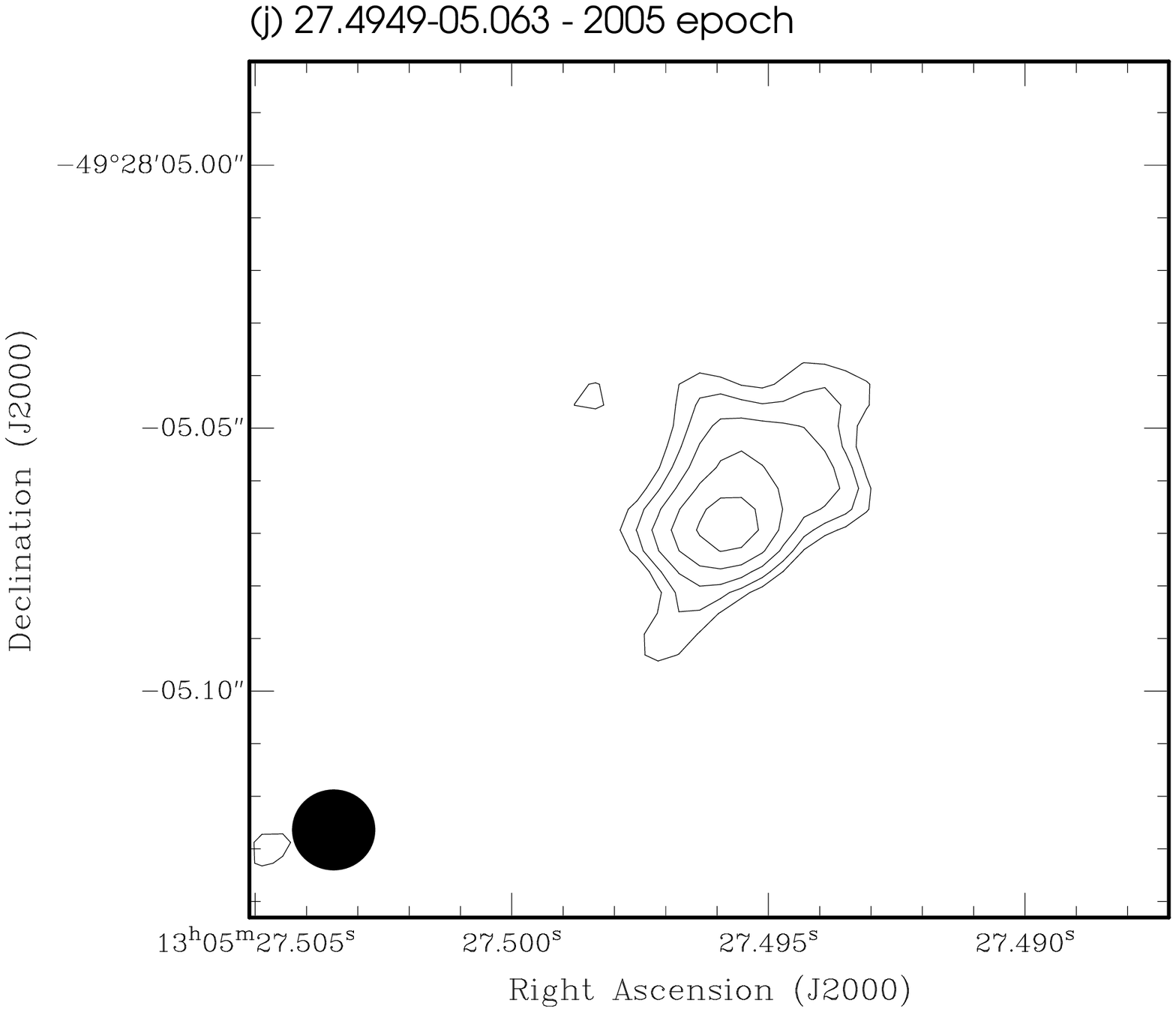} \quad
\plotone{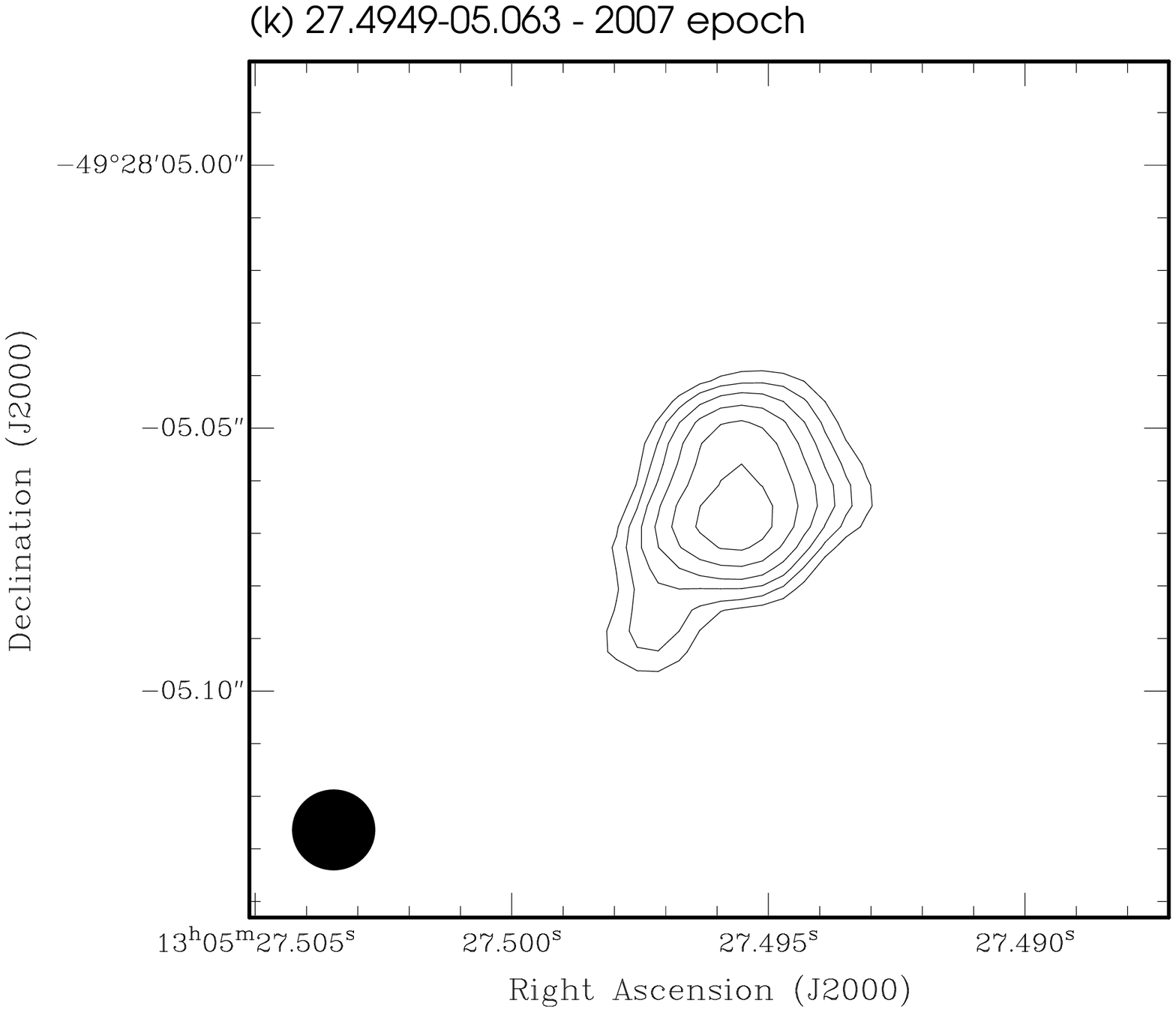} \quad
\plotone{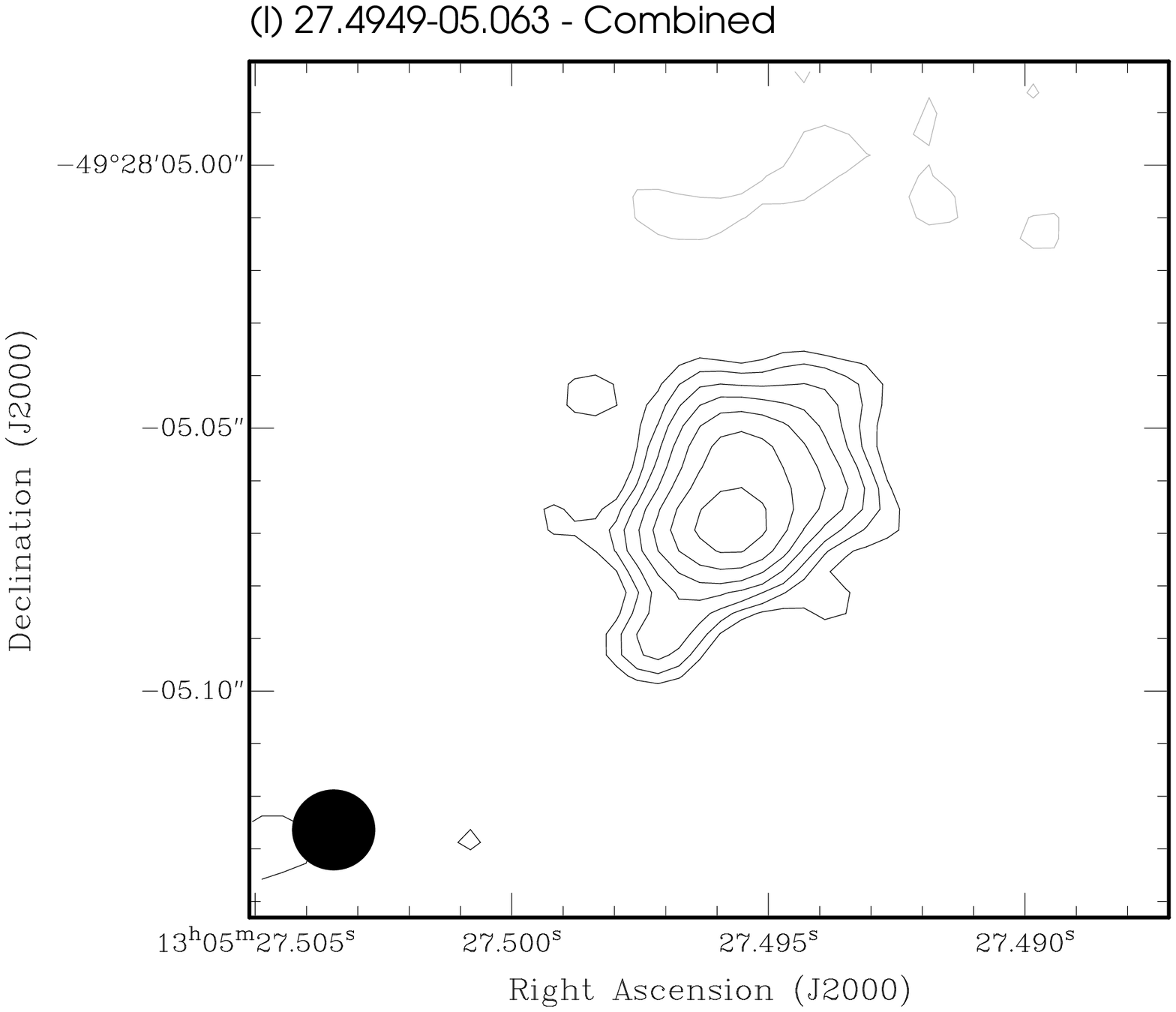}
}
\mbox{
\plotone{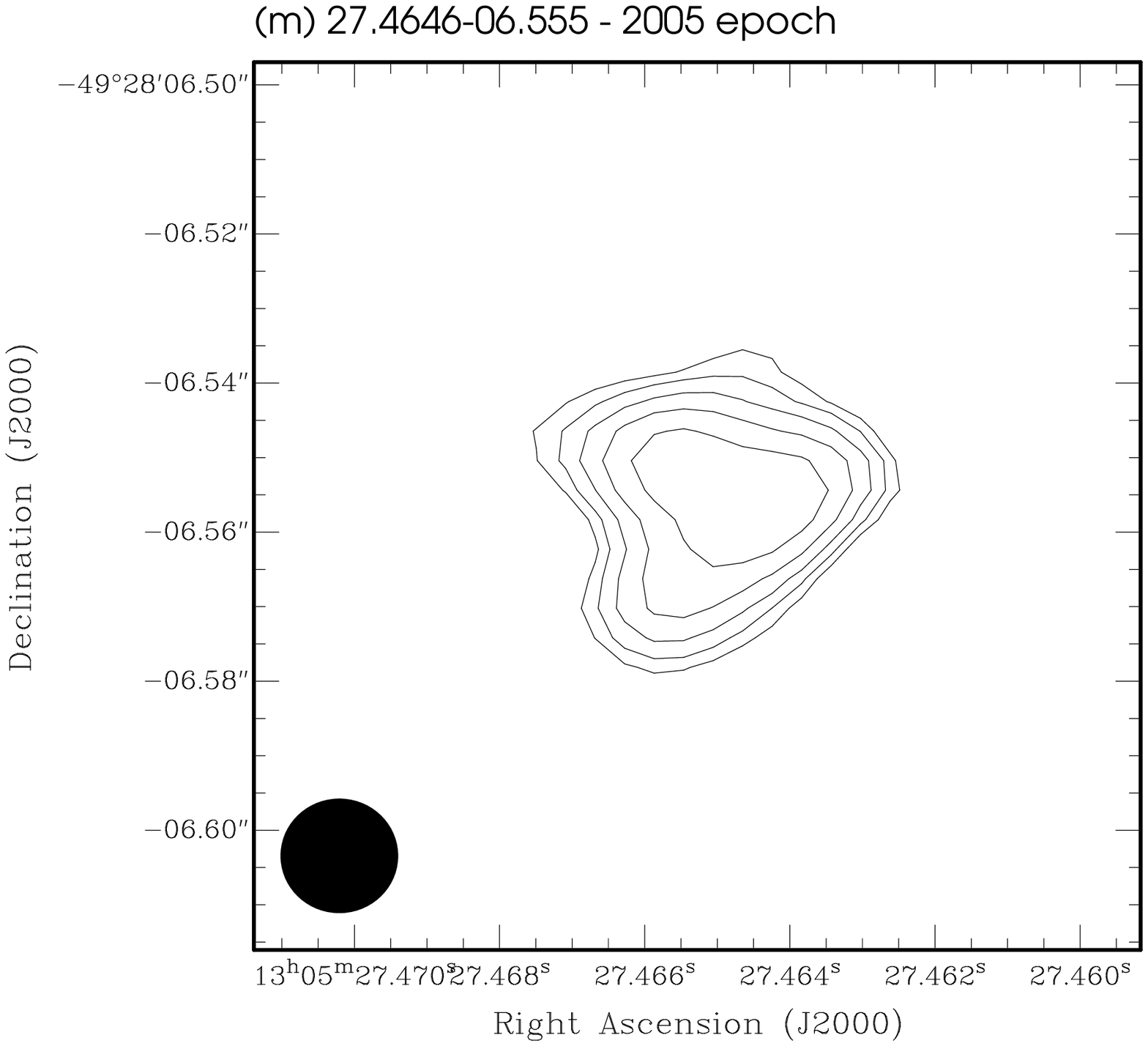} \quad
\plotone{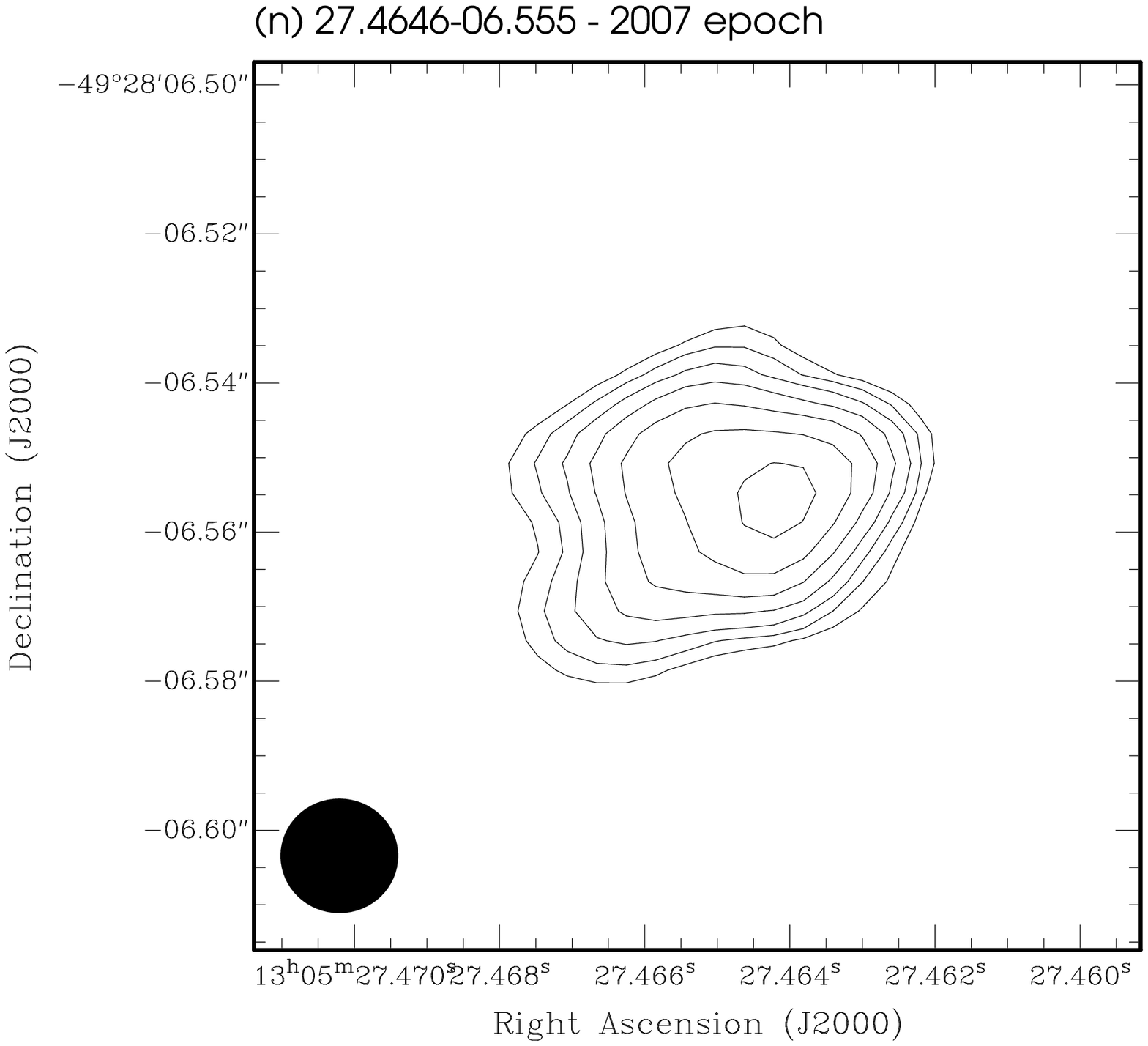} \quad
\plotone{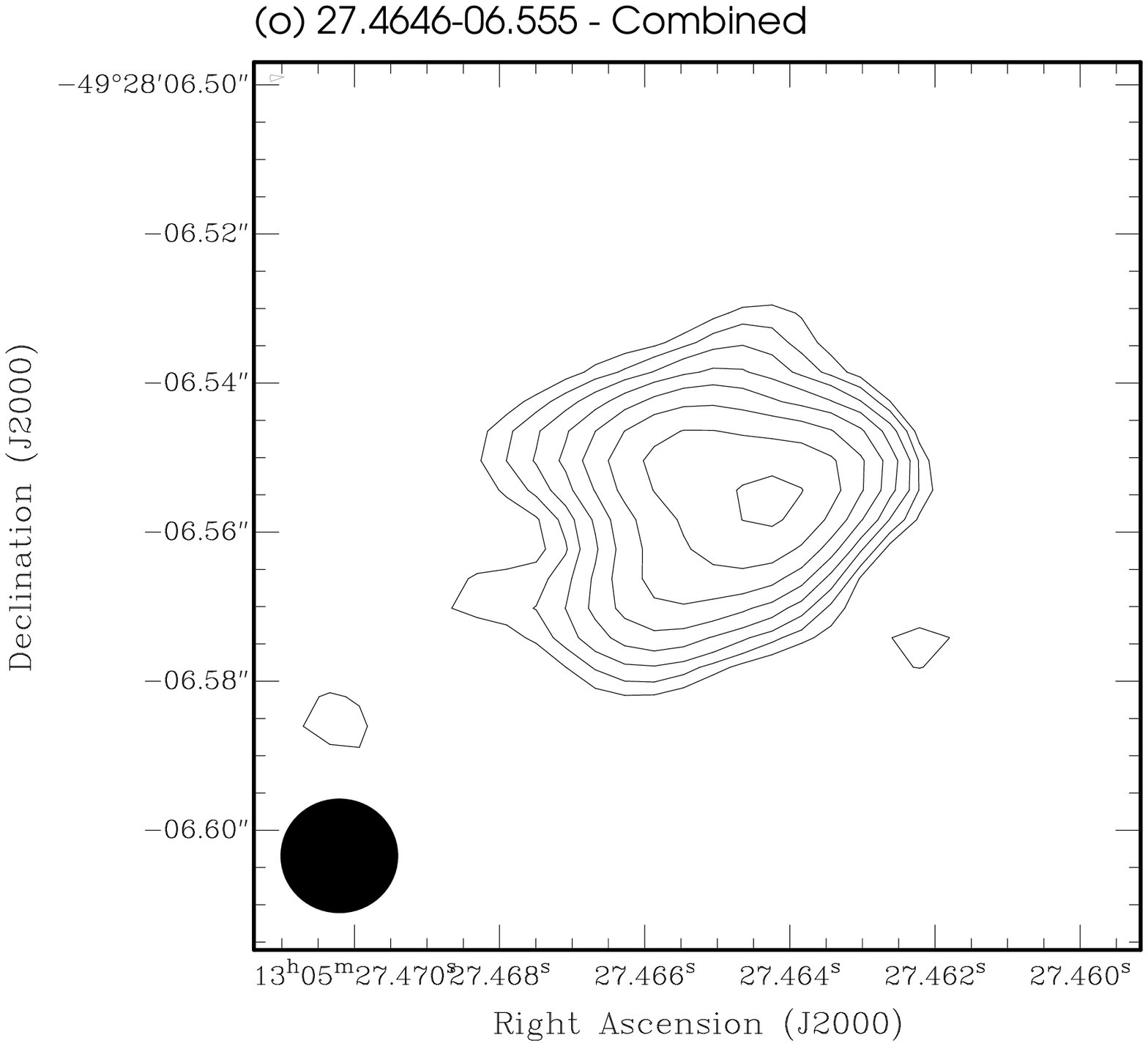}
}
\caption{Naturally-weighted total-power maps of compact sources detected in NGC 4945 using all six antennas of the LBA at 2.3 GHz. Map statistics for the individual maps are shown in Table \ref{tab:tabimage}. Contours are drawn at $\pm2^{0}, \pm2^{\frac{1}{2}}, \pm2^{1}, \pm2^{\frac{3}{2}}, \cdots$ times the $3\sigma$ rms noise for all maps.}
\label{fig:fighra}            
\end{center}
\end{figure}

\setcounter{figure}{5}
\begin{figure}
\epsscale{0.3}
\begin{center}
\mbox{
\plotone{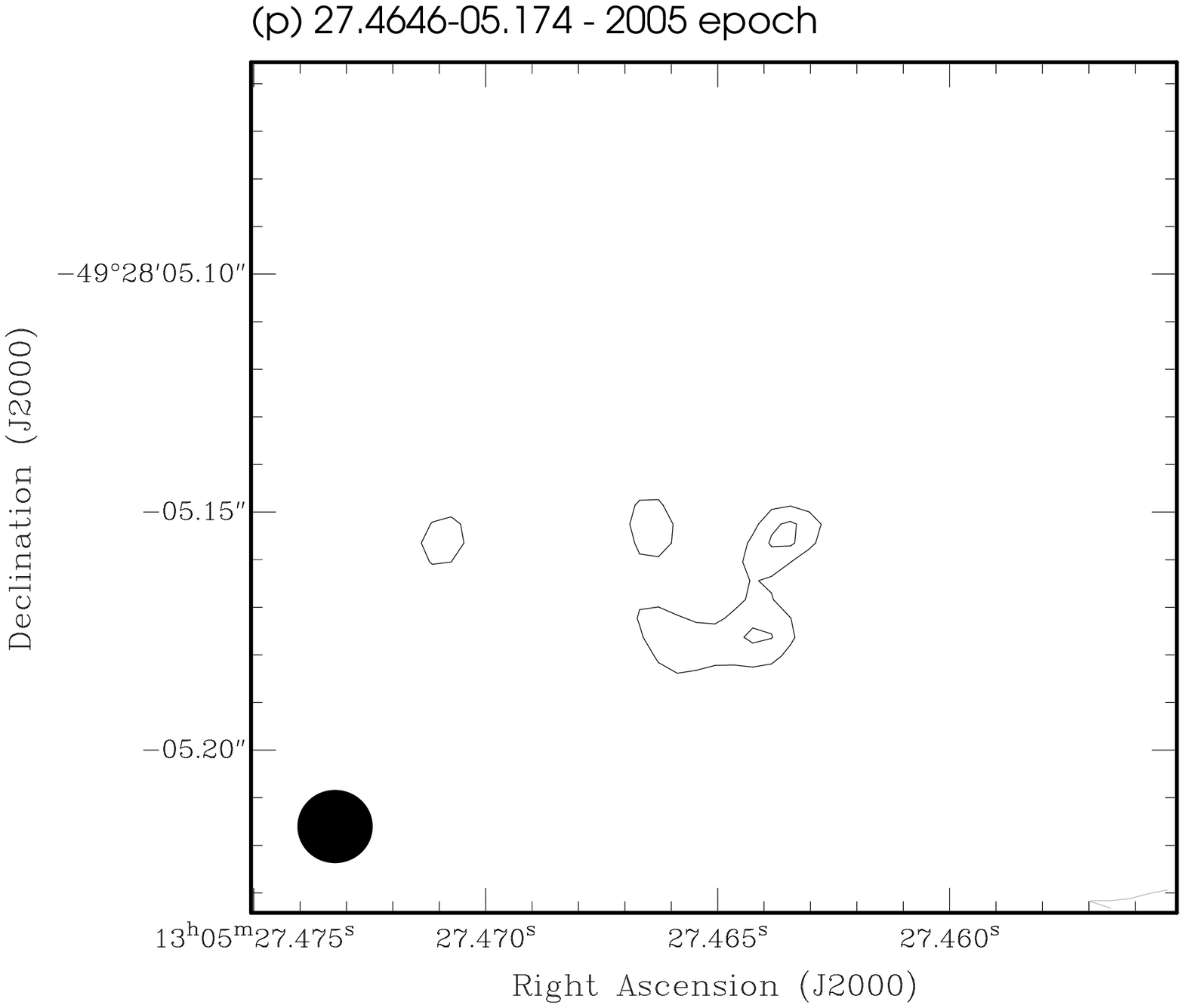} \quad
\plotone{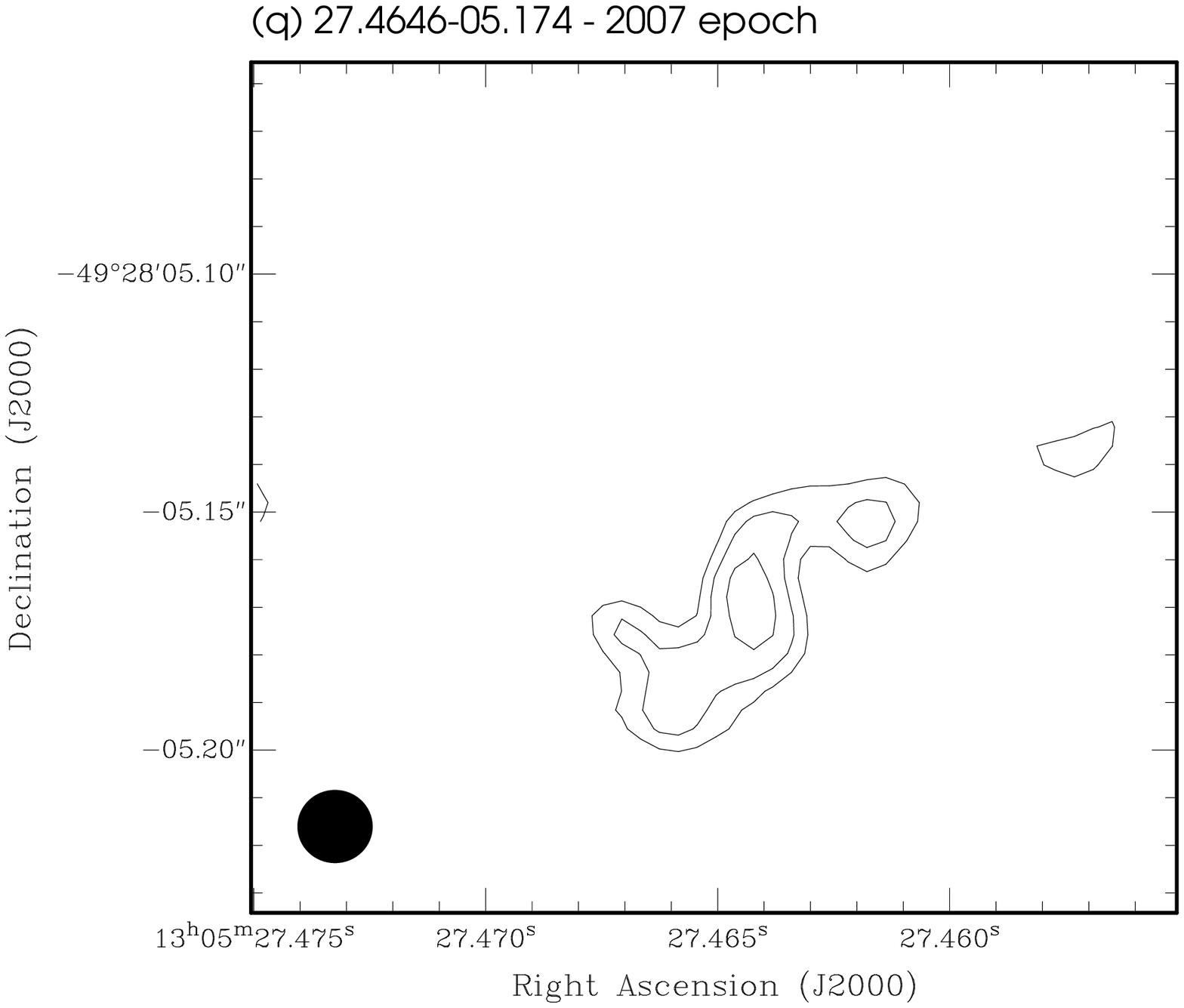} \quad
\plotone{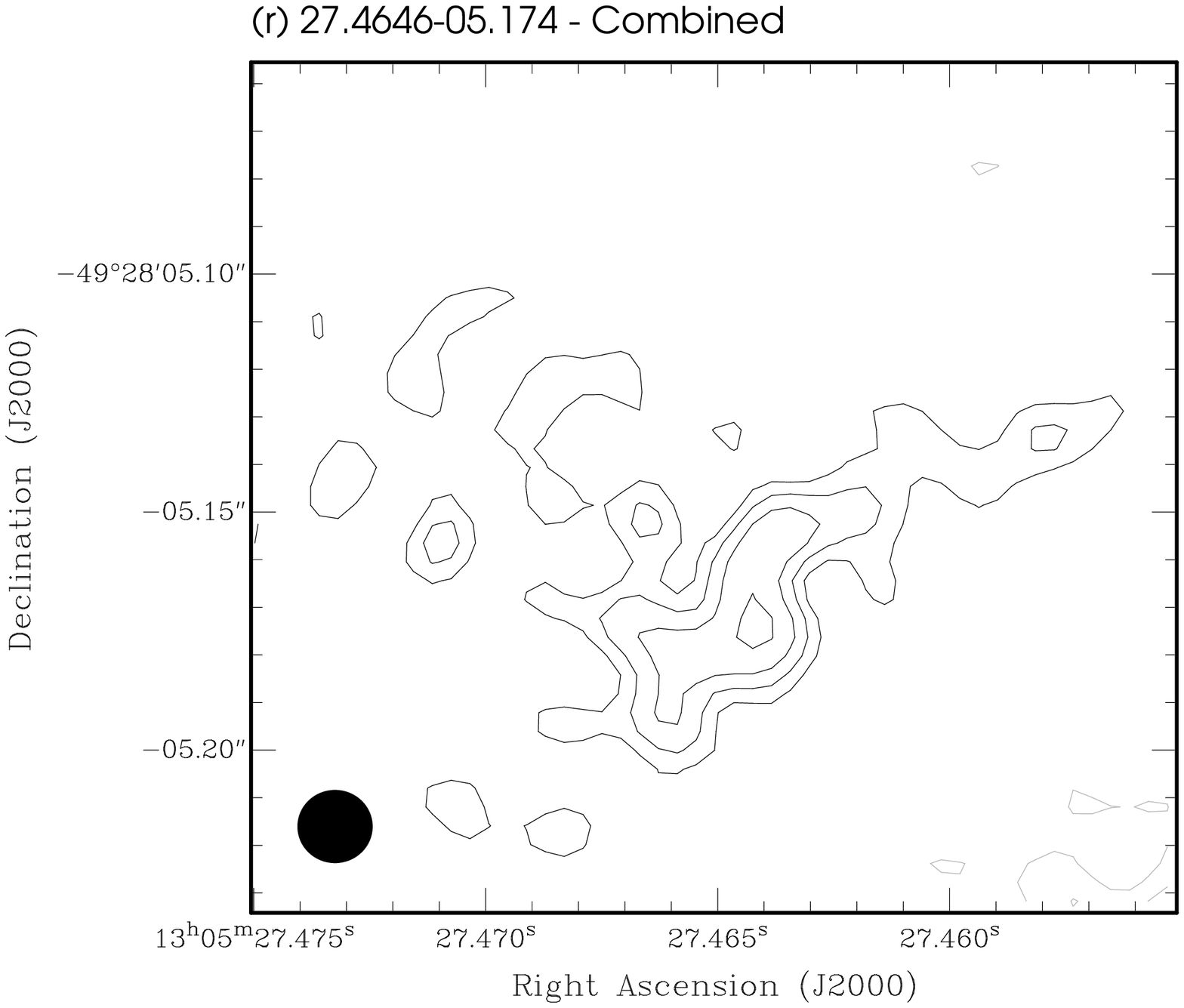}
}
\mbox{
\plotone{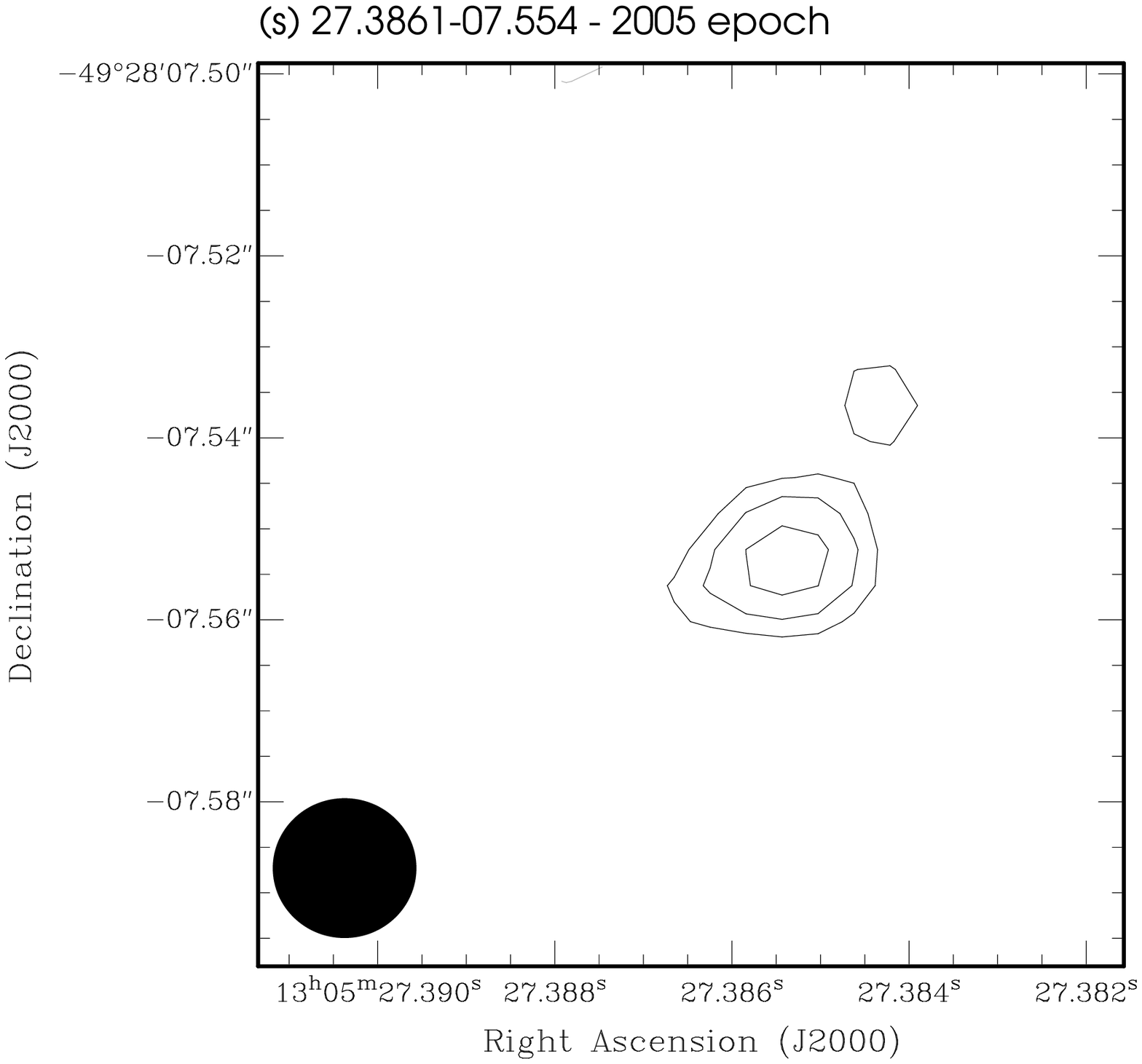} \quad
\plotone{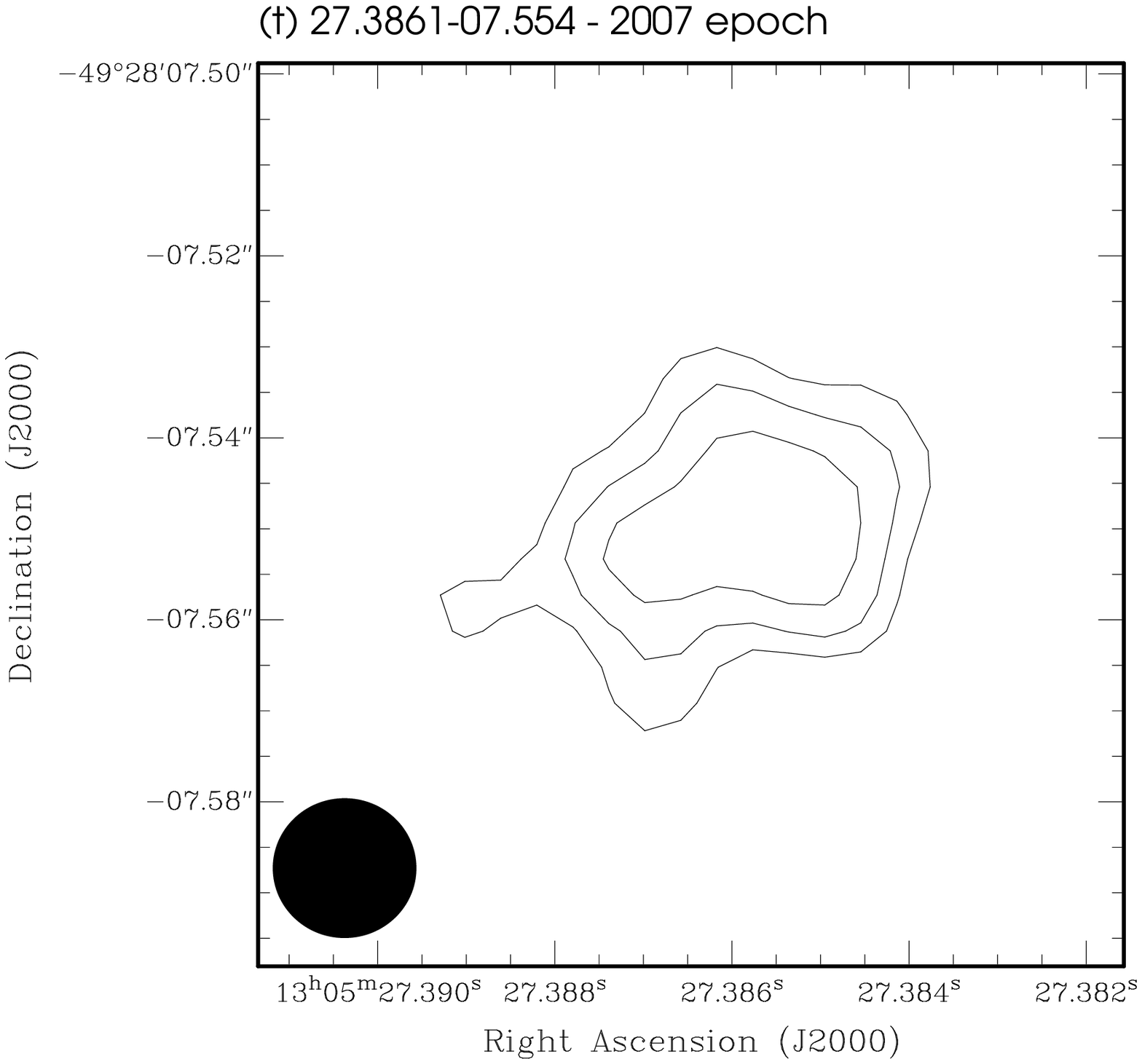} \quad
\plotone{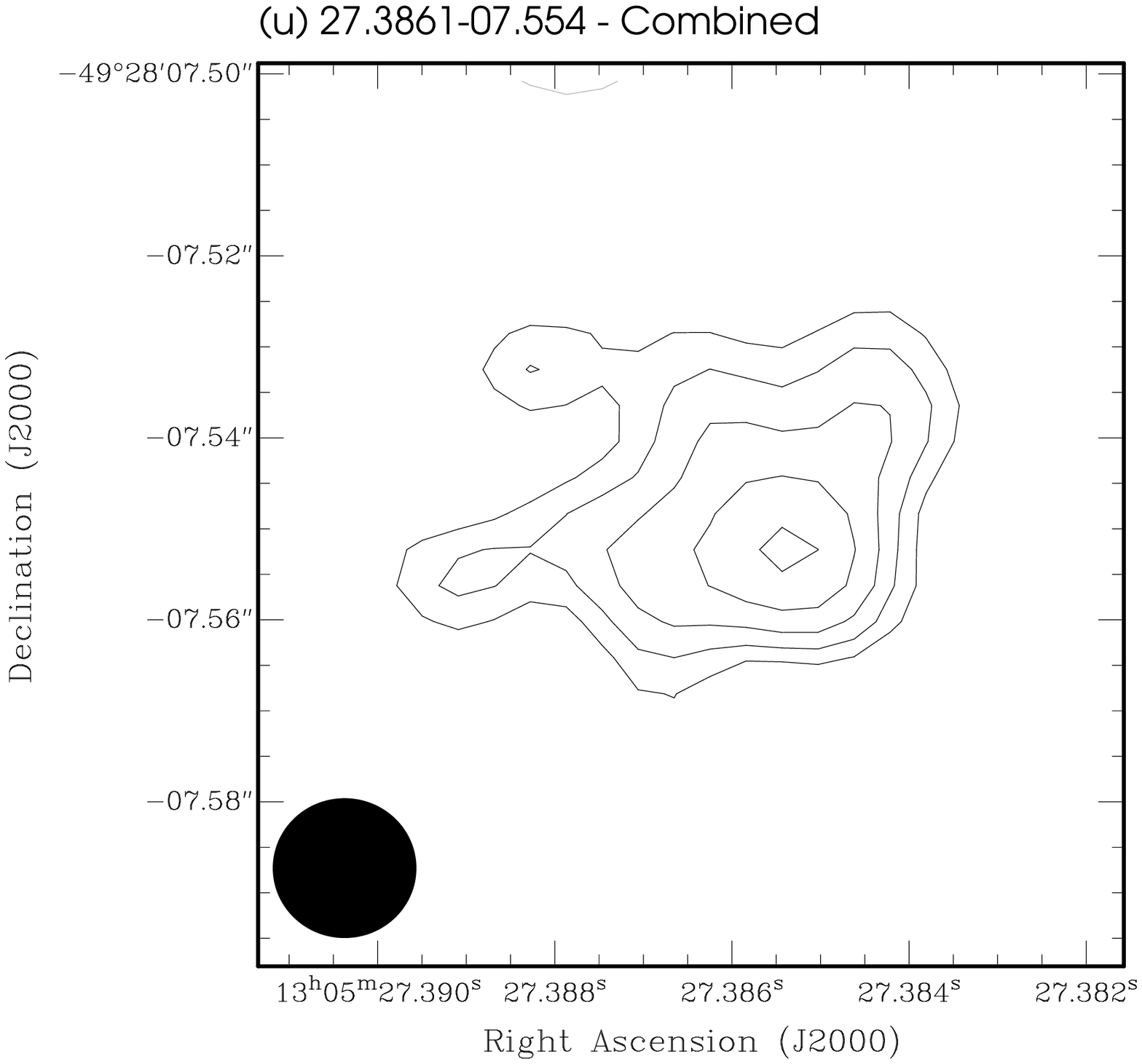}
}
\mbox{
\plotone{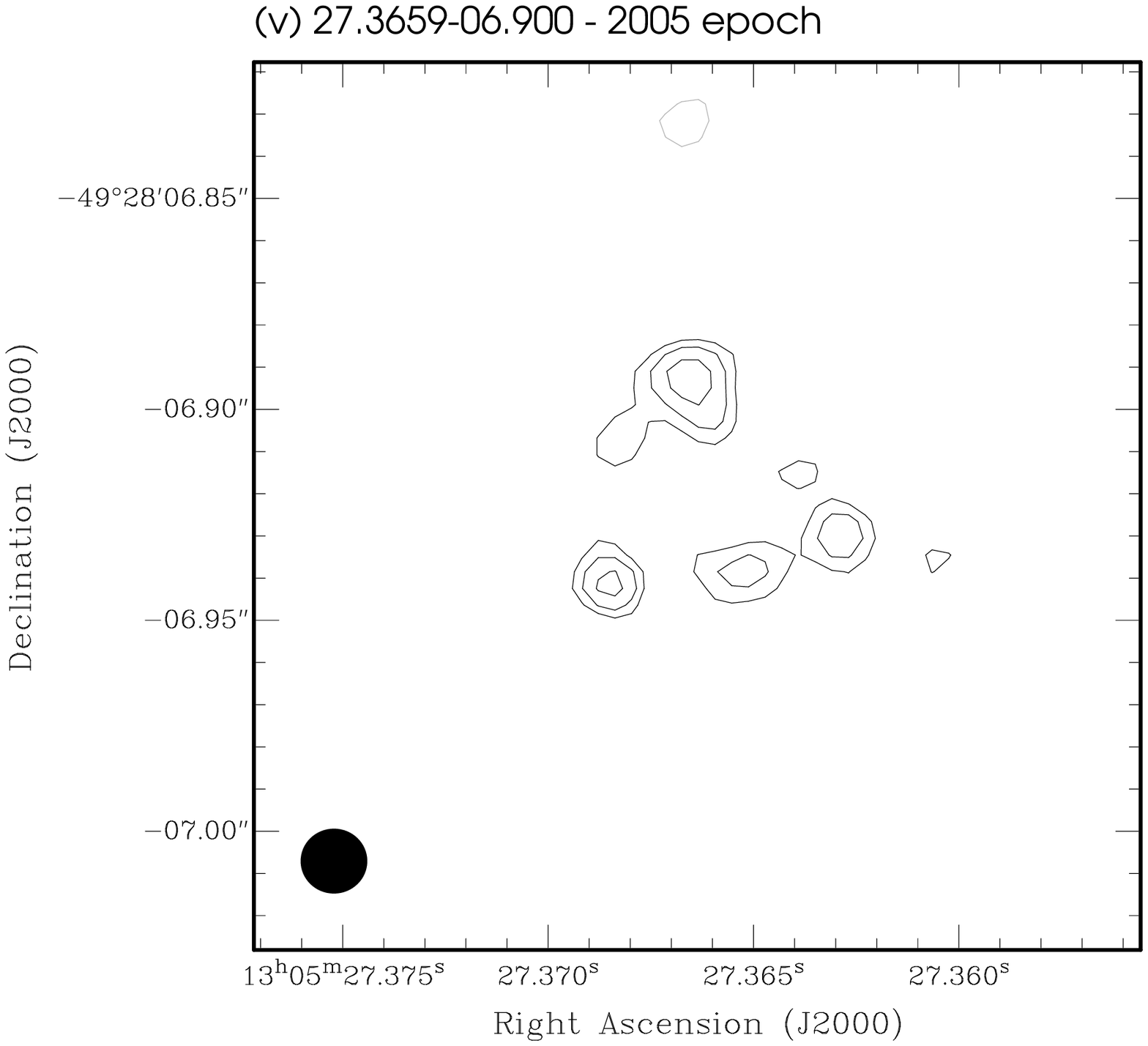} \quad
\plotone{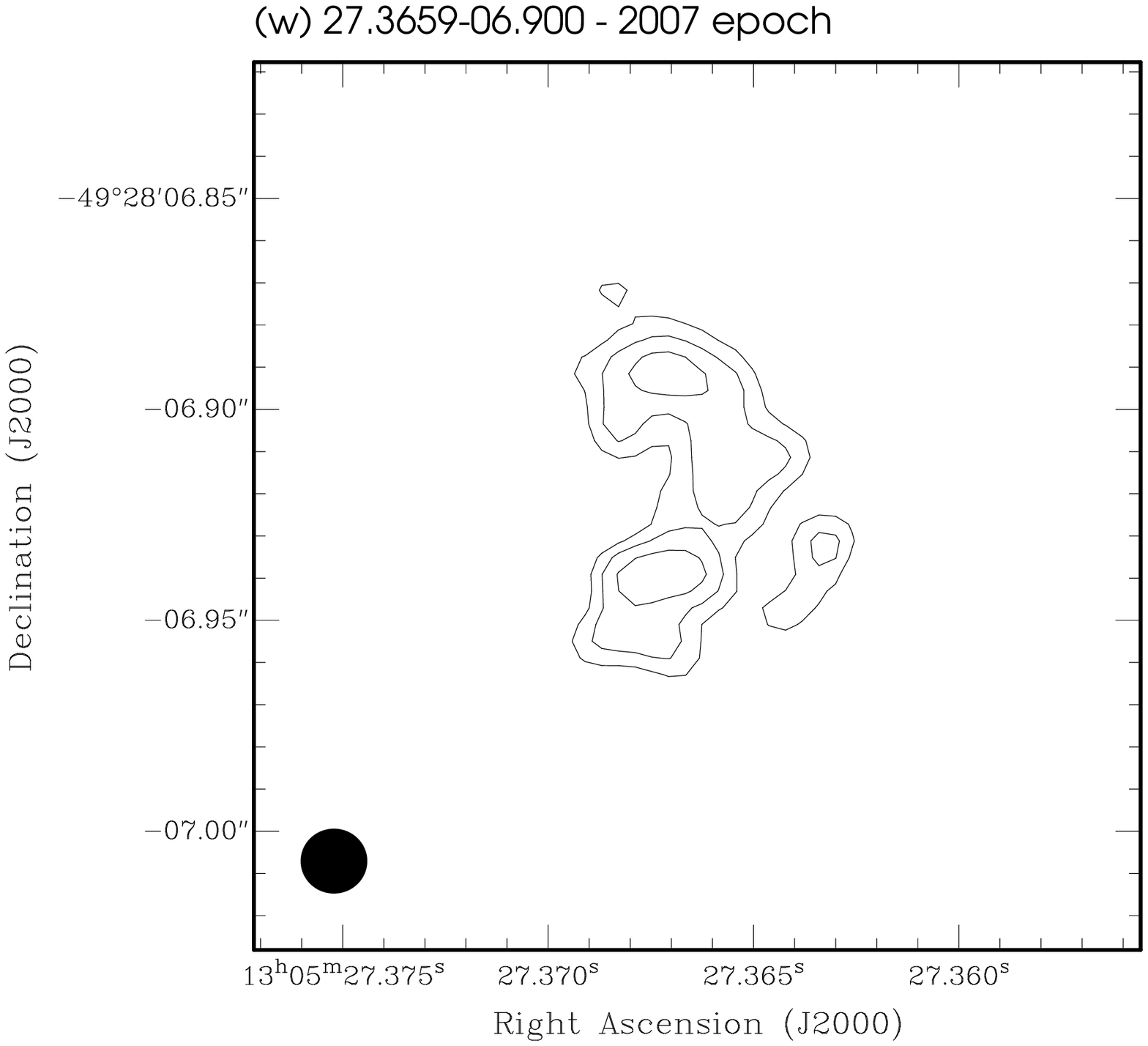} \quad
\plotone{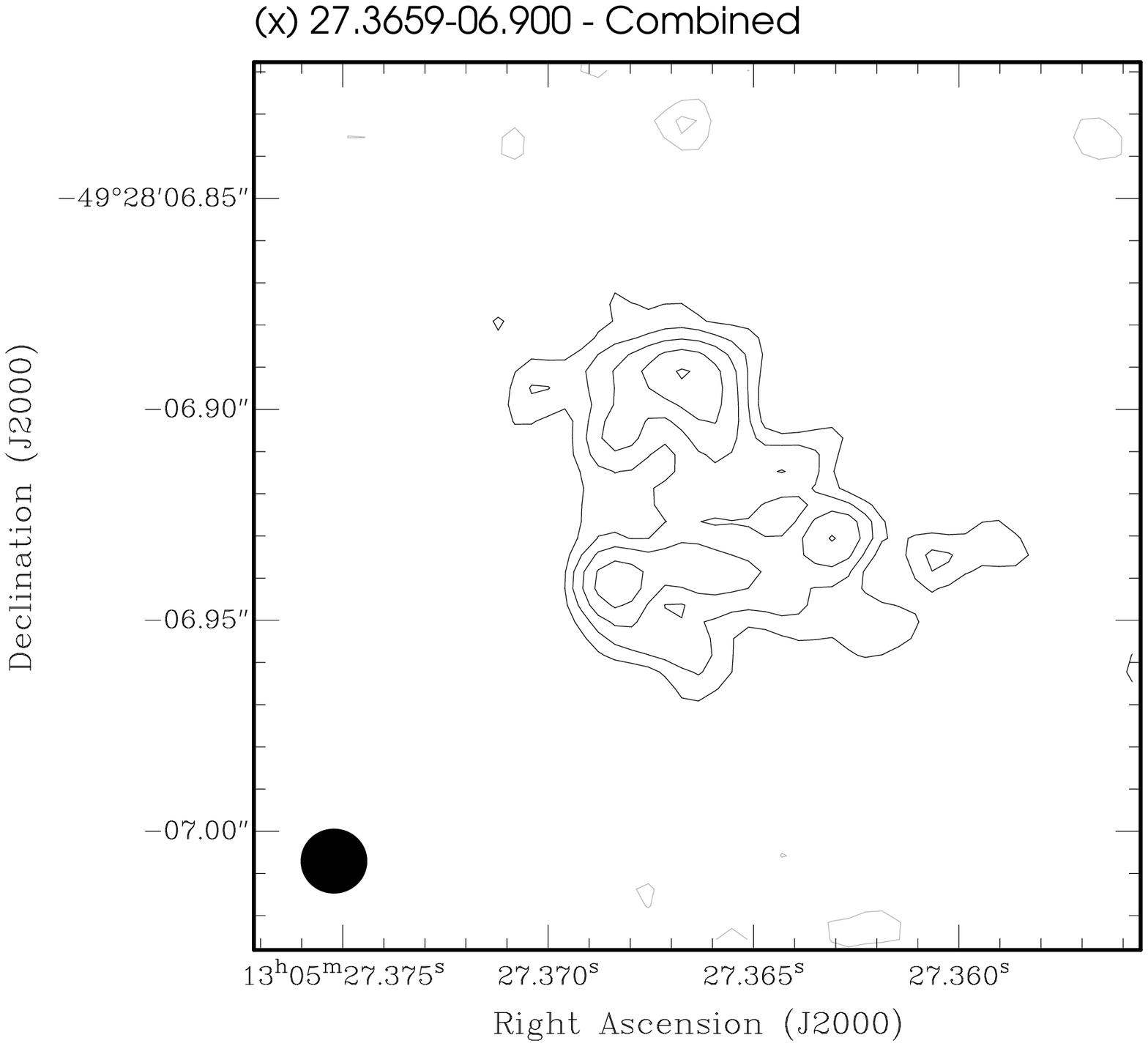}
}
\mbox{
\plotone{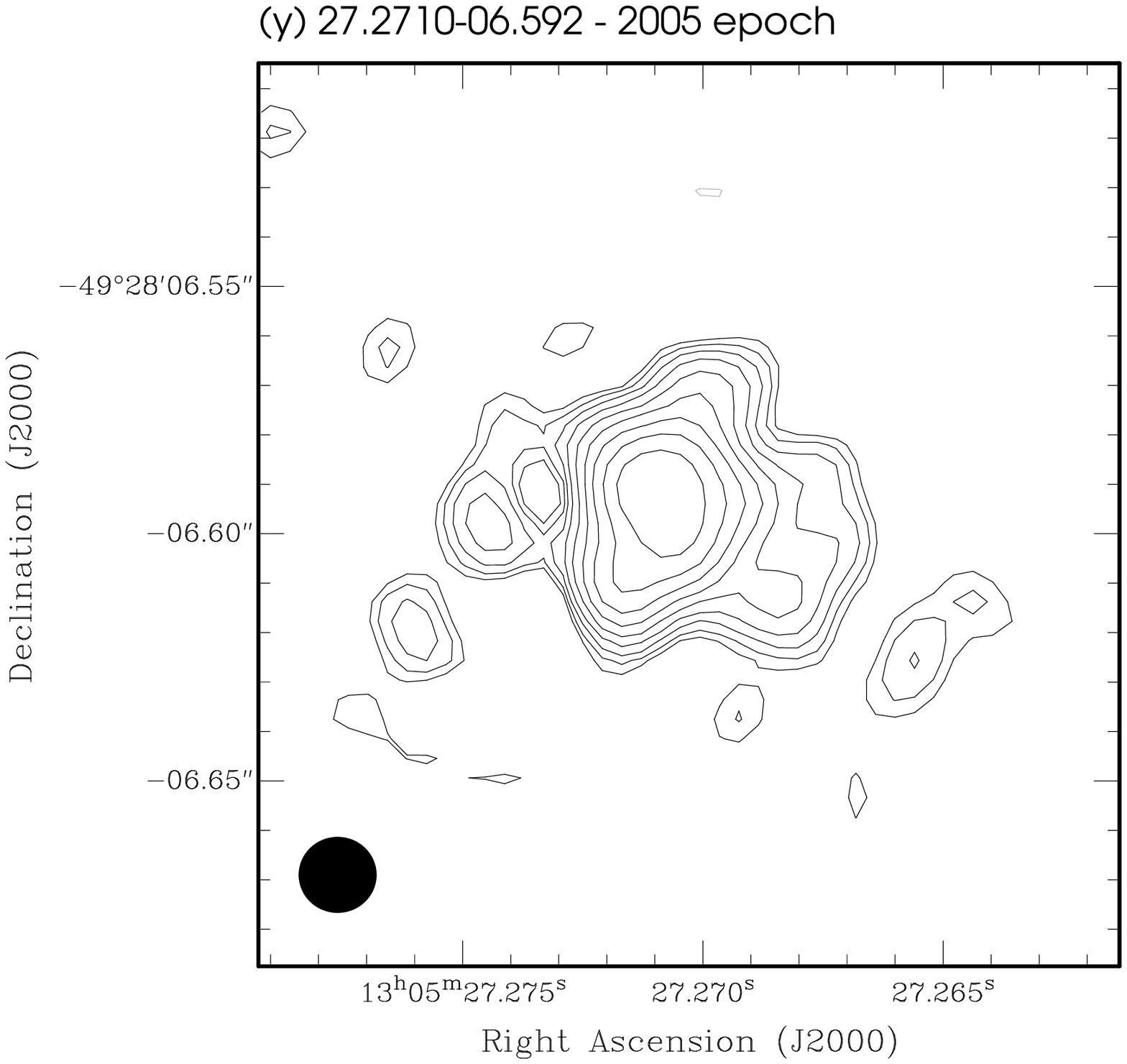} \quad
\plotone{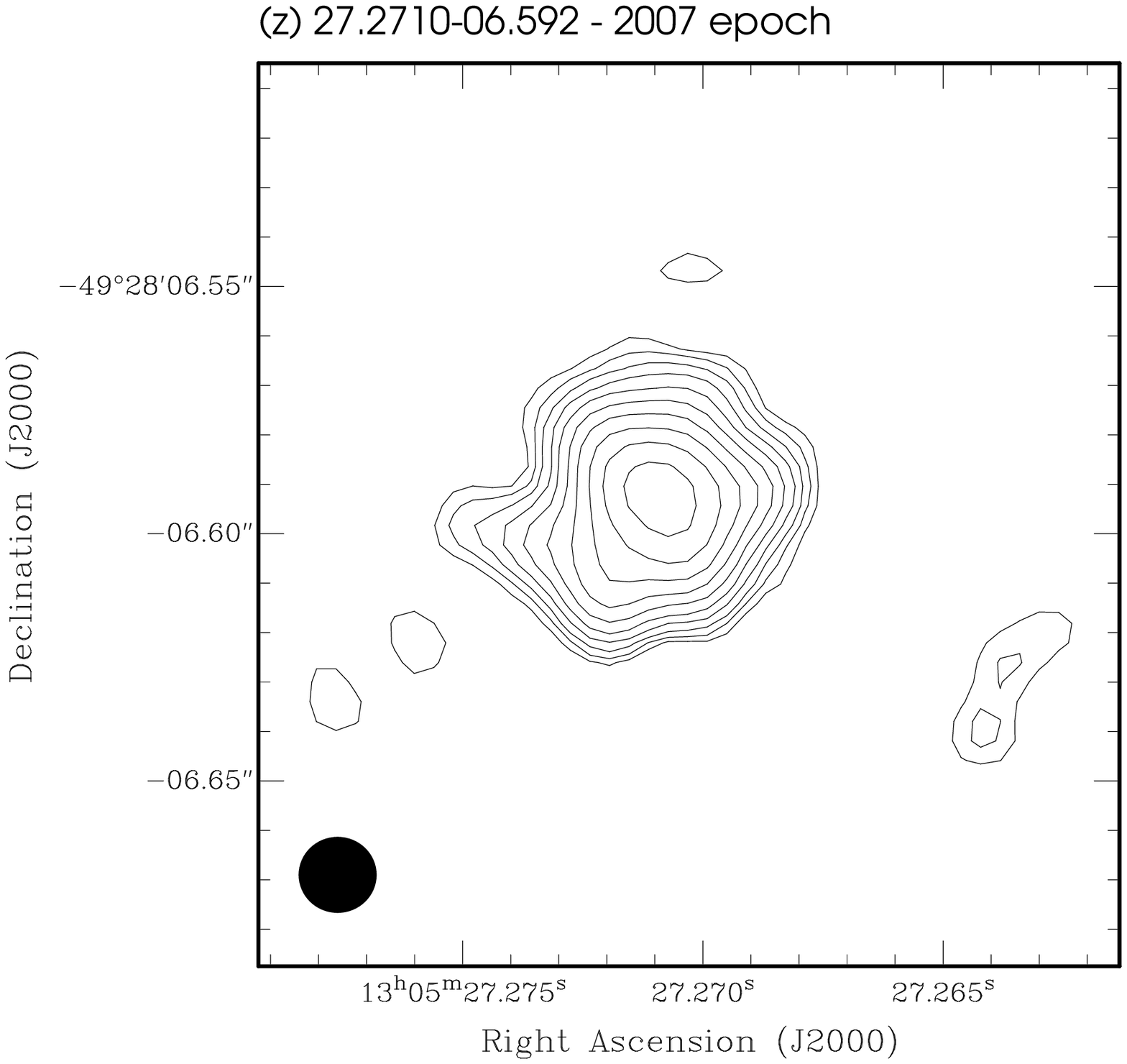} \quad
\plotone{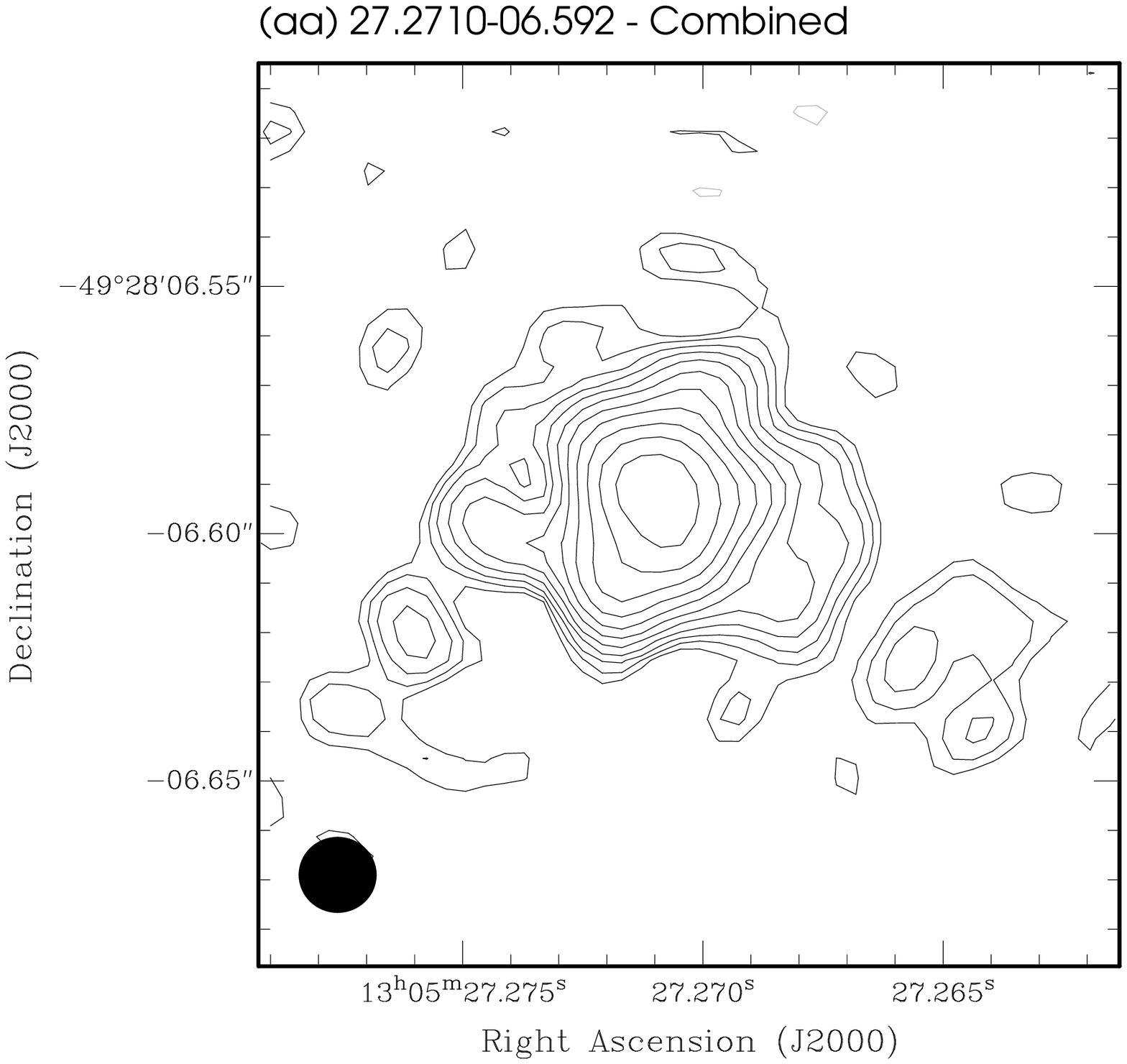}
}
\caption{Continued.}
\label{fig:fighrb}            
\end{center}
\end{figure}

\begin{figure}
\epsscale{0.9}
\begin{center}
\plotone{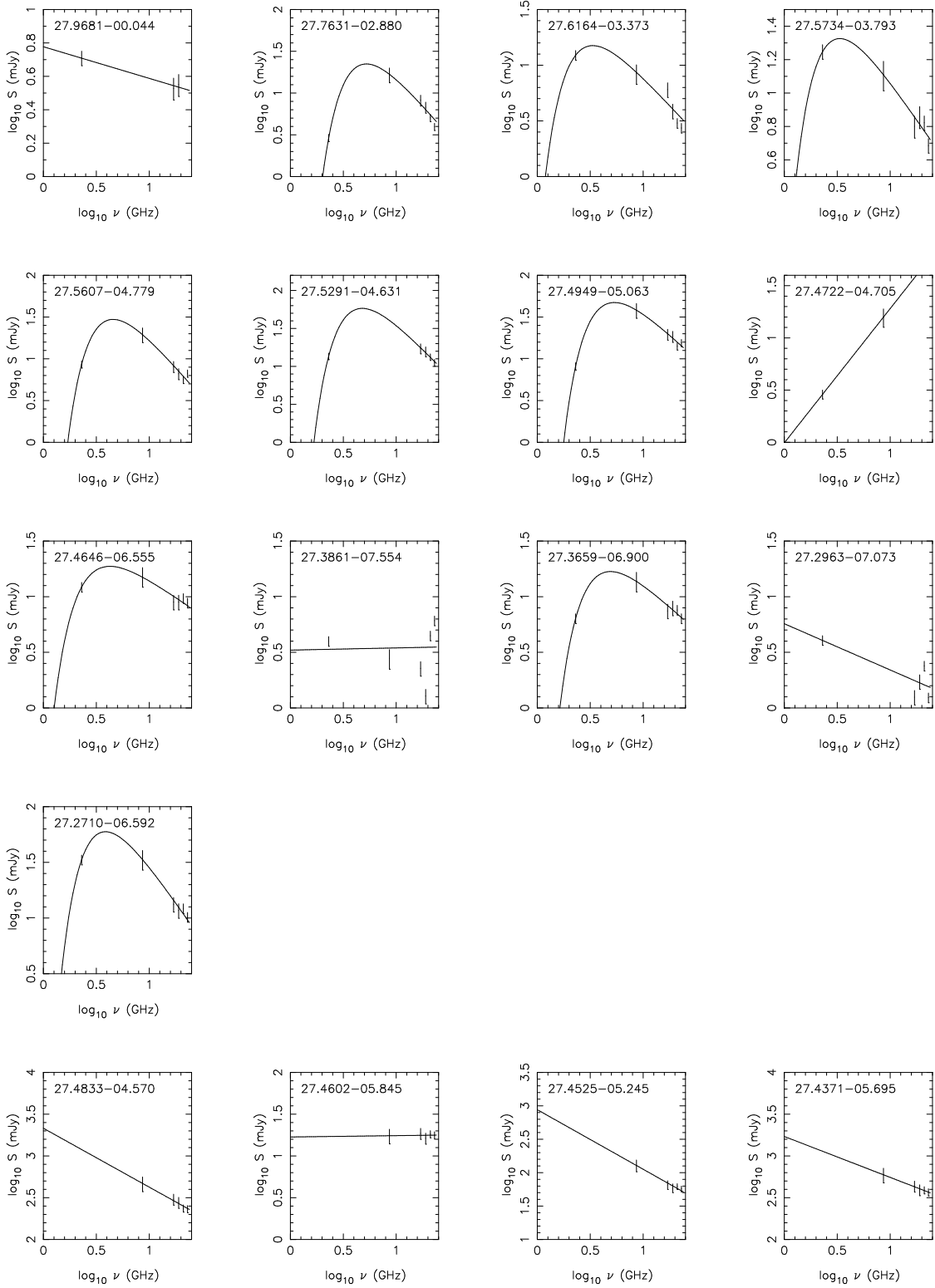}
\caption{Measured flux densities (symbols with error bars) and free-free absorption models (solid lines) for 17 detected sources (upper plots) and four diffuse sources detected with the ATCA data (lower plots). The error bars are 10\% of the measured flux densities except at 8.64 GHz where they are 20\% of the measured flux density.}
\label{fig:figff}
\end{center}
\end{figure}

\end{document}